\documentclass[a4paper,reqno]{amsart}

\usepackage[table,xcdraw,dvipsnames]{xcolor}

\usepackage{comment}
\usepackage{float}
\usepackage{colortbl}
\usepackage[T1]{fontenc}
\usepackage[margin=2cm]{geometry}
\usepackage[foot]{amsaddr}
\usepackage[colorlinks, linkcolor = blue, citecolor = blue, urlcolor = blue]{hyperref}

\usepackage{amsmath,amsthm,amssymb,amsfonts}
\usepackage{ytableau}
\usepackage{pifont}

\usepackage[normalem]{ulem}

\usepackage{mathtools}

\usepackage{thmtools}

\usepackage{mathrsfs}

\usepackage{cleveref}

\usepackage[shortlabels]{enumitem}

\usepackage{tabularx}
\usepackage{makecell}
\usepackage{multirow}
\usepackage{hhline}
\usepackage{diagbox}
\usepackage{booktabs}
\usepackage{slashbox}
\usepackage{cases}

\usepackage{caption}

\usepackage[group-separator={,},group-minimum-digits=4]{siunitx}

\usepackage{epigraph} 

\usepackage{subcaption}  

\usepackage{graphicx}
\usepackage{tikz}
\usetikzlibrary{quantikz2}
\usetikzlibrary{backgrounds}
\usetikzlibrary{calc}
\usetikzlibrary{shapes}
\usetikzlibrary{decorations.pathreplacing}
\usetikzlibrary{external}

\usepackage{nicematrix}
\NiceMatrixOptions{cell-space-top-limit=5pt,cell-space-bottom-limit=5pt}
\usepackage{nicefrac}

\usepackage{ytableau,varwidth}

\usepackage[backend=biber,style=alphabetic,maxnames=9,maxalphanames=5,isbn=false,backref]{biblatex}

\usepackage{adjustbox}

\newtheorem*{theorem*}{Theorem}
\newtheorem{theorem}{Theorem}
\newtheorem{lemma}[theorem]{Lemma}

\newtheorem{corollary}[theorem]{Corollary}
\newtheorem{remark}[theorem]{Remark}
\newtheorem{example}[theorem]{Example}

\crefname{lemma}{Lemma}{Lemmas}
\crefname{definition}{Definition}{Definitions}
\crefname{theorem}{Theorem}{Theorems}
\crefname{conjecture}{Conjecture}{Conjectures}
\crefname{section}{Section}{Sections}
\crefname{claim}{Claim}{Claims}
\crefname{appendix}{Appendix}{Appendices}
\crefname{figure}{Fig.}{Figs.}
\crefname{table}{Table}{Tables}
\crefname{proposition}{Proposition}{Propositions}
\crefname{corollary}{Corollary}{Corollaries}
\crefname{example}{Example}{Examples}
\crefname{remark}{Remark}{Remarks}

\providecommand\given{}
\newcommand\SetSymbol[1][]{%
    \nonscript\:#1\vert
    \allowbreak
    \nonscript\:
    \mathopen{}}
\DeclarePairedDelimiterX\Set[1]\{\}{%
    \renewcommand\given{\SetSymbol[\delimsize]}
    #1
}

\DeclarePairedDelimiter{\set}{\lbrace}{\rbrace}
\DeclarePairedDelimiter{\abs}{\lvert}{\rvert}

\DeclarePairedDelimiter{\of}{\lparen}{\rparen}
\DeclarePairedDelimiter{\sof}{\lbrack}{\rbrack}

\newcommand{\defeq}{\vcentcolon=}

\renewcommand{\leq}{\leqslant}
\renewcommand{\geq}{\geqslant}

\renewcommand{\bra}[1]{\langle{#1}\rvert}
\renewcommand{\ket}[1]{\lvert{#1}\rangle}

\newcommand{\ketbra}[2]{\ket{#1}\bra{#2}}
\renewcommand{\proj}[1]{\ketbra{#1}{#1}}

\newcommand{\x}{\otimes}
\newcommand{\xp}[1]{^{\otimes #1}}

\newcommand{\Z}{\mathbb{Z}} 
\newcommand{\C}{\mathbb{C}} 

\DeclareMathOperator{\poly}{poly} 

\newcommand{\0}{\varnothing}

\let\S\relax
\DeclareMathOperator{\S}{S} 
\DeclareMathOperator{\CS}{\mathbb{C}S} 
\DeclareMathOperator{\GL}{GL} 
\DeclareMathOperator{\End}{End} 
\DeclareMathOperator{\Rep}{Rep} 
\DeclareMathOperator{\Tr}{Tr} 
\DeclareMathOperator{\spn}{span} 
\DeclareMathOperator{\CG}{CG} 
\DeclareMathOperator{\Prep}{Prep} 
\DeclareMathOperator{\QFT}{QFT}

\newcommand{\SU}[1]{\mathrm{SU}_{#1}} 
\newcommand{\U}[1]{\mathrm{U}_{#1}} 

\newcommand{\cH}{\mathcal{H}}
\newcommand{\cW}{\mathcal{W}}
\newcommand{\cP}{\mathcal{P}}

\newcommand{\USch}{U_\mathrm{Sch}} 

\newcommand{\GT}{\mathrm{GT}} 

\newcommand{\pt}{\mathbin{\vdash}} 
\newcommand{\AC}{\mathrm{AC}} 

\newcommand{\SYT}{\mathrm{SYT}} 
\newcommand{\SSYT}{\mathrm{SSYT}} 

\newcommand{\w}{0.5cm}

\newcommand{\bx}[3]{
  \draw[fill = white] #3 (#1*\w-\w/2,-#2*\w-\w/2) rectangle (#1*\w+\w/2,-#2*\w+\w/2);
}

\newcommand{\yd}[2][0.4]{%
  \begin{tikzpicture}[scale = #1, baseline={([yshift=-0.6ex]current bounding box.center)}]
    \foreach \li [count = \y] in {#2} {
      \foreach \x in {1,...,\li} {
        \bx{\x}{\y}{}
      }
    }
  \end{tikzpicture}
}

\newcommand\restr[2]{{
  \left.\kern-\nulldelimiterspace 
  #1 
  \vphantom{\big|} 
  \right|_{#2} 
  }}

\newcommand{\syd}[1]{\yd[0.3]{#1}}

\newcommand{\lm}{\lambda}

\title[High-dimensional quantum Schur transforms]{High-dimensional quantum Schur transforms}

\author{Adam Burchardt$^{1,2}$}
\author{Jiani Fei$^{3}$}
\author{Dmitry Grinko$^{1,4,5}$}
\author{Martin Larocca$^{6}$}
\author{Maris Ozols$^{1,4,5}$}
\author{Sydney Timmerman$^{3}$}
\author{Vladyslav Visnevskyi$^{1,7}$}

\address[1]{QuSoft, Amsterdam}
\address[2]{Centrum Wiskunde \& Informatica, Amsterdam}
\address[3]{Stanford Institute for Theoretical Physics, Stanford University}
\address[4]{Institute for Logic, Language and Computation, University of Amsterdam}
\address[5]{Korteweg-de Vries Institute for Mathematics, University of Amsterdam}
\address[6]{Theoretical Division, Los Alamos National Laboratory}
\address[7]{Institute of Physics, University of Amsterdam}

\email{adam.burchardt.uam@gmail.com}
\email{jianif@stanford.edu}
\email{grinko.dimitry@gmail.com}
\email{laroccamartin@gmail.com}
\email{marozols@gmail.com}
\email{sydneyt@stanford.edu}
\email{vladislav.visnevskyy@gmail.com}

\begin{document}

\begin{abstract}

The quantum Schur transform has become a foundational quantum algorithm, yet even after two decades since the seminal 2005 paper by Bacon, Chuang, and Harrow (BCH), some aspects of the transform remain insufficiently understood.
Moreover, an alternative approach proposed by Krovi in 2018 was recently found to contain a crucial error.
In this paper, we present a corrected version of Krovi's algorithm along with a detailed treatment of the high-dimensional version of the BCH Schur transform.
This high-dimensional focus makes the two versions of the transform practical for regimes where the number of qudits $n$ is smaller than the local dimension $d$, with Krovi's algorithm scaling as $\widetilde{O}(n^4)$ and BCH as $\widetilde{O}(\min(n^5,nd^4))$. 
Our work addresses a key gap in the literature, strengthening the algorithmic foundations of a wide range of results that rely on Schur--Weyl duality in quantum information theory and quantum computation.
\end{abstract}

\maketitle
\setcounter{tocdepth}{1}
\tableofcontents

\section{Introduction}\label{sec:introduction}

\subsection{Background}

The \textit{Schur--Weyl duality} is a classic and fundamental result in representation theory. From an algebraic perspective, it is a theorem about two subalgebras of the full matrix algebra, one generated by the unitary group and another by the symmetric group. Specifically, Schur--Weyl duality says that the diagonal action of the unitary group $\U{d}$ on the tensor space $\mathcal{H} \defeq (\C^d)\xp{n}$ generates a matrix algebra that is the centralizer of the matrix algebra generated by the tensor representation of the symmetric group $\S_n$, and vice versa. In this sense, Schur--Weyl duality is a consequence of the double centralizer theorem \cite{etingof2011reptheory}.

However, from a representation-theoretic perspective the Schur--Weyl duality presents a much deeper argument about relating finite-dimensional irreducible representations (irreps) of the unitary and the symmetric groups. In particular, it claims that the tensor space $\mathcal{H}$ decomposes into a direct sum of tensor products, each formed between a pair of irreducible modules: one of the unitary group and one of the symmetric group. Each such pair carries a shared irrep label, thus relating the irreps of the two groups.

Any basis in which the tensor space $\mathcal{H}$ decomposes into such a direct sum over tensor products of irreducible modules of the unitary and symmetric groups is called a \textit{Schur basis}, and any unitary transformation from the standard computational basis to such basis is called a \textit{Schur transform}.

One notable consequence of the decomposition in a Schur basis is that the symmetric group and the unitary group act on different modules, and thus on different registers in the tensor product pairs. This separation makes Schur transform a powerful tool for quantum algorithms and protocols that
exploit either permutation or unitary symmetries, and is precisely the reason why it has found so many applications in quantum information and quantum computing~\cite{Harrow2005,Wright2016,Botero2017}.

\subsection{Applications}\label{sec:lit_overview}

A fundamental problem in quantum information theory is operationally estimating a quantum state, assuming access to identical copies of its density matrix. This naturally fits into the Schur--Weyl duality framework.
To exemplify, first, performing \textit{weak Schur sampling}, i.e.\ a measurement of the irrep label, can reveal the spectral information of the state, based on a quantum analogue of typical sequences~\cite{Harrow2005}. Instances include spectrum estimation~\cite{Keyl2001,Christandl2004}, quantum property testing~\cite{Montanaro2018,ODonnell2021,Soleimanifar2022}, entropy estimation~\cite{Acharya2020}, and an approach to the non-abelian hidden subgroup problem~\cite{Childs2007}.
Second, optimal protocols for full state tomography~\cite{KEYL2006,Haah2017,O’Donnell2016,O’Donnell2017,Hu2024} rely on \textit{strong Schur sampling}, i.e.\ measuring one of the irrep registers in a particular basis.
Third, in the context of quantum state discrimination, the asymptotic error exponents of quantum hypothesis testing~\cite{Cheng2024} are quantified by quantum Stein's lemma~\cite{Hayashi1997,Hayashi2002a}, quantum Chernoff bound~\cite{Audenaert2007} and quantum Sanov's theorem~\cite{Noetzel2014,Hayashi2024}. These minimum error probabilities can be expressed by entropic quantities and attained algorithmically by strong Schur sampling with measurement operators based on semi-definite programs~\cite{Cheng2024}, pretty-good measurement~\cite{Cheng2024} or representation-theoretical arguments~\cite{Hayashi1997,Hayashi2002a,Noetzel2014,Hayashi2024}.

Besides state characterization, strong Schur sampling can be used to compute expectation values $\Tr(O\rho)$ of permutation-invariant observables $O$ or permutation operators $O$. Virtual cooling~\cite{Cotler2019} and virtual distillation for error mitigation~\cite{Huggins2021} are applications in this spirit. Nevertheless, in specific cases, such as the moment $\Tr(O\sigma^{\otimes n})$ where $O$ is a cyclic shift and $\sigma$ a single-qudit state, the expectation can be computed much more efficiently by adding an ancilla and using the generalized swap test, see~\cite{Horodecki2002,Subasi2019} and in particular~\cite{Brun2004}.

In the remainder of this subsection we discuss applications to \textit{quantum machines}, i.e.\ subroutines with quantum outputs, rather than sampling tasks. A prominent class of examples are unitarily covariant quantum channels, including those for universal compression~\cite{Schumacher1995,Jozsa1998,Hayashi2002,Hayashi2003,Yang2016}, entanglement distillation~\cite{Matsumoto2007,BlumeKohout2014}, optimal pure state cloning~\cite{Gisin97,Keyl1999,Harrow2013}, purification of mixed states~\cite{Cirac1999,Keyl2001a,Li2024,Childs2025}, and quantum majority vote~\cite{Buhrman2022}. A common first step to all of these protocols is weak Schur sampling followed by \textit{holding} the post-measurement state in the Schur basis.

Since we improve the gate complexity for high-dimensional quantum Schur transform, the last two classes of examples concern this regime.
Quantum simulation of identical particles is a subfield where quantum Schur transforms become indispensable. Two bases of Hilbert spaces are often considered. The first one is the spin-orbital basis in first quantization~\cite{25Bastidas,Bravyi2017} (resp.\ the occupation basis in second quantization~\cite{Burkat2025}), in cases where a Hamiltonian is sparse in it. Any fermionic or bosonic state can then be prepared by an inverse quantum Schur transform (resp.\ quantum Paldus transform~\cite{Burkat2025}), applied to a state in the totally antisymmetric or symmetric sector in the total spin basis (resp.\ in any sector of the UGA basis~\cite{Burkat2025}). The second one is the irrep-decomposing basis, in cases where a Hamiltonian possesses favorable symmetries~\cite{Burkat2025,Gu2021,Lacroix2023,Alagic2019}. The whole simulation is then sandwiched by a pair of quantum Schur transforms (resp.\ quantum Paldus transforms) and guaranteed to be efficient, called fast-forwarding.

The optimal measurement in port-based teleportation (PBT) is highly symmetric and thus most conveniently expressed in Schur basis.  
It was recently shown that PBT can be implemented in time $\widetilde{O}(n d^4)$ \cite{23Grinko,grinko2023port}.
By first compressing the alphabet down to $n$, our Schur circuit removes the $d^4$ dependence, yielding a runtime $\widetilde{O}(n^4)$, which is dominated by Krovi's transform. Complexities of PBT protocols~\cite{23Fei} which directly use Schur transform as a subroutine are also improved accordingly.

Finally, variational quantum algorithms~\cite{Zheng2023} and quantum machine learning~\cite{Ragone2023,Schatzki2024,Zheng2025,Larocca2022,Nguyen2024} also take advantage of the quantum Schur transform.

\subsection{Prior work and our objective}
The seminal and first efficient quantum circuits for the quantum Schur transform were constructed by Bacon, Chuang, Harrow (BCH) in~\cite{Harrow2005,Bacon2006a}. The circuits use an iterative list of Clebsch--Gordan transforms of the unitary group $\U{d}$, each of which couples one tensor factor of the tensor space to the irreps of $\U{d}$ decomposing the preceding, say, $i$ factors, and decompose these tensor products into irreps in the $i+1$ factors. The circuits have a time complexity of $\poly(n,d,\log \epsilon^{-1})$ for accuracy $\varepsilon$. Recently this complexity was refined to $\widetilde{O}(d^4n)$\footnote{In the sequel, we will hide the generic $\poly(\log \epsilon^{-1})$ factors in complexity analyses, which come from compilations of numeric quantum gates that can be obtained using the Solovay--Kitaev theorem.} by \cite{Nguyen2024}.

Krovi enriched the collection by presenting an algorithm for the Schur transform~\cite{Krovi2019} whose framework is the representation theory of the symmetric group $\S_n$, compared to $\U{d}$ in BCH. In particular, representations of $\S_n$ induced from a certain subgroup of $\S_n$, called the \textit{Young subgroup}, are decomposed into irreps using the quantum Fourier transform over $\S_n$ as a subroutine. The time complexity was claimed to be $\poly(n,\log d)$. The noteworthy logarithmic factor of $d$ came from compressing the type vectors in the preparation step, following a description in the footnote in Section 8.1.2 of~\cite{Harrow2005}.

The objective of the current paper is two-fold. First, Krovi's paper~\cite{Krovi2019} was recently found~\cite{24Fei_QIP} to contain a crucial error in its final step. We address and correct it. Second, we work out the detailed circuit of the mentioned compression. Because it puts the subsequent operations on $n$ instead of $d$ wires, it should only be included when the local dimension $d$ is greater than the number of qudits $n$. Consequentially, we construct high-dimensional BCH and Krovi circuits, and perform in-depth analyses of their gate, time, and space complexities.

Other efficient quantum algorithms for the Schur transform include~\cite{Kirby2018} and~\cite{24Wills}. Recently, the Schur transform was generalized to tensors of mixed unitary symmetry, whose dual is, in analogy to the symmetric group, the partially transposed permutation matrix algebra. The corresponding basis change can also be efficiently implemented~\cite{23Grinko,23Nguyen,25Dmitry}.
Streaming versions of the Schur and mixed Schur transforms can be found in~\cite{Cervero2023,Cervero2024}.

\subsection{Main results}
\subsubsection{Revised Krovi approach}

Krovi's last step was erroneous in that although it accurately claimed that the copies of each distinct $\S_n$ irrep that decompose a target induced representation were labeled by Gelfand--Tsetlin (GT) patterns, it didn't really transform to this multiplicity space basis from one labeled by standard Young tableaux (SYT), which was a remnant of applying the $\S_n$ Fourier transforms. 

We first identify this change of basis block-diagonally as one from a split basis of an $\S_n$ irrep $\lambda$, i.e.\ subgroup-reduced down a tower
$\S_n \supset \S_{\mu_1+\cdots+\mu_{l-1}}\times \S_{\mu_l}\supset\cdots\supset \S_{\mu_1}\times\S_{\mu_2}\times\cdots\times\S_{\mu_l}=Y_\mu$, to the standard basis, i.e.\ down the tower $\S_n\supset\S_{n-1}\supset\cdots\supset \S_1$. Computing the entries of this entire block is neither practical, because of the generically highly degenerate branching down each level of the tower for the split basis, nor necessary. All we need is a submatrix, an isometry $V_{\lambda,\mu}$ mapped from the trivial sector of the split basis. Each basis element in this sector can be labeled by a multiplicity-free chain of irreps $\lambda\supset (\lambda_{l-1},(\mu_l)) \supset\cdots \supset((\mu_1),\dots,(\mu_l))=\mathrm{triv}$, a GT pattern, or a semistandard Young tableau (SSYT).

Our representation theoretic results therefore include, a formula for each matrix entry of each isometry $V_{\lambda,\mu}$, and a rigorous proof that appending $V^\dagger_{\lambda,\mu}$ to the Krovi circuit results in the Gelfand--Tsetlin bases for both $\S_n$ and $\U{d}$.
The strategies are deeply rooted in the Schur--Weyl duality: when embedded in the tensor space, the basis transformation $V_{\lambda,\mu}$ of the $\S_n$ irrep $\lambda$ occurs in the multiplicity space of its $\U{d}$ irrep counterpart $\lambda$, and turns out to be flips in the fusion order of the individual tensor factors which decompose into $\U{d}$ irreps. These flips are implementable in the tensor network representation by \textit{F-moves}, dubbed \textit{F-symbols}. In proving that the revised circuit yields the GT basis for $\U{d}$, the actions of the Lie algebra generators are covariantly transported through the preparation circuit, the $\S_n$ QFTs and the circuits for $V^\dagger$, reaching the $\U{d}$ irrep register and again manipulated by F-moves.

We then move on to constructing explicit quantum circuits for the revised Krovi's algorithm. It consists of three parts:  
\begin{itemize}
    \item a pre-processing circuit $\mathrm{P}$, which prepares coset states and does an alphabet compression of the ditstrings in the computational basis,
    \item quantum Fourier transforms $\mathrm{QFT}_{\S_n}$, which perform an irrep decomposition of every representation induced from the trivial irrep of a Young subgroup, and
    \item the inverse $V^\dagger$ composed of the $V^\dagger_{\lambda,\mu}$ blocks, which puts the multiplicity spaces of the $\S_n$ irreps, i.e.\ the irrep register for $\U{d}$, in the correct basis.
\end{itemize}
Notice that we have removed the Generalized Phase Estimation (GPE) part of the original Krovi's algorithm, which is unnecessary, as we will show.
A comprehensive complexity analysis will yield that the total time and gate complexity are both $\widetilde{O}(n^4)$, dominated by the $\mathrm{QFT}_{\S_n}$ step, and the space complexity is $\widetilde{O}(n^2)$.  

Our time complexity $\widetilde{O}(n^4)$ is the best in literature when $d>n$. Note that we have removed its polynomial dependency on $d$ entirely, leaving only polylogarithmic dependence hidden in the $\widetilde{O}$ notation. This would facilitate any applications which have a potentially high local dimension $d$.

\subsubsection{Description of the high-dimensional BCH circuit}

Equipped with the pre-processing compression circuit $\mathrm{P}$, we explain how to implement the high-dimensional BCH Schur transform. It consists of two parts:
\begin{itemize}
    \item a pre-processing circuit $\hat{\mathrm{P}}$ adapted from $\mathrm{P}$, which does an alphabet compression of the ditstrings in the computational basis,
    \item a sequence of $n$ Clebsch--Gordan transforms of $\U{n}$
\end{itemize}
Again, the resulting circuit has a high-dimensional focus, i.e.\ $d>n$, and in these cases improves the time complexity of the BCH Schur transform from $\widetilde{O}(d^4 n)$ to $\widetilde{O}(n^5)$, and space complexity from $\widetilde{O}(d^2)$ to $\widetilde{O}(n^2)$.

\section{Preliminaries}\label{sec:prelims}
In this section, we introduce necessary mathematical background on representation theory and notation. 
We assume that reader is familiar with basic concepts of algebra, combinatorics, group theory, Lie groups and algebras, and representation theory.

\subsection{Schur--Weyl duality}

Recall, that Schur--Weyl duality states that the diagonal action $\phi(U):=U^{\otimes n}$ of the unitary group $\U{d}$ on the tensor space $\mathcal{H} \defeq (\C^d)\xp{n}$ generates a matrix algebra $\mathcal{U}_n^d$ that is a commutant of the algebra $\mathcal{A}_n^d$ generated by the tensor representation $\psi(\sigma): \ket{x_1} \otimes \dots \otimes \ket{x_n} \mapsto \ket{x_{\sigma^{-1}(1)}} \otimes \dots \otimes \ket{x_{\sigma^{-1}(n)}}$ of the symmetric group $\S_n$. 
Restating this, there exists a \textit{Schur transform} unitary $U_\mathrm{Sch}$ that simultaneously decomposes the actions of $\phi$ and $\psi$ on $\mathcal{H}$, viewed as representations of $\U{d}$ and $\S_n$, into their respective irreducible representations, as follows:
\begin{equation}
    \begin{split}
        &\forall u \in \U{d}, \; \USch \, \phi(u) \, U_\mathrm{Sch}^{\dagger} = \bigoplus_{\lambda \pt_d n} I_{d_\lambda} \otimes \phi_{\lambda}(u), \\
        &\forall \sigma \in \S_n, \; \USch \, \psi(\sigma) \, U_\mathrm{Sch}^{\dagger} = \bigoplus_{\lambda \pt_d n} \psi_{\lambda}(\sigma) \otimes I_{m_\lambda},
    \end{split}
\end{equation}
where $\phi_{\lambda}:\U{d} \rightarrow \mathrm{End}(\cW_\lambda)$ and $\psi_{\lambda} : \S_n \rightarrow \mathrm{End}(\cH_\lambda)$ are irreps of $\U{d}$ and $\S_n$ respectively, with $m_\lm\coloneqq \dim(\cW_\lambda) = {\rm deg}(\phi_\lm)$ and  $d_\lm \coloneqq \dim(\cH_\lambda) = {\rm deg}(\psi_\lm)$, with $\lambda$ ranging over all partitions of $n$ into $d$ parts, with each partition uniquely corresponding to an irrep of $\U{d}$ and $\S_n$. 

Usually we think of $\lambda$ combinatorially as a \emph{Young diagram}.
On the level of vector spaces, we say that $\mathcal{H}$ decomposes into irreps as 
\begin{equation}
    \mathcal{H} = \bigoplus_{\lambda \pt_d n} \cH_{\lambda} \otimes \cW_{\lambda}
\end{equation}
with $\cH_\lambda$ an irrep of $\S_n$ (Specht module) and $\cW_\lambda$ an irrep of $\U{d}$ (Weyl module). 
Note that we can equivalently consider representations of $\GL_d$ or $\mathfrak{gl}_d$ instead of $\U{d}$ since it does not make a difference in the context of Schur--Weyl duality: tensor representations of $\GL_d$, $\U{d}$ and $\mathfrak{gl}_d$ generate the same associative matrix algebra on $\cH$. 

\subsection{Young--Yamanouchi basis}
For irreducible representations of the symmetric group $\S_n$, there are several natural choices of orthonormal bases.
A convenient basis for Specht modules is the \emph{Young--Yamanouchi basis}.
It is constructed recursively along the chain of subgroups
\begin{equation}
    \S_1 \subset \S_2 \subset \cdots \subset \S_n,
\end{equation}
using the branching rule that restricts an irreps of $\S_k$ to $\S_{k-1}$.  
Basis vectors are in bijection with \emph{standard Young tableaux} of shape $\lambda \pt n$.

Equivalently, basis vectors correspond to Yamanouchi words or paths in the Young graph, which is an example of a \emph{Bratteli diagram}.
For example, a basis label $T$ can have three equivalent forms:
\ytableausetup{boxsize=1em}
\begin{equation*}
    T \equiv \of[\big]{\0,\yd{1},\yd{1,1},\yd{2,1},\yd{2,2}\,} \equiv 
\begin{ytableau}
    1 & 3 \\
    2 & 4
\end{ytableau} \equiv (1,2,1,2),
\end{equation*}
so one should keep in mind that \emph{Bratteli diagram path} $\cong$ \emph{standard Young tableau} $\cong$ \emph{Yamanouchi word}. 

The action of adjacent transpositions $\sigma_i = (i,\,i{+}1)$ is particularly simple in this basis:
\begin{equation}
    \psi_\lambda(\sigma_i) \, \ket{T} 
    = \frac{1}{r_i(T)} \ket{T} + \sqrt{1 - \frac{1}{r_i(T)^2}} \, \ket{T'},
\end{equation}
where $T'$ is obtained from $T$ by swapping $i$ and $i+1$ (if $T'$ is also standard), 
and $r_i(T)$ is the \emph{axial distance} between $i$ and $i+1$ in $T$, defined as
\begin{equation}
    r_i(T) = (c(i) - c(i+1)) - (r(i) - r(i+1)),
\end{equation}
with $r(j),c(j)$ denoting the row and column of $j$ in $T$.  
This formula completely determines the $\S_n$ action.
Sometimes, Young--Yamanouchi basis is also called \emph{Gelfand--Tsetlin basis} for $\S_n$, but we reserve this name for similar basis of the unitary group $\U{d}$ and Lie algebra $\mathfrak{gl}_d$.

\subsection{Gelfand--Tsetlin basis}\label{sec:gt_action_def}

Irreps of $\U{d}$ (or equivalently of $\mathfrak{gl}_d$) are generally labeled by the highest weights $\lambda = (\lambda_1 \ge \cdots \ge \lambda_d)$, which are integers. Similarly to the Specht modules of the symmetric group, the choice of basis for the Weyl modules is also not unique. One such natural choice is the \textit{Gelfand--Tsetlin (GT)} basis, which is subgroup-adapted to the following chain:
\begin{equation}
    \U{1} \subset \U{2} \subset \dots \subset \U{d}.
\end{equation}
The GT basis is indexed by triangular arrays (GT patterns)
\begin{equation}
    M = \{ M_{k,i} \mid 1 \le i \le k \le d \},
\end{equation}
satisfying the interlacing conditions
\begin{equation}
    M_{k+1,i} \ge M_{k,i} \ge M_{k+1,i+1}.
\end{equation}
Here the top row $M_d = (M_{d,1},\dots,M_{d,d})$ equals $\lambda$. Equivalently, GT pattern can represented as \emph{Semistandard Young Tableau}.
For example, here is a GT pattern with top row $\lambda = (3,2,0,0,0)$:
\begin{equation}
    \sof*{\,
\begin{smallmatrix}
    3 & & 2 & & 0 & & 0 & & 0\\
    & 2 & & 1 & & 0 & & 0 \\
    & & 2 & & 0 & & 0 & \\
    & & & 2 & & 0 & & \\
    & & & & 0 & & & \\
\end{smallmatrix}} 
\equiv 
\begin{ytableau}
    2 & 2 & 5 \\
    4 & 5 \\
\end{ytableau} 
\end{equation}
A \emph{weight} $w$ of GT pattern $M$ is defined as a tuple:
\begin{align}
    w(M) &= (w_1(M),\dotsc,w_d(M)) \\
    w_k(M) &\defeq \sum_{i=1}^k M_{k,i} - \sum_{i=1}^{k-1} M_{k-1,i}
\end{align}
The action of the $\mathfrak{gl}_d$ generators $E_{k,k+1}, E_{k+1,k}, E_{k,k}$ on the GT basis is given by explicit formulas \cite{Vilenkin1992}.  
For example,
\begin{align}
    \label{eq:gl_gen_action_kk}
    E_{k,k} \, \ket{M} &=  w_k(M) \ket{M}, \\
    \label{eq:gl_gen_action_k-1k}
    E_{k-1,k} \, \ket{M} 
    &= \sum_{i=1}^{k-1} \gamma_{k-1,i}^+(M) \, \ket{M + \delta_{k-1,i}}, \\
    \label{eq:gl_gen_action_kk-1}
    E_{k,k-1} \, \ket{M} 
   & = \sum_{i=1}^{k-1} \gamma_{k-1,i}^-(M) \, \ket{M - \delta_{k-1,i}},
\end{align}
where $\delta_{k-1,i}$ is a triangular pattern with $1$ at position $(k-1,i)$ and zero everywhere else, and the coefficients $\gamma_{k-1,i}^\pm(M)$ are explicit square-root rational functions of the entries of $M$:
\begin{align}
    \gamma_{k-1,i}^+(M) &= \abs*{\frac{\prod_{i=1}^k (M_{k,i} - i - M_{k-1,j} + j) \prod_{i=1}^{k-2} (M_{k-2,i} - i - M_{k-1,j} + j - 1)}{\prod_{i=1, 1 \neq j}^k (M_{k-1,i} - i - M_{k-1,j} + j) (M_{k-1,i} - i - M_{k-1,j} + j - 1)}}^{1/2} \\ 
    \gamma_{k-1,i}^-(M) &= \abs*{\frac{\prod_{i=1}^k (M_{k,i} - i - M_{k-1,j} + j + 1) \prod_{i=1}^{k-2} (M_{k-2,i} - i - M_{k-1,j} + j)}{\prod_{i=1, 1 \neq j}^k (M_{k-1,i} - i - M_{k-1,j} + j) (M_{k-1,i} - i - M_{k-1,j} + j + 1)}}^{1/2}
\end{align}

\subsection{Compression of Gelfand--Tsetlin patterns}\label{sec:compression_of_GT}

A weight $w$ can be equivalently represented by a composition\footnote{For the concept of \emph{weight} $w$, entries $w_i$ can be zero. We assume that all entries of $\mu$ are non-zero for composition. Sometimes we also refer to composition by \emph{type}.} $\mu$ and an \textit{alphabet map} $p$, that is, $w \equiv (\mu,p)$. A given Gelfand--Tsetlin pattern $M \in \GT(\lambda,w)$ can be equivalently represented by a smaller GT pattern $\widetilde{M} \in \GT(\lambda,\mu)$ of length $\ell(\mu)$, where $\ell(\mu)$ is the length of the composition $\mu$, together with the alphabet map $p$, that is, $M \cong (\widetilde{M},p)$. For example, a Gelfand--Tsetlin pattern $M = ((0),(2,0),(2,0,0),(2,1,0,0),(3,2,0,0,0))$ corresponds to $\widetilde{M}=((2),(2,1),(3,2,0))$ and $p = (2, 4, 5)$:
\begin{equation}
\sof*{\,
\begin{smallmatrix}
    3 & & 2 & & 0 & & 0 & & 0\\
    & 2 & & 1 & & 0 & & 0 \\
    & & 2 & & 0 & & 0 & \\
    & & & 2 & & 0 & & \\
    & & & & 0 & & & \\
\end{smallmatrix}} 
\quad \equiv \quad 
\of*{
\sof*{\begin{smallmatrix}
    3 & & 2 & & 0 \\
    & 2 & & 1 & \\
    & & 2 & & \\
\end{smallmatrix}}, \; (2, 4, 5)
},
\end{equation}
which can be restated in the language of SSYT as
\begin{equation}
\ytableausetup{centertableaux}
    \begin{ytableau}
    2 & 2 & 5 \\
    4 & 5 \\
    \end{ytableau}    
    \quad \equiv \quad 
    \of*{\;
    \begin{ytableau}
    1 & 1 & 3 \\
    2 & 3 \\
    \end{ytableau}
    \, , \; (2, 4, 5)
    }.
\end{equation}

\subsection{Induction, restriction and Permutation modules}

Let $G$ be a finite group and $H \subseteq G$ a subgroup. 
Suppose $\rho: H \to \End(V)$ is a representation of $H$ on a vector space $V$.  
The \emph{induced representation} $\widetilde{\rho}$ is the representation of $G$ on the space
\begin{equation}
    V \Big\uparrow^{G}_{H} \defeq \spn_\C \Set{ \ket{t_i} \otimes \ket{v} \given t_i \in G/H, \, \ket{v} \in V},
\end{equation}
where every group element $g \in G$ acts as
\begin{equation}
    \widetilde{\rho}(g) \of{\ket{t_i} \otimes \ket{v}} \defeq \ket{t_j} \otimes \rho(h)\ket{v}
\end{equation}
where we used unique decomposition of $g t_i = t_j h$ with $t_i, \, t_j \in G/H,$ are transversals and $h \in H$ ($j$ and $h$ depend on $i$).
In the literature, this induced represention is denoted by $\C[G] \otimes_{\C[H]} V$. 
Dimension of the induced representation is $\tfrac{\abs{G}}{\abs{H}} \dim \, V$.

Consider now a composition $\mu = (\mu_1,\dots,\mu_\ell)$ of $n$, and define the \emph{Young subgroup}
\begin{equation}
    Y_\mu \defeq \S_{\mu_1} \times \cdots \times \S_{\mu_\ell}.
\end{equation}
Restriction of an irrep $\cH_\lambda$ of $\S_n$ to $Y_\mu$ decomposes as
\begin{equation}
    \cH_\lambda \Big\downarrow^{S_n}_{Y_\mu}  \, \cong \, \bigoplus_{\substack{\alpha = (\alpha_1,\dots,\alpha_l) \\ \alpha_i \pt \mu_i}} \cH_{\alpha_1} \otimes \dotsb \otimes \cH_{\alpha_l} \otimes \C^{m^\lambda_\alpha}
\end{equation}
where $m^\lambda_\alpha$ are multiplicities.  For the trivial choice $\alpha = ((\mu_1),\dots,(\mu_\ell)) \equiv \mathrm{triv}$, this multiplicity is the \emph{Kostka number} $K_{\lambda\mu}$.

Conversely, \emph{permutation modules} are induced representations from trivial irreps of Young subgroups:
\begin{equation}
    \cP_\mu \defeq \mathrm{triv} \Big\uparrow^{\S_n}_{Y_\mu} \equiv \cH_{(\mu_1)} \otimes \dotsb \otimes \cH_{(\mu_l)} \Big\uparrow^{S_n}_{Y_\mu}.
\end{equation}
They admit the decomposition
\begin{equation}
    \cP_\mu \;\cong\; \bigoplus_{\lambda \vdash n} \cH_\lambda \otimes \C^{K_{\lambda\mu}},
\end{equation}
again with multiplicities given by Kostka numbers. 

\subsection{Clebsch--Gordan transformations}

A fundamental structure in the representation theory of groups is the decomposition of tensor products of irreducible representations.
Let $G$ be a compact Lie group (e.g.\ $G = \U{d}$) and let $\cW_\lambda, \cW_\mu$ denote two irreducible representations (irreps) of $G$, labeled by highest weights $\lambda$ and $\mu$. 
The tensor product representation $\cW_\lambda \otimes \cW_\mu$ is generally reducible and decomposes into a direct sum of irreps:
\begin{equation}
    \cW_\lambda \otimes \cW_\mu \;\cong\; \bigoplus_{\nu} \cW_\nu \otimes \C^{c_{\lambda,\mu}^{\nu}},
\end{equation}
where the multiplicities $c_{\lambda,\mu}^\nu \in \Z_{\geq 0}$ are known as \emph{Littlewood--Richardson coefficients}. 
These determine how many times the irrep $\cW_\nu$ appears in the tensor product.

A \emph{Clebsch--Gordan (CG) transformation} is a unitary change of basis implementing the above decomposition. 
Concretely, it is the isometry
\begin{equation}
   \CG : \cW_\lambda \otimes \cW_\mu \;\longrightarrow\; \bigoplus_{\nu} \cW_\nu \otimes \C^{c_{\lambda,\mu}^{\nu}}.
\end{equation}
Relative to fixed bases in the domain and codomain, the entries of $\CG$ are called \emph{Clebsch--Gordan coefficients}. 
These coefficients resolve the tensor product basis (adapted to $\cW_\lambda \otimes \cW_\mu$) into a \emph{coupled basis}, adapted to the irreps $\cW_\nu$ in the decomposition.

\textit{Special case: $\SU{2}$.}
For the group $\SU{2}$, irreps are labeled by spins $j \in \tfrac{1}{2}\Z_{\geq 0}$, with $\dim \cW_j = 2j+1$. 
The decomposition simplifies to
\begin{equation}
    \cW_{j_1} \otimes \cW_{j_2} \;\cong\; \bigoplus_{j = |j_1 - j_2|}^{j_1 + j_2} \cW_j,
\end{equation}
where each summand appears with multiplicity one, i.e.\ $c_{j_1,j_2}^j \in \{0,1\}$. 
The corresponding CG coefficients are exactly the angular momentum coupling coefficients.

Clebsch--Gordan transformation can be thought as being ``post-selected'' on the output irrep $\nu$ and represented in the tensor network notation as follows:
\begin{equation}
    \adjincludegraphics[valign=c, width=0.15\textwidth, page=1]{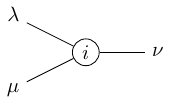}
\end{equation}
where $i \in [c_{\lambda \mu}^{\nu}]$ is the basis label in the multiplicity space.

In the language of tensor categories, the CG transformation realizes the so-called \emph{fusion rules},
where the fusion multiplicities coincide with the Littlewood--Richardson coefficients \cite{simon2023topological}.
When we tensor to irreps, we can say that we ``fuse'' them. 

\subsection{Split basis}

For symmetric groups $\S_n$, one may also work in the \emph{split basis}, adapted to the restriction to a Young subgroup chain:
\begin{equation}
    \S_n \supset \S_{\mu_1+\cdots+\mu_{l-1}}\times \S_{\mu_l}\supset\cdots\supset \S_{\mu_1}\times\S_{\mu_2}\times\cdots\times\S_{\mu_l} \equiv Y_\mu,
\end{equation}
where in each individual component $\S_{\mu_k}$ the basis is Young--Yamanouchi.
This basis differs from the full Young--Yamanouchi basis, which is adapted to the full chain $\S_n\supset\S_{n-1}\supset\cdots\supset \S_1$. 
The transformation between these bases can be interpreted as a generalized CG transformation, since it reorganizes permutation modules into irreducible components.
This transformation will be of key importance to us, and to describe it later we need to introduce the concept of $F$-symbols and $F$-moves in the next section.

\subsection{Tensor category structures and $F$-symbols}
The representation categories of both $G = \U{d}$ and $G =\S_n$ naturally carry the structure of a \emph{tensor category} $\Rep(G)$ \cite{simon2023topological}. 
Objects are finite-dimensional representations, and the tensor product of representations is again a representation. 
For irreps $\cW_a,\cW_b$, their tensor product decomposes as
\begin{equation}
    \cW_a \otimes \cW_b \;\cong\; \bigoplus_{c} \cW_c \otimes \C^{N_{ab}^c},
\end{equation}
where $N_{ab}^c \in \Z_{\geq 0}$ are the \emph{fusion multiplicities}.  
For $\S_n$ the $N_{ab}^c$ are given by Kronecker coefficients, and for $\U{d}$ they coincide with the same combinatorial rule.

\begin{figure}[H]
    \centering
    $$\includegraphics[width=0.5\textwidth, page=1]{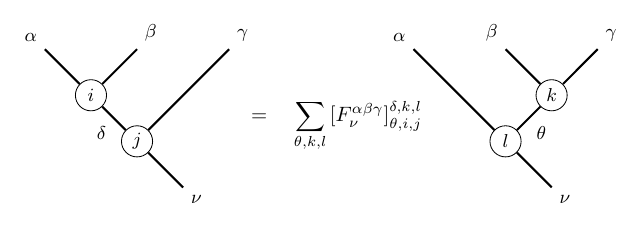} \quad
    \includegraphics[width=0.5\textwidth, page=2]{figures/f_move.pdf}$$
    \caption{A change in the fusion order is described by $F$-symbols. 
    On the left we see a general situation when fusion happens with multiplicities, while on the right--fusion is multiplicity free, which is relevant for our application.} 
    \label{fig:F_move}
\end{figure}

The tensor product is associative, but not strictly associative: when decomposing $\cW_a \otimes \cW_b \otimes \cW_c$, one may first fuse $(\cW_a \otimes \cW_b)$ and then $\cW_c$, or first fuse $(\cW_b \otimes \cW_c)$ and then $\cW_a$.  
Each choice yields a (possibly different) orthonormal basis of the same vector space.  
The change of basis between these two fusion orders is given by a unitary matrix called the \emph{$F$-symbol}: 
\begin{equation}
    F^{\alpha \beta \gamma}_{\nu} : \bigoplus_{\delta} \C^{N_{\alpha \beta}^\delta} \otimes \C^{N_{\delta \gamma}^\nu}
    \;\;\longrightarrow\;\;
    \bigoplus_{\theta} \C^{N_{\beta \gamma}^\theta} \otimes \C^{N_{\alpha \theta}^\nu}.
\end{equation}
Graphically, $F$-symbols implement the \emph{recoupling move} or \emph{$F$-move} that is, the change of parenthesization (see \cref{fig:F_move}):
\begin{equation}
(\cW_a \otimes \cW_b) \otimes \cW_c \;\cong\; \cW_a \otimes (\cW_b \otimes \cW_c).
\end{equation}

In the case of $\Rep(\U{2})$, the $F$-symbols coincide (up to normalization and phase conventions) with the classical \emph{Racah $W$-coefficients} and \emph{$6j$-symbols}.  
More generally, for semisimple categories like $\Rep(\U{d})$ or $\Rep(\S_n)$, the $F$-symbols are unitary matrices satisfying the \emph{pentagon identity} \cite{mac1998categories}, ensuring coherence of associativity.

When multiplicities $N_{ab}^c > 1$ occur, the $F$-symbol is not just a number but a block unitary matrix whose indices label the different fusion channels. 
In many of the cases relevant to Schur--Weyl duality (e.g.\ when fusing fundamental representations), multiplicities are absent, and $F$-symbols reduce to scalar phases or small unitary matrices, see \cref{fig:F_move}.

In practice, $F$-symbols provide the algebraic data needed to manipulate tensor network diagrams built from representation-theoretic building blocks.  
This makes them a key ingredient in both diagrammatic recoupling theory and computational approaches to representation theory.

\section{Two approaches to the Schur transform}
In general, the Schur transform is not uniquely defined, as it is specified up to the choice of bases on the unitary and symmetric registers. The most common conventional choice of bases for the Schur transform are the Gelfand--Tsetlin basis and the Young--Yamanouchi basis (together commonly referred to as the \textit{Gelfand--Tsetlin bases}), which are described in the previous section. However, even for a fixed choice of bases, the approaches to implementing the Schur transform can be fundamentally different. In this section, we describe the main ideas behind two conceptually different approaches to the Schur transform: the BCH approach \cite{Bacon2006a}, and Krovi's approach \cite{Krovi2019}.
\subsection{BCH} The main idea behind BCH approach to Schur transform is to treat the full Hilbert space $\cH = (\C^d)^{\otimes n}$ as a tensor product of $n$ defining irreps of $\U{d}$ \cite{Bacon2006a}:
\begin{equation}
    \cH = \cW_\square \otimes \cW_\square \otimes \dotsc \otimes \cW_\square.
\end{equation}
In that case, we deal with a special case of fusion of irreps, which is called \emph{Pieri rule}:
\begin{equation}
    \cW_{\lambda} \otimes \cW_{\square} \simeq^{\CG_d} \bigoplus_{a \in \AC_d(\lambda)} \cW_{\lambda^{+a}},
\end{equation} 
where $\AC_d(\lambda)$ is a set of addable rows of $\lambda$ and $\lambda^{+a}$ is the Young diagram $\lambda$ with box added at the row $a$. 
The Clebsch--Gordan transformation $\CG_d$ is multiplicity free in this case, which we denote in tensor network notation as
\begin{equation}
    \adjincludegraphics[valign=c, width=0.15\textwidth, page=2]{figures/cg_tn.pdf}
\end{equation}
Therefore, according to BCH one can construct full Schur transform $\USch$ as a sequence of CG transforms $\CG_d$:
\begin{equation*}
    ((\cW_{\square} \otimes \cW_{\square}) \otimes  \cW_{\square}) \otimes \dotsb \;\overset{\CG_d}{\longrightarrow}\; ((\cW_{\yd[0.3]{1,1}} \oplus \cW_{\yd[0.3]{2}}) \otimes  \cW_{\square}) \otimes \dotsb 
    \;\overset{\CG_d}{\longrightarrow}\; (\cW_{\yd[0.3]{1,1,1}} \oplus \cW_{\yd[0.3]{2,1}} \oplus \cW_{\yd[0.3]{2,1}} \oplus \cW_{\yd[0.3]{3}}) \otimes \dotsb 
\end{equation*}
Multiplicity space of a given $\cW_{\lambda}$ in such decomposition process is naturally identified with $\cH_{\lambda}$, so
\begin{equation*}
    (\C^d)\xp{n} \simeq \bigoplus_{\lambda \pt_d n} \cH_{\lambda} \otimes \cW_{\lambda}.
\end{equation*}

\subsection{Krovi} On the other hand, Krovi's approach views the full Hilbert space differently \cite{Krovi2019}:
\begin{equation}
    \cH = \bigoplus_{w} \cP_w,
\end{equation}
where the direct sum is taken over all possible weights $w$ of strings $x=(x_1,\dotsc,x_n) \in [d]^n$ (there are $\binom{n+d-1}{d-1}$ different weights) and 
\begin{equation}
    \cP_w \defeq \spn_{\C} \Set{ \, \psi(\pi) \, | \underbrace{1,\dotsc,1}_{w_1},\underbrace{2,\dotsc,2}_{w_2},\dotsc,\underbrace{d,\dotsc,d}_{w_d} \rangle \; \given \; \pi \in \S_n \,}.
\end{equation}
However, it is clear that different $\cP_w$ with $w = (p,\mu)$ but with the same composition $\mu$ are in fact all isomorphic to the permutation module $\cP_\mu$. Recall that $p$ here is the alphabet map introduced in \cref{sec:compression_of_GT}. So we can write
\begin{equation}
    \cH = \bigoplus_{(p,\mu)}  \cP_\mu.
\end{equation}
Now recall decomposition of permutation modules:
\begin{equation}
    \cP_\mu = \mathrm{triv} \Big\uparrow^{S_n}_{Y_\mu} \simeq \bigoplus_{\lambda \pt n} H_{\lambda} \otimes \C^{K_{\lambda,\mu}},
\end{equation}
where $K_{\lambda,\mu}$ are Kostka numbers. 
This number is also the dimension of the weight $w = (p,\mu)$ subspace inside Weyl module $W_\lambda$, and all such weight subspaces comprise the full unitary group irrep $\lambda$:
\begin{equation}
    \cW_\lambda = \bigoplus_{(p,\mu)} \C^{K_{\lambda,\mu}}.
\end{equation}
So, by putting everything together, we get the following decomposition of the full Hilbert space of $n$ qudits:
\begin{equation}
    \cH = \bigoplus_{(p,\mu)} \mathrm{triv} \Big\uparrow^{S_n}_{Y_\mu} \cong \bigoplus_{(p,\mu)} \bigoplus_{\lambda \pt n} \cH_{\lambda} \otimes \C^{K_{\lambda,\mu}} \cong \bigoplus_{\lambda \pt n} \cH_{\lambda} \otimes \of[\big]{\bigoplus_{(p,\mu)} \C^{K_{\lambda,\mu}}} = \bigoplus_{\lambda \pt n} \cH_{\lambda} \otimes \cW_{\lambda}.
\end{equation}

\section{Revised Krovi's Schur transform}\label{sec:krovi_def}

Krovi’s original construction of the high-dimensional Schur transform \cite[Section~4, Theorem~3]{Krovi2019} contains two critical issues. 
The first concerns the final stage of the algorithm. 
After applying the quantum Fourier transform over $\S_n$, Krovi claims that the subsequent step preparing the unitary register is a \emph{classical} unitary, i.e., a permutation matrix with entries restricted to $\{0,1\}$. 
This assumption is incorrect: the required transformation is not a classical gate but rather a genuinely quantum operation. 
In fact, we show that the correct map is the inverse of a well-defined isometry, which we denote
\begin{equation}
    V = \sum_\lambda \proj{\lambda} \otimes V_{\lambda} 
    = \sum_\lambda \sum_\mu  \proj{\lambda} \otimes \proj{\mu} \otimes V_{\lambda,\mu}.
\end{equation}
This isometry acts blockwise on the multiplicity spaces of Quantum Fourier transform over symmetric group and plays a fundamental role in ensuring the unitarity of the overall transform. 
The second issue concerns the use of Generalized Phase Estimation (GPE). Although GPE is included as a separate component of Krovi’s circuit, we prove that it is in fact unnecessary: the same information is already encoded by the $\QFT_{\S_n}$ stage and the isometry $V$. 

\begin{figure}[H]
    \centering
    \includegraphics[width=0.7\textwidth,page=1]{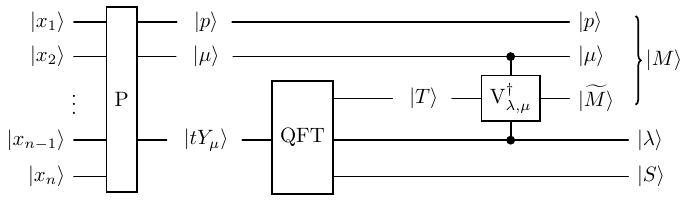}
    \caption{Krovi's quantum Schur transform consists of three steps: preprocessing $\mathrm{P}$, Quantum Fourier Transform $\mathrm{QFT}$ over $\S_n$ and compression (inverse) isometry $V^\dagger$. The state after  $\mathrm{QFT}$ lies in ${\rm im} V$ (Lemma 9), so $V^\dagger$ is reversible.
    }
    \label{fig:krovi_schur_transform}
\end{figure}

Taken together, these observations allow us to present a corrected and fully consistent version of Krovi’s algorithm. 
Our revised circuit is shown in \cref{fig:krovi_schur_transform}, where the isometry $V$ is made explicit. 
Crucially, because the state after $\QFT_{\S_n}$ always lies inside the image of $V$, the transformation is reversible. 
This establishes the corrected version of the high-dimensional quantum Schur transform, which we analyze and implement in detail in the following sections. 

\begin{theorem}[Corrected version of Krovi’s Schur transform algorithm]\label{thm:krovi}
    The quantum circuit presented in \cref{fig:krovi_schur_transform}
    performs a (high-dimensional) quantum Schur transform, and implements Gelfand--Tsetlin bases for the
    symmetric and unitary registers. This algorithm has total gate complexity $\widetilde{O}(n^4)$ and total space
    complexity $\widetilde{O}(n^2)$ in terms of the number of qudits $n$ in the input.
\end{theorem}

\begin{proof}
    First, we provide detailed definitions of each operation in \cref{fig:krovi_schur_transform} throughout \cref{sec:krovi_def}.
    Second, we argue in \cref{sec:gt_basis} that the defined circuit achieves Gelfand--Tsetlin bases for the symmetric and unitary registers.
    Finally, we discuss implementation of operations $\mathrm{P}$ and $V$ isometry in detail in \cref{sec:step_1_circuits} and \cref{sec:step_3_circuits} respectively. 
    Our analysis shows that the dominating operation is in fact symmetric group QFT with complexity $\widetilde{O}(n^4)$ \cite{kawano2016quantum}.
    Therefore, the total gate and depth complexities are $\widetilde{O}(n^4)$, and space complexity is $\widetilde{O}(n^2)$, see \cref{sec:step_1_circuits,sec:step_3_circuits}.
\end{proof}

We now detail each stage—Preprocessing $P$, the symmetric-group Fourier transform $\mathrm{QFT}_{S_n}$, and the isometry $V$; a stage-wise resource summary can be found in Table~\ref{tab:stage-resources}.

\begin{table}[t]
  \centering
  \setlength{\tabcolsep}{8pt}
  \renewcommand{\arraystretch}{1.2}
  \begin{tabular}{@{}lccc@{}}
    \toprule
    \textbf{Stage} & \textbf{Time/Depth} & \textbf{Workspace ancillas} & \textbf{I/O registers} \\
    \midrule
    $P$ (preprocess)      & $\tilde O(n^{3})$ & $\tilde O(1)$   & $\tilde O(n)$   \\
    $\mathrm{QFT}_{S_n}$  & $\tilde O(n^{4})$ & $\tilde O(n^{2})$ & $\tilde O(n)$   \\
    $V$ (isometry)        & $\tilde O(n^{3})$ & $\tilde O(n)$   & $\tilde O(n^{2})$ \\
    \midrule
    \textbf{Total}        & $\tilde O(n^{4})$ & $\tilde O(n^{2})$ & $\tilde O(n^{2})$ \\
    \bottomrule
  \end{tabular}
  \caption{Stage-wise resource summary for the revised Krovi Schur transform (cf.\ Fig.~2). Here, $\tilde{O}$ hides polylogarithmic factors in $n,d,1/\varepsilon$. “Workspace ancillas” are transient registers recycled within each stage; “I/O registers” are persistent outputs such as $|p\rangle$, $|\mu\rangle$, $|tY_\mu \rangle$, and Schur labels (YY/GT).}
  \label{tab:stage-resources}
\end{table}

\subsection{Step 1: Preprocessing}

The idea behind preprocessing step is to embed a given qudit string $x$ into a pair specified by the weight $w_x$ of $x$ and transversal element $t_x$ of $\S_n / Y_\mu$, which sorts $x$ into increasing order string $s_w$:

We have the following definitions and properties:
\begin{align}
    s_{w} &\defeq (\underbrace{1,\dotsc,1}_{w_1},\underbrace{2,\dotsc,2}_{w_2},\dotsc,\underbrace{d,\dotsc,d}_{w_d}) \\
    \psi(t_x) \ket{s_{w_x}} &\defeq \ket{x}, \text{ i.e.\ we can think } x \equiv (t_x, w_x) 
\end{align}
So we define isometry $\mathrm{P}$ by the following action:
\begin{align}
    \mathrm{P} \ket{x} &\defeq \ket{w_x} \otimes \ket{t_x Y_{w_x}}, \text{ where } \ket{t Y_{\mu}} \defeq L(t) \ket{Y_{\mu}} \\
    \mathrm{P} \ket{s_{w}} &= \ket{w} \otimes \ket{Y_{w}},
\end{align}
where coset states are elements of the regular representation of $\S_n$:
\begin{equation}
    \ket{Y_w} \defeq \frac{1}{\sqrt{\abs{Y_w}}} \sum_{h \in Y_w} \ket{h} 
    = \frac{1}{\sqrt{\abs{Y_w}}} \sum_{h \in Y_w} L(h) \ket{e} 
    = \frac{1}{\sqrt{\abs{Y_w}}} \sum_{h \in Y_w} R(h^{-1}) \ket{e}
    = \frac{1}{\sqrt{\abs{Y_w}}} \sum_{h \in Y_w} R(h) \ket{e} 
\end{equation}
Note that when you push an action of arbitrary permutation $\pi$ through $\mathrm{P}$ then it transforms into left regular action on $\CS_n$:
\begin{align}
    \mathrm{P} \cdot \psi(\pi) &= (I \otimes L(\pi)) \cdot \mathrm{P}.
\end{align}
For example, when $n=5$ then
\begin{equation*}
    \mathrm{P} \,: \, \ket{1,2,1,5,2} \to |\underbrace{(1, 2, 5)}_p, \; \underbrace{(2,2,1)}_{\mu} \rangle \otimes |\underbrace{(23)(45)}_{t} \rangle .
\end{equation*}
\subsection{Step 2: Quantum Fourier Transform over $\S_n$}

Quantum Fourier Transform (QFT) over $\S_n$ is a well known primitive \cite{97Beals}. Recall that
\begin{equation}
    \QFT \defeq \sum_{\lambda \pt n} \sqrt{\frac{d_\lambda}{n!}} \sum_{T,S \in \SYT(\lambda)} \sum_{\pi \in S_n} \psi_\lambda(\pi)_{S,T} \ket{\lambda,S,T}\bra{\pi} = 
    \sum_{\lambda \pt n} \frac{d_\lambda}{\sqrt{n!}} \sum_{\pi \in S_n} \of{I \otimes \psi_\lambda(\pi) \otimes I_{d_\lambda}} \ket{\lambda,\Psi^+_\lambda}\bra{\pi},
\end{equation}
where we defined a maximally entangle state $\Psi^+_\lambda$ in the irrep $\lambda$ as
\begin{align}
    \ket{\Psi^+_\lambda} &\defeq \frac{1}{\sqrt{d_\lambda}} \sum_{T \in \SYT(\lambda)} \ket{T,T}.
\end{align}
This $\QFT$ achieves decomposition of arbitrary linear combinations of permutations, i.e.\ elements of the group algebra $\CS_n$ viewed as left and right regular representations, into irreducible components.

It satisfies the following equivariance property:
\begin{equation}
    \QFT \cdot L(\pi) = \of*{\sum_{\lambda \pt n} \proj{\lambda} \otimes \psi_\lambda(\pi) \otimes I_{d_\lambda}} \cdot \QFT
\end{equation}

Recently, the complexity of implementation of $\QFT$ was studied in detail \cite{kawano2016quantum}. Below we present the result without proof.
\begin{lemma}[{\cite{kawano2016quantum}}]
    Quantum Fourier Transform over $\S_n$ has total gate and depth complexity $\widetilde{O}(n^4)$. 
    The space complexity is $\widetilde{O}(n^2)$.
    Moreover, this QFT achieves Young--Yamanouchi basis on both output registers, corresponding to left and right regular actions.
\end{lemma}

\subsection{Step 3: Compression isometry $V$}

Contrary to Krovi’s claim, the final map from YY‑labelled multiplicities to GT multiplicities is not a classical permutation: it’s the inverse of an isometry 
$V$ that stitches together split‑basis blocks via $F$‑moves. We define the isometry $V$ as follows. We can think of $V$ as being a block matrix:
\begin{equation}\label{eq:V_blocks}
    V = \sum_\lambda \proj{\lambda} \otimes V_{\lambda} 
    = \sum_\lambda \sum_\mu  \proj{\lambda} \otimes \proj{\mu} \otimes V_{\lambda,\mu},
\end{equation}
where domain and range of isometries is $V_{\lambda} \, : \, \C^{\GT(\lambda)} \rightarrow \C^{\SYT(\lambda)}$ and $V_{\lambda,\mu} \, : \, \C^{\GT(\lambda,\mu)} \rightarrow \C^{\SYT(\lambda)}$.
Matrix elements of $V_{\lambda,\mu}$ are defined for every irrep $\lambda \pt n$, composition $\mu = (\mu_1,\dotsc,\mu_l)$, map $p$, $M=(M_0,M_1,\dotsc,M_n) \in \GT(\lambda,(\mu,p))$ and $T=(T_0,T_1,\dotsc,T_n) \in \SYT(\lambda)$\footnote{We assume $T_0 = M_0 = () = \0$.} as 
\begin{align}
    \label{def:V_matrix_elements}
    \bra{T} V_{\lambda,\mu} \ket{M} & \defeq \prod_{k=1}^l \prod_{j=1}^{\mu_k} \sof*{F^{(\mu_k-j),(1),T^{k-1}_{j-1}}_{\widetilde{M}_k}}^{T^{k-1}_j}_{(\mu_k-j+1)} = \prod_{k=1}^l \of*{ \of*{ \prod_{j=1}^{\mu_k-1} \sof*{F^{(\mu_k-j),(1),T^{k-1}_{j-1}}_{\widetilde{M}_k}}^{T^{k-1}_j}_{(\mu_k-j+1)}} \delta_{T^k, \widetilde{M}_k} }, 
\end{align}
where $\sof{F^{a,b,c}_{d}}^{f}_{e}$ are \emph{F-symbols} or \emph{recoupling coefficients}\footnote{Note, that we have dropped multiplicity indices, see \cref{fig:F_move} for a more general $F$-symbol.}, and the convention
\begin{align}\label{eq:T_convention}
       T^k &\defeq T_{\widetilde{\mu}_k}, \qquad T^k_j \defeq T_{\widetilde{\mu}_k + j}.
\end{align}
In the last equality of \cref{def:V_matrix_elements} we have used $T^k = T^{k-1}_{\mu_k}$ and $\sof*{F^{\0,\square,T^{k-1}_{\mu_k-1}}_{\widetilde{M}_k}}^{T^{k-1}_{\mu_k}}_{\square} = \delta_{T^k,\widetilde{M}_k}$.

\begin{figure}[H]
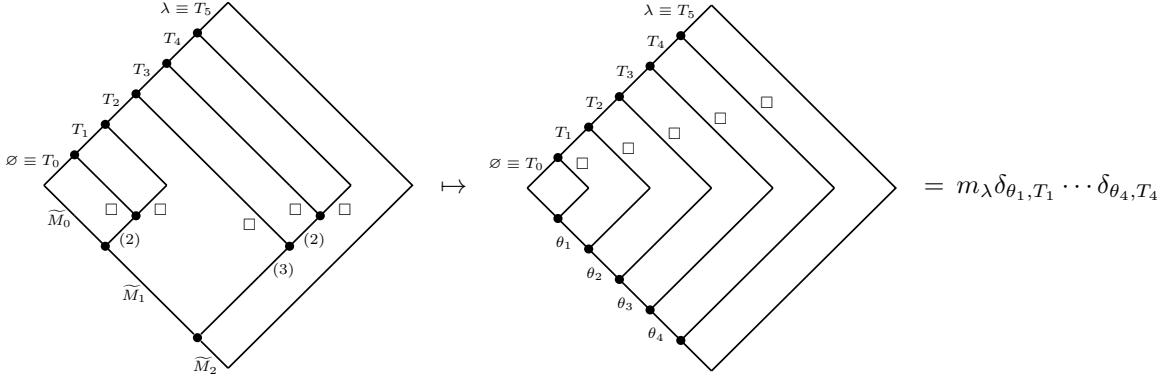

    \centering
    \adjincludegraphics[valign=c,width=0.35\textwidth,page=10]{figures/f_move.pdf} 
    $\mapsto \hspace{-5pt}$ 
    \adjincludegraphics[valign=c, width=0.35\textwidth,page=11]{figures/f_move.pdf}
    $ = \, m_\lambda \delta_{\theta_1,T_1} \dotsb \delta_{\theta_4,T_4}$
    \caption{Tensor network contraction between Young--Yamanouchi basis vector labelled by $T \in \SYT(\lambda)$ and a split basis vector $\widetilde{M} \in \SSYT(\lambda,\mu)$.
    The example is drawn for $n=5$ qudit system, where $T^5 = \widetilde{M}_2 = \lambda$ with $\mu = (2,3)$. The sequence of $F$-moves transforms the diagram on the left to the diagram on the right, which corresponds to trivial contraction.}
    \label{fig:V_tn_def}
\end{figure}

The motivation for this definition comes from tensor network contraction and $F$-moves, see \cref{fig:V_tn_def}. Namely, starting from inner product between Young--Yamanouchi basis vector $T$ and split basis vector $\widetilde{M}$ (tensor network on the left in \cref{fig:V_tn_def}), we get with a sequence of $F$ moves the diagram on the right, which is trivial.

The fact, that this is the correct definition for the isometry is proven in \cref{sec:gt_basis}, where it is shown that this definition achieves the Gelfand--Tsetlin basis on the unitary group register of revised Krovi Schur transform.

\begin{example}
    Consider $n = d = 5$, $\lambda = (3,2)$, composition $\mu = (2,1,2)$ and a map $p = (2, \, 4, \, 5)$ ($(\mu,p)$ corresponds to the weight $w = (0,2,0,1,2)$) a SYT $T = (\0,(1),(2),(3),(3,1),(3,2))$, and a Gelfand--Tsetlin pattern $M = (\0,(0),(2,0),(2,0,0),(2,1,0,0),(3,2,0,0,0))$. Then
    \begin{equation}
        \bra{T} V_{\lambda,\mu} \ket{M} = \of*{\sof*{F^{\syd{1},\syd{1},\0}_{\syd{2}}}^{\syd{1}}_{\syd{2}} \cdot  \delta_{\syd{2},\syd{2}}} \cdot \of*{\delta_{\syd{3},\syd{2,1}}} \cdot \of*{\sof*{F^{\syd{1},\syd{1},\syd{3}}_{\syd{3,2}}}^{\syd{3,1}}_{\syd{2}} \cdot \delta_{\syd{3,2},\syd{3,2}}} = 0
    \end{equation}
\end{example}

\begin{remark}
    If $\mu=(1,1,\dotsc,1)$ and $p=(1, \dotsc, n)$, then SSYT $M$ is actually an SYT. In that case, according to \cref{def:V_matrix_elements}, we can easily see that
    \begin{equation}
        \bra{T} V \ket{M} = \delta_{T,M},
    \end{equation}
    i.e.\ the gate $V$ is classical. Moreover, the state $\ket{Y_\mu} = \ket{e}$ and therefore $\ket{tY_\mu} = \ket{t}$. This means that the only non-classical gate inside the Schur transform circuit is $\mathrm{QFT_{S_n}}$, which explains an observation that $\mathrm{QFT_{S_n}}$ can be found as a submatrix inside $\USch$ when $d \geq n$. 
    More formally, our circuit easily explains the following property for the case $\mu = (1,\dots,1)$:
    \begin{equation}
        \bra{\lambda,T',M} \USch \ket{x} = \bra{\lambda,T',M} \mathrm{QFT_{S_n}} \ket{t},
    \end{equation}
    where $M \in \SYT(\lambda)$ and $t \in S_n / Y_{(1,\dotsc,1)}$ is a transversal which is in one-to-one correspondence with a dit-string $x = (x_1,\dots,x_n)$, i.e.\ $t \equiv x$.

    Note that this property is difficult to observe for BCH implementation of the Schur transform.
\end{remark}

It turns out that the specific $F$-symbols which we need were already computed before, which is the content of the next lemma:

\begin{lemma}[{\cite[Eqs. (A9)--(A10)]{le1987new}}]\label{lemma:main_F}
    For $k$, $k'$, $\lambda \, \pt \, k'$, $\nu \, \pt \, k+k'$ with $\lambda \subset \nu$, $a \in \AC(\lambda)$, where $a$ is a row number of the addable box and $\lambda^{+a}$ denotes a Young diagram obtained from $\lambda$ by adding a box to the row $a$, 
    \begin{equation}
    \label{eq:lemma_F}
        \sof*{F^{(k-1),\square,\lambda}_{\nu}}_{(k)}^{\lambda^{+a}} = \sqrt{\frac{1}{k} \frac{\prod_{i=1}^{\ell(\nu)} \of{\nu_i-i-\lambda_a+a}}{\prod_{\substack{i = 1, i \neq a}}^{\ell(\nu)} \of{\lambda_i-i-\lambda_a+a} }}, \quad
        \sof*{F^{\square,(k-1),\lambda}_{\nu}}_{(k)}^{\nu^{-r}} = \sqrt{\frac{1}{k} \frac{\prod_{i=1}^{\ell(\nu)} \of{\nu_r-r-\lambda_i+i}}{\prod_{\substack{i = 1, i \neq r}}^{\ell(\nu)} \of{\nu_r-r-\nu_i+i} }}
    \end{equation}
    where $\ell(\nu)$ is the number of rows in Young diagram $\nu$.
\end{lemma}

\section{Revised Krovi's approach implements Gelfand--Tsetlin bases for $\S_n$ and $\U{d}$} \label{sec:gt_basis}

In this section, we prove that our corrected version of Krovi's Schur transform achieves the Gelfand--Tsetlin basis on the unitary register, and the Young--Yamanouchi basis on the symmetric register. For simplicity of notation we treat two registers $\ket{p} \otimes \ket{\mu}$ as one register $\ket{w}$. We also define $Y_w \defeq \S_{w_1} \times \dotsc \times \S_{w_d}$, where if $w_k = 0$ for some $k$ then we simply omit $\S_{w_k}$ from the product.

On the one hand, it is easy to see that an action of permutation $\pi$ on $\cH$ translates into left regular representation, and therefore after $\QFT$ it acts on the register $\ket{S}$, which by construction implements Young--Yamanouchi type action.
Therefore, it is trivial in Krovi's Schur transform that
\begin{equation}
    V \QFT \mathrm{P} \psi(\pi) \mathrm{P}^\dagger \QFT^\dagger V^\dagger = \sum_{\lambda \pt n} \proj{\lambda} \otimes \psi_\lambda(\pi) \otimes I_{d_\lambda},
\end{equation}
which basically follows from definitions in \cref{sec:krovi_def}.

On the other hand, to see that the action of Lie algebra generators $E_{k,l} \in \mathfrak{gl}_d$ is of Gelfand--Tsetlin basis type according to \cref{sec:gt_action_def} is completely nontrivial.
In the following, we prove directly that tensor representation $\phi$ of $E_{k,l}$ satisfies GT formulas of \cref{sec:gt_action_def}.

First we define tensor representation $\phi$:
\begin{align} 
    \phi(E_{k,l}) &\defeq \sum_{i=1}^n \underbrace{I \otimes \dotsc \otimes I}_{i-1} \otimes \ket{k}\bra{l} \otimes \underbrace{I \otimes \dotsc \otimes I}_{n-i} 
\end{align}
It is obvious that this tensor action $\phi$ commutes with the tensor representation $\psi$ of permutations:
\begin{align}
    \phi(E_{k,l}) \cdot \psi(\pi) &= \psi(\pi) \cdot \phi(E_{k,l}).
\end{align}

Consider now the action of $\QFT$ on the coset vector $\ket{Y_{\mu}}$:
\begin{align}
    \QFT \ket{Y_{\mu}} &= 
    \QFT \frac{1}{\sqrt{\abs{Y_{\mu}}}} \sum_{h \in Y_\mu} L(h) \ket{e} = 
    \frac{1}{\sqrt{\abs{Y_{\mu}}}} \of*{\sum_{\lambda \pt n} \proj{\lambda} \otimes \sum_{h \in Y_{\mu}} \psi_\lambda(h) \otimes I_{d_\lambda}} \QFT \ket{e} \\
    &= \sum_{\lambda \pt n} d_\lambda \sqrt{\frac{\abs{Y_{\mu}}}{n!}}  \ket{\lambda} \otimes \of{\Pi_{\lambda,\mu} \otimes I} \ket{\Psi^+_\lambda} 
    = \sum_{\lambda \pt n} d_\lambda \sqrt{\frac{\abs{Y_{\mu}}}{n!}}  \ket{\lambda} \otimes \of{I \otimes \Pi_{\lambda,\mu}} \ket{\Psi^+_\lambda},
\end{align}
where
\begin{align}
    \Pi_\mu &\defeq \Pi_{Y_\mu} = \frac{1}{\abs{Y_{\mu}}} \sum_{h \in Y_{\mu}} h \in \CS_n \\
     \Pi_{\lambda,\mu} &\defeq \psi_\lambda(\Pi_\mu) = \frac{1}{\abs{Y_{\mu}}} \sum_{h \in Y_{\mu}} \psi_\lambda(h) \in {\rm End}(\cH_\lm)
\end{align}

In the following, we need to use an important consequence of the Schur's lemma: 
\begin{lemma}[Grand orthogonality relations]
\label{lemma:GOR}
For any group $G$, its irreps $\lambda,\lambda' \in \hat{G}$ and basis vectors $\ket{T}, \ket{T'}$ the following holds:
\begin{align}
    \frac{1}{\abs{G}}\sum_{g \in G} \psi_\lambda(g) \ket{T}\bra{T'} \psi_{\lambda'}(g^{-1}) = \delta_{\lambda,\lambda'} \delta_{T,T'} \frac{1}{d_\lambda} I_{d_\lambda}.
\end{align}
\end{lemma}

In particular, grand orthogonality relations (\cref{lemma:GOR}) imply
\begin{corollary}
\begin{equation}
    \sum_{\lambda, \lambda'} \ket{\lambda}\bra{\lambda'} \otimes \frac{1}{\abs{G}}\sum_{g \in G} \of{ \psi_{\lambda}(g) \otimes A_\lambda} \ket{\Psi^+_\lambda}\bra{\Psi^+_{\lambda'}} 
    \of{\psi_{\lambda'}(g^{-1}) \otimes B_{\lambda'}} 
    = \sum_{\lambda} \frac{1}{d_\lambda^2} \proj{\lambda} \otimes I_{d_\lambda} \otimes A_\lambda B_\lambda.
\end{equation}
\end{corollary}

\subsection{$E_{k,k}$ generator}

Consider the $E_{k,k}$ generator. It is easy to see that
\begin{equation}
    \phi(E_{k,k}) \ket{x} = w_{x,k} \ket{x},
\end{equation}
therefore using previous formulas we get
\begin{align}
    \mathrm{P} \phi(E_{k,k}) \mathrm{P}^\dagger &= \mathrm{P} \phi(E_{k,k}) \of[\bigg]{\sum_{x \in [d]^n} \ket{x}\bra{x}} \mathrm{P}^\dagger = \sum_{x \in [d]^n} w_{x,k} \ket{w_x}\bra{w_x} \otimes \of*{ L(t_x) \ket{Y_{w_x}}\bra{Y_{w_x}} L(t_x^{-1}) } \\
    &= \sum_{w} w_k \ket{w}\bra{w} \otimes 
    \of*{
    \sum_{t \in \mathcal{T}(w)}
    L(t) \ket{Y_{w}}\bra{Y_{w}} L(t^{-1}) 
    } 
\end{align}

\begin{lemma}
\label{lemma:qft_twirl}
We have
\begin{equation}
    \QFT \of[\bigg]{
    \sum_{t \in \mathcal{T}(\mu)}
    L(t) \ket{Y_{\mu}}\bra{Y_{\mu}} L(t^{-1}) 
    } \QFT^\dagger = \sum_{\lambda \pt n} \proj{\lambda} \otimes I \otimes \Pi_{\lambda,\mu},
\end{equation}
or, equivalently,
\begin{equation}
    \sum_{t \in \mathcal{T}(\mu)}
    L(t) \ket{Y_{\mu}}\bra{Y_{\mu}} L(t^{-1}) = R(\Pi_{\mu})
\end{equation}
\end{lemma}
\begin{proof}
Consider action of $\QFT$ on $\ket{t Y_{\mu}}$:
\begin{equation}
    \QFT L(t) \ket{Y_{\mu}} = \of*{\sum_{\lambda \pt n} \proj{\lambda} \otimes \psi_\lambda(t) \otimes I_{d_\lambda}} \QFT \ket{Y_{\mu}} = \sum_{\lambda \pt n} d_\lambda \sqrt{\frac{\abs{Y_{\mu}}}{n!}}  \ket{\lambda} \otimes \of{\psi_\lambda(t) \otimes \Pi_{\lambda,\mu}} \ket{\Psi^+_\lambda}
\end{equation}
Note that for every $h \in Y_\mu$ we can write
\begin{equation}
    \psi_\lambda(h) \Pi_{\lambda,\mu} = \Pi_{\lambda,\mu}  \psi_\lambda(h) = \Pi_{\lambda,\mu},
\end{equation}
then together with transpose trick this implies
\begin{align}
    \QFT &\of[\bigg]{
    \sum_{t \in \mathcal{T}(\mu)}
    L(t) \ket{Y_{\mu}}\bra{Y_{\mu}} L(t^{-1}) 
    } \QFT^\dagger = \\
    &= \frac{\abs{Y_{\mu}}}{n!} \sum_{\substack{\lambda \pt n \\ \lambda' \pt n}} d_\lambda d_{\lambda'}
    \ket{\lambda}\bra{\lambda'} \otimes \sum_{t \in \mathcal{T}(w)} \of{\psi_\lambda(t) \otimes \Pi_{\lambda,\mu}} \ket{\Psi^+_\lambda} \bra{\Psi^+_{\lambda'}} \of{\psi_{\lambda'}(t^{-1}) \otimes \Pi_{\lambda',\mu}} \\
    &=\frac{1}{n!} \sum_{\substack{\lambda \pt n \\ \lambda' \pt n}} d_\lambda d_{\lambda'}
    \ket{\lambda}\bra{\lambda'} \otimes \sum_{t \in \mathcal{T}(w)} \sum_{h \in Y_{\mu}} \of{\psi_\lambda(th) \otimes \Pi_{\lambda,\mu}} \ket{\Psi^+_\lambda} \bra{\Psi^+_{\lambda'}} \of{\psi_{\lambda'}((th)^{-1}) \otimes \Pi_{\lambda',\mu}} \\
    &= \frac{1}{n!} \sum_{\substack{\lambda \pt n \\ \lambda' \pt n}} d_\lambda d_{\lambda'}
    \ket{\lambda}\bra{\lambda'} \otimes \sum_{g \in \S_n} \of{\psi_\lambda(g) \otimes \Pi_{\lambda,\mu}} \ket{\Psi^+_\lambda} \bra{\Psi^+_{\lambda'}} \of{\psi_{\lambda'}(g^{-1}) \otimes \Pi_{\lambda',\mu}} \\
    &= \sum_{\lambda \pt n} \proj{\lambda} \otimes I \otimes \Pi_{\lambda,\mu}^2 = \sum_{\lambda \pt n} \proj{\lambda} \otimes I \otimes \Pi_{\lambda,\mu}
\end{align}
\end{proof}
So using \cref{lemma:qft_twirl} we see that
\begin{align}
    (I \otimes \QFT) \mathrm{P} \phi(E_{k,k}) \mathrm{P}^\dagger (I \otimes \QFT^\dagger) &= 
    \sum_{\lambda \pt n} \proj{\lambda} \otimes I \otimes \sum_{w} w_k \ket{w}\bra{w} \otimes \Pi_{\lambda,w},
\end{align}
which implies
\begin{align}
    V^\dagger (I \otimes \QFT) \mathrm{P} \phi(E_{k,k}) \mathrm{P}^\dagger (I \otimes \QFT^\dagger)V 
    &= \sum_{\lambda \pt n} \proj{\lambda} \otimes I \otimes  \sum_{w} w_k \proj{w} \otimes V^\dagger_{\lambda, w}\Pi_{\lambda,w} V_{\lambda,w}.
\end{align}
Recall, that we want to show
\begin{align}
    V^\dagger (I \otimes \QFT) \mathrm{P} \phi(E_{k,k}) \mathrm{P}^\dagger (I \otimes \QFT^\dagger)V  &= \sum_{\lambda} \proj{\lambda} \otimes I_{d_\lambda} \otimes \sum_{w} w_k \proj{w} \otimes \of[\bigg]{\sum_{M \in \GT(\lambda,w)} \proj{M}} \\ 
    \label{eq:want_to_show_k_k}
    &= \sum_{\lambda} \proj{\lambda} \otimes I_{d_\lambda} \otimes \sum_{w} w_k \proj{w} \otimes I_{K_{\lambda, w}},
\end{align}
where $I_{K_{\lambda, w}}$ is the identity acting on the space spanned by $\GT(\lambda,w)$, which has dimension equal to Kostka number $K_{\lambda, w}$. This easily follows from \cref{lemma:Pi_proj}.

\begin{lemma}\label{lemma:Pi_proj}
For every $\lambda$ and $w$, we have
\begin{equation}
    \Pi_{\lambda,w} = V_{\lambda,w} V_{\lambda,w}^\dagger.
\end{equation}
\end{lemma}
\begin{proof}
Note that $\mathrm{rank} \, \Pi_{\lambda,w} = \dim \, \mathrm{im} V_{\lambda,w} = K_{\lambda, w}$. At the same time, $\bra{M'} V^\dagger_{\lambda, w}\Pi_{\lambda,w} V_{\lambda,w} \ket{M} = \delta_{M,M'}$. To see that, note that $\Pi_{\lambda,w} V_{\lambda,w} \ket{M} = V_{\lambda,w} \ket{M}$ follows easily from tensor network identification of $V_{\lambda,w} \ket{M}$ with split basis tensor network, since each individual permutation of $\Pi_{\lambda,w}$ acts trivially on such tensor network. Therefore, $ V^\dagger_{\lambda, w}\Pi_{\lambda,w} V_{\lambda,w} = I_{K_{\lambda, w}}$, which implies $ \mathrm{im} V_{\lambda,w} \subseteq \mathrm{im} \Pi_{\lambda,w}$, but since $\dim \, \mathrm{im} \Pi_{\lambda,w} = \dim \, \mathrm{im} V_{\lambda,w}$ we must have $\mathrm{im} \Pi_{\lambda,w} = \mathrm{im} V_{\lambda,w}$. Therefore, the claim $\Pi_{\lambda,w} = V_{\lambda,w} V_{\lambda,w}^\dagger$ follows.
\end{proof}

\subsection{$E_{k,k-1}$ and $E_{k-1,k}$ generators}
To describe the derivation of the formula in that case, we need to define
\begin{align}
    w^{(k)} &\defeq (w_1,\dotsc,w_{k-2},w_{k-1}-1,w_{k}+1,w_{k+1},\dotsc,w_d) \\
    \widetilde{w}_k &\defeq \sum_{i=1}^k w_i.
\end{align}

Consider now a generator $E_{k,k-1} \in \mathfrak{gl}_d$. Observe that
\begin{align}
    \mathrm{P} \phi(E_{k,k-1}) \ket{x} &= \mathrm{P} \phi(E_{k,k-1}) \psi(t_x) \ket{s_{w_x}} = \mathrm{P} \psi(t_x) \phi(E_{k,k-1}) \ket{s_{w_x}} = (I \otimes L(t_x)) \mathrm{P} \phi(E_{k,k-1}) \ket{s_{w_x}} \\
    &= (I \otimes L(t_x)) \mathrm{P} \of*{\sum_{j=0}^{w_{x,k-1}} \psi((\widetilde{w}_{x,k-1}-j+1,\widetilde{w}_{x,k-1}))} \ket{s_{w^{(k)}_x}} \\
    &= \sum_{j=1}^{w_{x,k-1}} \ket{w^{(k)}_x} \otimes \of*{ L(t_x) L((\widetilde{w}_{x,k-1}-j+1, \widetilde{w}_{x,k-1})) \ket{Y_{\mu(w^{(k)}_x)}} }
\end{align}
Then it is easy to compute the following:
\begin{align}
    \mathrm{P} \phi(E_{k,k-1}) \mathrm{P}^\dagger &= \mathrm{P} \phi(E_{k,k-1}) \of[\bigg]{\sum_{x \in [d]^n} \ket{x}\bra{x}} \mathrm{P}^\dagger \\ 
    &= \sum_{x \in [d]^n} \sum_{j=1}^{w_{x,k-1}} \ket{w^{(k)}_x}\bra{w_x} \otimes \of*{ L(t_x) L((\widetilde{w}_{x,k-1}-j+1, \widetilde{w}_{x,k-1})) \ket{Y_{\mu(w^{(k)}_x)}}\bra{Y_{\mu(w_x)}} L(t_x^{-1}) } \\
    \label{eq:E_kk-1_1}
    &= \sum_{w} \ket{w^{(k)}}\bra{w} \otimes 
    \of*{
    \sum_{t \in \mathcal{T}(w)} \sum_{j=1}^{w_{k-1}} 
    L(t) L((\widetilde{w}_{k-1}-j+1,\widetilde{w}_{k})) \ket{Y_{w^{(k)}}}\bra{Y_{w}} L(t^{-1}) 
    }
\end{align}
To proceed further, we need to prove several lemmas first.
\begin{lemma}
    \label{lemma:h1_h2_k}
    For any subgroups $H_1, H_2 \subseteq G$ and any subgroup $K \subset G$ such that $K \subseteq H_1$, $K \subseteq H_2$ we have
    \begin{equation}
    \of[\bigg]{ \, \sum_{t \in H_1 / K} t} \of[\bigg]{\, \sum_{h \in H_2} h} = \frac{\abs{H_1} \abs{H_2}}{\abs{K}} \Pi_{H_1} \Pi_{H_2},
    \end{equation}
    where $\Pi_{H} \defeq \frac{1}{\abs{H}} \sum_{h \in H} h$.
\end{lemma}
\begin{proof}
    Notice that $k \Pi_{K} = \Pi_{K}$ for every $k \in K$. Then
    \begin{align}
         \of[\bigg]{ \, \sum_{t \in H_1 / K} t} \of[\bigg]{\, \sum_{h \in H_2} h} 
         &= 
         \of[\bigg]{ \, \sum_{t \in H_1 / K} t}
         \of[\bigg]{\, \sum_{k \in K} k} 
         \of[\bigg]{ \, \sum_{s \in K / H_2} s}
         \\ 
         &=
         \frac{1}{\abs{K}}\of[\bigg]{ \, \sum_{t \in H_1 / K} t}
         \of[\bigg]{\, \sum_{k \in K} k} \of[\bigg]{\, \sum_{k \in K} k}
         \of[\bigg]{ \, \sum_{s \in K / H_2} s} \\
         &=
         \frac{1}{\abs{K}}
         \of[\bigg]{\, \sum_{h \in H_1} h} \of[\bigg]{\, \sum_{h \in H_2} h}
         = \frac{\abs{H_1} \abs{H_2}}{\abs{K}} \Pi_{H_1} \Pi_{H_2},
    \end{align}
    which finishes the proof.
\end{proof}
Now we can prove the following lemma:
\begin{lemma} 
\label{lemma:L_R_trick}
For every weight $w$ the following holds:
    \begin{equation}
        \sum_{j=1}^{w_{k-1}} L((\widetilde{w}_{k-1}-j+1,\widetilde{w}_{k})) \ket{Y_{w^{(k)}}}
        = \sqrt{w_{k-1}(w_k+1)} R(\Pi_{w^{(k)}}) \ket{Y_w}
    \end{equation}
\end{lemma}
\begin{proof}
Note that
\begin{align}
    \sum_{j=1}^{w_{k-1}}
    (\widetilde{w}_{k-1}-j+1,\widetilde{w}_{k}) = \sum_{t \in Y_w / K} t,
\end{align}
where a Young subgroup $K$ is defined as intersection of both groups $Y_{w}$ and $Y_{w^{(k)}}$:
\begin{equation}
    K \defeq Y_{w} \cap Y_{w^{(k)}} =  \S_{w_1} \times \dotsc \times \S_{w_{k-2}} \times \S_{w_{k-1}-1} \times \S_1 \times \S_{w_k} \times \S_{w_{k+1}} \times \dotsc \times \S_{w_d}.
\end{equation} 
Using \cref{lemma:h1_h2_k} we can rewrite
\begin{align}
    \sum_{j=1}^{w_{k-1}}&
    L((\widetilde{w}_{k-1}-j+1,\widetilde{w}_{k})) \ket{Y_{w^{(k)}}} = \frac{1}{\sqrt{\abs{Y_{w^{(k)}}}}} \of[\bigg]{\sum_{t \in Y_w / K} L(t)} \of[\bigg]{\sum_{h \in Y_{w^{(k)}}} L(h)} \ket{e} \\
    &= \frac{1}{\sqrt{\abs{Y_{w^{(k)}}}}} \frac{\abs{Y_{w}} \abs{Y_{w^{(k)}}}}{\abs{K}} L(\Pi_{w}) L(\Pi_{w^{(k)}}) \ket{e} = w_{k-1} \sqrt{\abs{Y_{w^{(k)}}}}  L(\Pi_{w}) L(\Pi_{w^{(k)}}) \ket{e} \\
    &= w_{k-1} \sqrt{\abs{Y_{w^{(k)}}}}  L(\Pi_{w}) R(\Pi_{w^{(k)}}) \ket{e} = w_{k-1} \sqrt{\abs{Y_{w^{(k)}}}}  R(\Pi_{w^{(k)}}) L(\Pi_{w}) \ket{e} \\
    &= w_{k-1} \sqrt{\frac{\abs{Y_{w^{(k)}}}}{\abs{Y_{m}}}} R(\Pi_{w^{(k)}}) \ket{Y_w} = \sqrt{w_{k-1}(w_k+1)} R(\Pi_{w^{(k)}}) \ket{Y_w},
\end{align}
which finishes the proof.
\end{proof}

Now using \cref{lemma:L_R_trick} we can further rewrite \cref{eq:E_kk-1_1} as follows:
\begin{align}
\mathrm{P} \phi(E_{k,k-1}) \mathrm{P}^\dagger &= \sum_{w} \ket{w^{(k)}}\bra{w} \otimes 
    \of*{
    \sum_{t \in \mathcal{T}(w)} \sum_{j=1}^{w_{k-1}} 
    L(t) L((\widetilde{w}_{k-1}-j+1,\widetilde{w}_{k})) \ket{Y_{w^{(k)}}}\bra{Y_{w}} L(t^{-1}) 
    } \\
    &= \sum_{w} \sqrt{w_{k-1}(w_k+1)} \ket{w^{(k)}}\bra{w} \otimes 
    R(\Pi_{w^{(k)}})
    \of*{\sum_{t \in \mathcal{T}(w)} L(t)  \ket{Y_{w}}\bra{Y_{w}} L(t^{-1}) }\\
    &= \sum_{w} \sqrt{w_{k-1}(w_k+1)} \ket{w^{(k)}}\bra{w} \otimes 
    R(\Pi_{w^{(k)}})
    R(\Pi_{w}),
\end{align}
which implies
\begin{align}
(I \otimes \QFT) \mathrm{P} \phi(E_{k,k-1}) \mathrm{P}^\dagger (I \otimes \QFT^\dagger) = 
    \sum_{\lambda \pt n} \proj{\lambda} \otimes I \otimes \sum_{w} \sqrt{w_{k-1}(w_k+1)} \ket{w^{(k)}}\bra{w} \otimes \Pi_{\lambda,w^{(k)}} \Pi_{\lambda,w}.
\end{align}
Consequently, using \cref{lemma:Pi_proj} we get
\begin{align}
V^\dagger(I \otimes \QFT) \mathrm{P}& \phi(E_{k,k-1}) \mathrm{P}^\dagger (I \otimes \QFT^\dagger) V = \nonumber \\
    &= 
    \sum_{\lambda \pt n} \proj{\lambda} \otimes I \otimes \sum_{w} \sqrt{w_{k-1}(w_k+1)} \ket{w^{(k)}}\bra{w} \otimes V_{\lambda,w^{(k)}}^\dagger \Pi_{\lambda,w^{(k)}} \Pi_{\lambda,w} V_{\lambda,w}\\
    &= 
    \sum_{\lambda \pt n} \proj{\lambda} \otimes I \otimes \sum_{w} \sqrt{w_{k-1}(w_k+1)} \ket{w^{(k)}}\bra{w} \otimes V_{\lambda,w^{(k)}}^\dagger V_{\lambda,w^{(k)}} V_{\lambda,w^{(k)}}^\dagger V_{\lambda,w} V_{\lambda,w}^\dagger V_{\lambda,w}\\
    &= 
    \sum_{\lambda \pt n} \proj{\lambda} \otimes I \otimes \sum_{w} \sqrt{w_{k-1}(w_k+1)} \ket{w^{(k)}}\bra{w} \otimes  V_{\lambda,w^{(k)}}^\dagger V_{\lambda,w}.
\end{align}
That means we have to verify the following lemma to obtain the desired result: 
\begin{lemma}
    For every $M \in \GT(\lambda, w)$ and $M' \in \GT(\lambda, w^{(k)})$ such that if $\exists j \in [k-1] \, : \, M'_{k-1} = M_{k-1}^{-j}$ then
    \begin{equation}
         \sqrt{w_{k-1}(w_k+1)} \bra{M'} V_{\lambda,w^{(k)}}^\dagger  V_{\lambda,w} \ket{M} =  \abs*{\frac{\prod_{i=1}^k (M_{k,i} - i - M_{k-1,j} + j + 1) \prod_{i=1}^{k-2} (M_{k-2,i} - i - M_{k-1,j} + j)}{\prod_{i=1, 1 \neq j}^k (M_{k-1,i} - i - M_{k-1,j} + j) (M_{k-1,i} - i - M_{k-1,j} + j + 1)}}^{1/2},
    \end{equation}
    and if $ \forall j \in [k-1] \, : \, M'_{k-1} \neq M_{k-1}^{-j}$ then $\bra{M'} V_{\lambda,w^{(k)}}^\dagger  V_{\lambda,w} \ket{M} = 0$,
    where $M_{k-1}^{-j}$ is GT pattern defined such that it is different from $M$ only in one position: $M_{k-1,j}^{-j} \defeq M_{k-1,j} - 1$.
\end{lemma}
\begin{proof}
   Note that vector $V_{\lambda,w} \ket{M}$ can be embedded into $(\C^d)^{\otimes n} \otimes \cW_\lambda$ as a tensor network:
    \begin{align}
        V_{\lambda,w} \ket{M} &\equiv \adjincludegraphics[valign=c,page=3]{figures/f_move.pdf} \\
                              &= \adjincludegraphics[valign=c,page=4]{figures/f_move.pdf},
    \end{align}
   where $M = (p,\widetilde{M})$, and in the second equality we used the fact that each tree is invariant under the action of any permutation, so we can orient the tree as much as we like.
   Similarly,
    \begin{align}
        V_{\lambda,w^{(k)}} \ket{M'} &\equiv \adjincludegraphics[valign=c,page=5]{figures/f_move.pdf} \\
                                     &= \adjincludegraphics[valign=c,page=6]{figures/f_move.pdf}
    \end{align}
   Now the overlap $\bra{M'} V_{\lambda,w^{(k)}}^\dagger  V_{\lambda,w} \ket{M}$ can be computed as a tensor network contraction:
    \begin{align}
        \bra{M'} V_{\lambda,w^{(k)}}^\dagger  V_{\lambda,w} \ket{M} &= \frac{1}{m_\lambda}\adjincludegraphics[valign=c,page=7]{figures/f_move.pdf}
    \end{align} 
    Due to the properties of $F$ symbols and CG coefficients, we can contract the tensor network from left and right. As a result we get a simpler tensor network:
    \begin{align}
        &\bra{M'} V_{\lambda,w^{(k)}}^\dagger  V_{\lambda,w} \ket{M}  = \frac{1}{m_{M_k}} \adjincludegraphics[valign=c,page=8]{figures/f_move.pdf} \\
        &= \sum_{\theta,\theta'} \sof*{F^{\square,(\mu_{k-1}-1),M_{k-2}}_{M_{k-1}}}_{(\mu_{k-1})}^{\theta} \sof*{F^{(\mu_k),\square,M'_{k-1}}_{M_k}}_{(\mu_k+1)}^{\theta'} \frac{1}{m_{M_k}} \adjincludegraphics[valign=c,page=9]{figures/f_move.pdf}
    \end{align}
    where $w = (p,\mu)$, and red vertices indicate the positions of two $F$-moves which were applied in the second equality. The resulting tensor network trivially contracts to $m_{M_k} \delta_{\theta,M'_{k-1}} \delta_{\theta',M_{k-1}}$:
    \begin{align}
        \bra{M'} V_{\lambda,w^{(k)}}^\dagger  V_{\lambda,w} \ket{M}  &= \sum_{\theta,\theta'} \sof*{F^{\square,(\mu_{k-1}-1),M_{k-2}}_{M_{k-1}}}_{(\mu_{k-1})}^{\theta} \sof*{F^{(\mu_k),\square,M'_{k-1}}_{M_k}}_{(\mu_k+1)}^{\theta'} \delta_{\theta,M'_{k-1}} \delta_{\theta',M_{k-1}}  \\ 
        &= \sof*{F^{\square,(\mu_{k-1}-1),M_{k-2}}_{M_{k-1}}}_{(\mu_{k-1})}^{M'_{k-1}} \sof*{F^{(\mu_k),\square,M'_{k-1}}_{M_k}}_{(\mu_k+1)}^{M_{k-1}} \\
        &= \abs*{\frac{1}{w_k+1}\frac{1}{w_{k-1}}\frac{\prod_{i=1}^k (M_{k,i} - i - M_{k-1,j} + j + 1) \prod_{i=1}^{k-2} (M_{k-2,i} - i - M_{k-1,j} + j)}{\prod_{i=1, 1 \neq j}^k (M_{k-1,i} - i - M_{k-1,j} + j) (M_{k-1,i} - i - M_{k-1,j} + j + 1)}}^{1/2},
    \end{align}
    where in the third step we used \cref{lemma:main_F}.
\end{proof}

The similar calculation can be done for the generator $E_{k-1,k}$ with the same strategy as we just did for $E_{k-1,k}$. For brevity, we will not repeat it here.

\section{Quantum circuits for step 1: preprocessing circuits}
\label{sec:step_1_circuits}

In this section, we present a quantum circuit for an isometry
\begin{equation}
    \label{eq:prep}
    \mathrm{P}: \ket{x}
    \mapsto\ket{p}\ket{\mu}\ket{tY_\mu}
    ,
\end{equation}
which transforms arbitrary computational basis vector $\ket{x} = \ket{x_1,\ldots,x_n} \in \cH_d^{n}$ into a triplet $(\ket{p},\ket{\mu},\ket{tY_\mu})$, where $\ket{p} = \ket{p_1,\ldots,p_n}\in \cH_d^{n}$ is related alphabet vector, $\ket{\mu} = \ket{\mu_1,\ldots,\mu_n} \in \cH_{n}^{n}$ is related type vector, and $\ket{tY_\mu} \in \cH_n^{n}$ related transversal element. 
To be more specific, recall that the transversal element $\ket{tY_\mu}$ has the following form:
\begin{equation}
    \label{eq:tY_ket}
    \ket{tY_\mu}=
    \tfrac{1}{\sqrt{|Y_\mu |}}
   \sum_{\substack{\sigma\in \S_n:\\ x_{\sigma (1)}\leq\cdots\leq x_{\sigma (n)}}}
    \ket{\sigma}.
\end{equation}
At first, we shall use the following encoding for any permutation $\sigma\in \S_n$:
\begin{equation}
    \label{eq:sigma_ket}
    \ket{\sigma}= 
    \ket{\sigma (1),\ldots,\sigma (n) } \in \cH_n^n ,
\end{equation}
and hence (\ref{eq:tY_ket}) become:
\begin{equation}
    \label{eq:tY_ket_2}
    \ket{tY_\mu}=
    \tfrac{1}{\sqrt{|Y_\mu |}}
    \sum_{\substack{\sigma\in \S_n:\\ x_{\sigma (1)}\leq\cdots\leq x_{\sigma (n)}}}
    \ket{\sigma (1),\ldots,\sigma (n) } .
\end{equation}

\begin{figure}[!ht]
    \centering
      \adjincludegraphics[valign=c,width=0.17\textwidth,page=1]{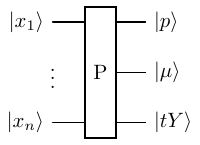}
    \quad$=$\quad
    \adjincludegraphics[valign=c,width=0.5\textwidth,page=2]{figures/Prep_circuit.pdf}
    \caption{Inductive decomposition of preparation isometry $\mathrm{P} \equiv \Prep$. 
    Notice that some auxiliary registers are omitted for clarity of the picture.}
    \label{fig:Prep}
\end{figure}

Notice that isometry (\ref{eq:prep}), can be defined recursively. Indeed, supposed that we have a transformation:
\begin{equation}
    \label{eq:prep_prim}
    \Prep': \ket{x'}
    \mapsto\ket{p'}\ket{\mu'}\ket{tY_\mu'}
\end{equation}
which transforms computational basis vector $\ket{x'}=\ket{x_1,\ldots,x_{n-1}}\in \cH_d^{n-1}$ into related alphabet vector $\ket{p}=\ket{p_1',\ldots,p_{n-1}'}\in \cH_d^{n-1}$, type vector $\ket{\mu'}=\ket{\mu_1',\ldots,\mu_{n-1}'}\in \cH_n^{n-1}$, and transversal element $\ket{tY_\mu'}\in \cH_n^{n-1}$. 
We shall define the following map: 
\begin{equation}
    \label{eq:prep_prim}
    \Prep_n: 
    \ket{p'}\ket{\mu'}\ket{tY_\mu'} \otimes \ket{x_n}
    \mapsto\ket{p}\ket{\mu}\ket{tY_\mu}
\end{equation}
where $\ket{p}$ is an alphabet vector, $\ket{\mu}$ type vector and $\ket{tY_\mu}$ transversal corresponding to $\ket{x}=\ket{x'}\ket{x_n}=\ket{x_1,\ldots,x_n}\in \cH_d^n$. In that way, transformation (\ref{eq:prep}) can be achieved by
\begin{equation}
    \label{eq:Perp_induction}
    \mathrm{P} = \Prep_1 \Prep_2 \cdots \Prep_n
\end{equation}
as presented on \cref{fig:Prep}. 
In order to provide a quantum circuit for isometry $\Prep$, we shall use the above decomposition. 
Indeed, we will present a quantum circuit for $\Prep_n$. In that way, the quantum circuit achieving transformation $\Prep$ will be simply given by composing $\Prep_i$ for $i\in [n]$. Circuits for $\Prep_i$ with $i < n$ are constructed in a completely analogous manner as the circuit for $\Prep_n$.

\begin{figure}[!ht]
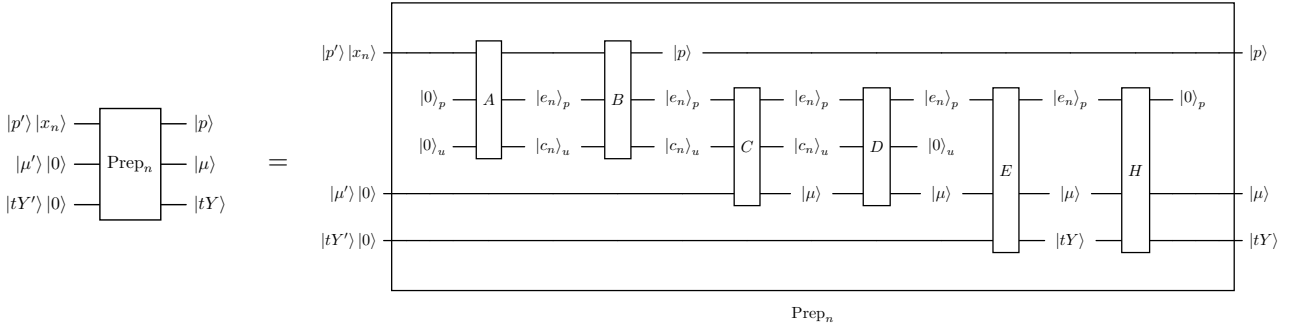

    \centering
    \adjincludegraphics[valign=c,width=0.18\textwidth,page=3]{figures/Prep_circuit.pdf}
    \quad$=$\quad
    \adjincludegraphics[valign=c,width=0.75\textwidth,page=4]{figures/Prep_circuit.pdf}
    \caption{A quantum circuit for an isometry $\Prep_n$ utilizes two auxiliary registers ($\ket{\cdot }_{p}\in \cH_d$ and $\ket{\cdot }_{u}\in \cH_2$) that store information about the position $e_n$ and uniqueness $c_n$ of $x_n$. It can be further decomposed into six subroutines: $A, B, C, D, E,$ and $H$. \Cref{fig:subroutibesABCD} presents the quantum circuits for the aforementioned subroutines and provides an explanation of their roles in implementing $\Prep_n$.}
    \label{fig:subroutibesABCDE}
\end{figure}

In the remaining part of this section, we shall present a quantum circuit for an isometry $\Prep_n$. We will do it by using two auxiliary registers ($\ket{\cdot }_{p}\in \cH_d$ and $\ket{\cdot }_{u}\in \cH_2$), and further decomposing it into five subroutines ($A,B,C,D,E$, and $F$), see \cref{fig:subroutibesABCDE}. 

We begin with introducing two integer quantities, which will be stored in auxiliary registers. Suppose that $\ket{p'}\ket{\mu}\ket{tY_\mu'}$ are alphabet vector, type vector and transversal corresponding to $\ket{x'}=\ket{x_1,\ldots,x_{n-1}}\in \cH_d^{n-1}$. For arbitrary $\ket{x_n}$, we define a binary number:
\begin{equation}
    \label{eq:c_n}
    c_n :=
    \begin{cases}
    1 & \text{if } x_n \in \{x_1,\ldots,x_{n-1}\}
    \\0 &\text{otherwise}
    \end{cases}
\end{equation}
which is determined by the \textit{uniqueness} of the value $x_n$, i.e., whether value $x_n$ already appeared in the sequence $x_1,\cdots,x_{n-1}$. Furthermore, we assign a number $e_n \in [n]$:
\begin{equation}
    \label{eq:s_n}
    e_n : = i, \; \text{ s.t. } \; x_n = p_i
\end{equation}
which represents a \textit{position} of a value $x_n$ in the sequence $p=(p_1,\cdots,p_{n})$ while sorted ascending. 

\begin{figure}[htbp]
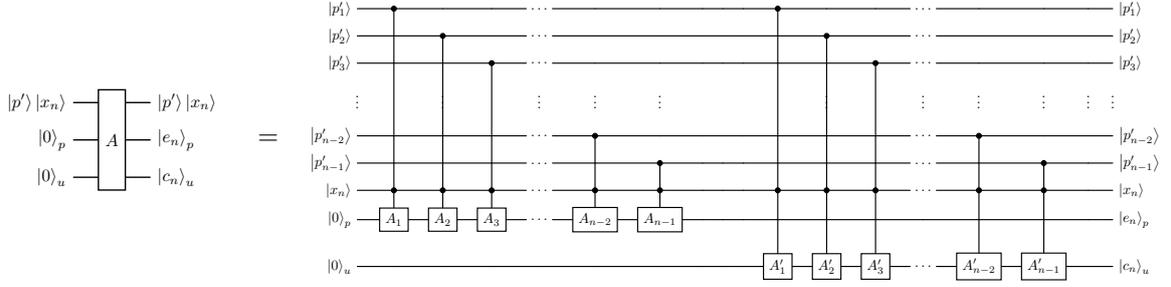
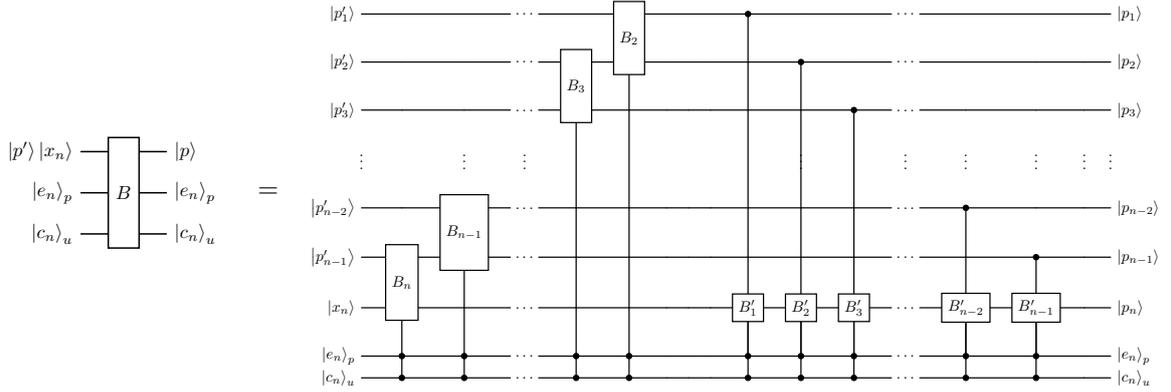
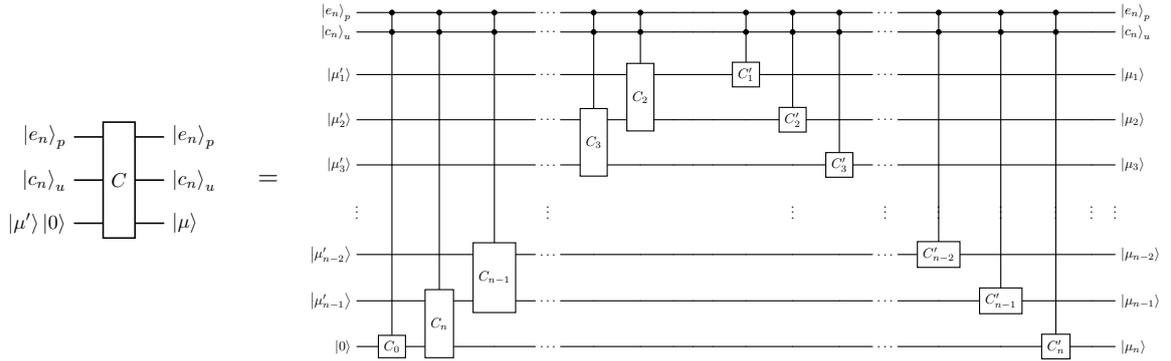
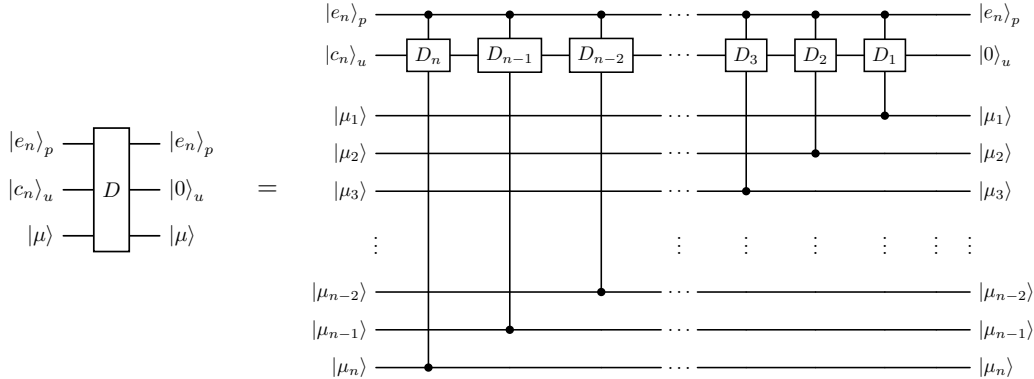

    \centering
    \begin{subfigure}{0.95\textwidth}
        \centering
        \adjincludegraphics[valign=c,width=0.18\textwidth,page=5]{figures/Prep_circuit.pdf}
    \quad$=$\quad
        \adjincludegraphics[valign=c,width=0.7\textwidth,page=6]{figures/Prep_circuit.pdf}
        \caption{Subroutine $A$ computes the value of the position $e_n$ and uniqueness $c_n$ of $x_n$ stored in registers $\ket{\cdot }_{p}$ and $\ket{\cdot }_{u}$ respectively. All the gates $A_k,A_k'$ are controlled $X$ gates defined in (\ref{eq:Ak}) and (\ref{eq:Ak2}) respectively.}
        \label{fig:subroutibe_A}
    \end{subfigure}
    \hfill
    \begin{subfigure}{0.95\textwidth}
        \centering
        \adjincludegraphics[valign=c,width=0.18\textwidth,page=7]{figures/Prep_circuit.pdf}
    \quad$=$\quad
        \adjincludegraphics[valign=c,width=0.7\textwidth,page=8]{figures/Prep_circuit.pdf}
        \caption{Subroutine $B$ updates register $\ket{p'}$ into a new value $\ket{p}$ based on previousely computed values $e_n$ and $c_n$. Operators $B_k$ are controlled SWAP-gates, and $B_k'$ are controlled $X$-gates defined in (\ref{eq:Bk}) and (\ref{eq:Bk2}) respectively.}
        \label{fig:subroutibe_B}
    \end{subfigure}
    \hfill
    \begin{subfigure}{0.95\textwidth}
        \centering
        \adjincludegraphics[valign=c,width=0.18\textwidth,page=9]{figures/Prep_circuit.pdf}
    \quad$=$\quad
        \adjincludegraphics[valign=c,width=0.7\textwidth,page=10]{figures/Prep_circuit.pdf}
        \caption{Subroutine $C$ updates register $\ket{\mu'}$ into a new value $\ket{\mu}$ based on previously computed values $e_n$ and $c_n$. Operators $C_0$, and $C_k'$ are controlled $X$-gates defined in (\ref{eq:C0}), and (\ref{eq:Ck2}) respectively, while $C_k$ are controlled SWAP-gates defined in (\ref{eq:Ck}).}
        \label{fig:subroutibe_C}
    \end{subfigure}
    \hfill
    \begin{subfigure}{0.95\textwidth}
        \centering
        \adjincludegraphics[valign=c,width=0.18\textwidth,page=11]{figures/Prep_circuit.pdf}
    \quad$=$\quad
        \adjincludegraphics[valign=c,width=0.6\textwidth,page=12]{figures/Prep_circuit.pdf}
        \caption{Subroutine $D$ uncomputes the value $c_n$ on register $\ket{\cdot }_{u}$. Operators $D_k$ are controlled $X$-gates defined in (\ref{eq:Dk}), and at most one of them will be triggered.}
        \label{fig:subroutibe_D}
    \end{subfigure}
    \caption{First four subroutines in  the preparation circuit $\Prep_n$ from \cref{fig:subroutibesABCDE} are sequences of controlled $X$-, and SWAP-gates.  Subroutine $A$ computes values $e_n$ and $c_n$. Subroutine $B$ transforms $\ket{p'}\ket{x_n}$ into $\ket{p}$. Subroutine $C$ transforms $\ket{\mu'}$ into $\ket{\mu}$, while subroutine $D$ uncomputes the value $c_n$.}
    \label{fig:subroutibesABCD}
\end{figure}

\textbf{Subroutine $A$}. At first, we define unitary
\begin{equation}
    \label{eq:unitary_A}
    A: 
    \ket{p'}\ket{x_n} \ket{0 }_{u}\ket{0}_{p}
    \mapsto 
    \ket{p'}\ket{x_n} \ket{c_n }_{u}\ket{e_n}_{p}
\end{equation}
which computes position $e_n$ and uniqueness $c_n$ controlled on the registers $\ket{p'}$ and $\ket{e_n}$, see \cref{fig:subroutibe_A}. In order to compute $e_n$, we apply a sequence of unitary gates $A_1,\ldots,A_{n-1}$, where
\begin{equation}
\label{eq:Ak}
    A_k : \ket{p'_k}\ket{x_n} \ket{i}_{p}
    \mapsto 
    \begin{cases}
    \ket{p'_k}\ket{x_n} \ket{i+1}_{p} & \text{if } p'_k < x_n,
    \\
    \ket{p'_k}\ket{x_n} \ket{i}_{p} &\text{otherwise},
    \end{cases}
\end{equation}
which is a controlled $X$-gate acting on register $\ket{\cdot }_{p}$, see \cref{fig:subroutibe_A}. Observe that from (\ref{eq:s_n}), we have
\begin{equation}
\label{eq:A_1}
    A_1 A_2\cdots A_{n-1} :
    \ket{p'} \ket{x_n} \ket{0}_{p} 
    \mapsto
    \ket{p'} \ket{x_n} \ket{e_n}_{p}
\end{equation}
correctly computes the value $e_n$. 

Similarly to (\ref{eq:Ak}), we introduce gates 
\begin{equation}
\label{eq:Ak2}
   A_k' : \ket{p'_k}\ket{x_n} \ket{i}_{c}
    \mapsto 
    \begin{cases}
    \ket{p'_k} \ket{x_n} \ket{i+1}_{c} & \text{if } p'_k = x_n
    \\
    \ket{p'_k} \ket{x_n} \ket{i}_{c} &\text{otherwise}
    \end{cases}
\end{equation}
which is a controlled $X$-gate acting on register $\ket{\cdot }_{c}$, triggered only if there is an index $k$ such that $p'_k = x_n$. Notice that 
\begin{equation}
\label{eq:A_2}
    A_1' A_2'\cdots A_{n-1} ':
    \ket{p'} \ket{x_n} \ket{0}_{u} 
    \mapsto
    \ket{p'} \ket{x_n} \ket{c_n}_{u}
\end{equation}
correctly computes the value $c_n$ written on register $\ket{\cdot }_{c}$. Combining (\ref{eq:A_1}) and (\ref{eq:A_2}), we observe that 
\[
A=A_1\cdots A_{n-1}A_1'\cdots A_{n-1}',
\]
achieves transformation (\ref{eq:unitary_A}), see \cref{fig:subroutibesABCDE} for details.

\textbf{Subroutine $B$}. In the following step, we shall update register $\ket{p'}$ to its new value $\ket{p}$ in accordance to $\ket{x_n}$. We will do it based on the precomputed values $e_n$, and $c_n$. Namely, we will construct the following unitary transformation: 
\begin{equation}
    \label{eq:unitary_B}
    B: 
    \ket{p'}\ket{x_n} \ket{c_n }_{u}\ket{e_n}_{p}
    \mapsto 
    \ket{p}\ket{c_n }_{u}\ket{e_n}_{p}.
\end{equation}
Note that $\ket{p'}\in \cH_d^{n-1}$, $\ket{x_n}\in \cH_d^{}$, and $\ket{p}\in \cH_d^{n}$. Note that depending on the binary value $c_n$, we have one of the two cases. If $c_n=0$, the value $x_n\not\in \{p_1,\ldots,p_{n-1}\} $ is a new value. In that case, we shall swap vector $\ket{x_n}$ into the right position in the sequence $\ket{p_1},\ldots,\ket{p_{n-1}}$ in accordance to the recomputed value $e_n$. In order to do it, we introduce the sequence of gates $B_{n}, B_{n-1},\ldots, B_2$, where
\begin{equation}
\label{eq:Bk}
    B_k : \ket{p'_k}\ket{p'_{k-1}}\ket{c_n}_{u}\ket{e_n}_{p}
    \mapsto 
    \begin{cases}
    \ket{p'_{k-1}}\ket{p'_k}\ket{c_n}_{u}\ket{e_n}_{p}
     & \text{if } c_n= 0, \text{ and } k > n-e_n,
    \\
    \ket{p'_k}\ket{p'_{k-1}}\ket{c_n}_{u}\ket{e_n}_{p} &\text{otherwise},
    \end{cases}
\end{equation}
is a controlled SWAP-gate permuting registers $\ket{p'_k}$ and $\ket{p'_{k-1}}$ triggered if $c_n=0$ and  $k > n - e_n$. Here we use a convention $\ket{p'_n}:=\ket{x_n}$, see \cref{fig:subroutibe_B}. Notice that unitary operators $B_k$ do not commute, and are triggered only if $c_n=0$. Moreover, we have the following: 
\begin{align}
\label{eq:B_1}
    B_{n} B_{n-1}\cdots B_{2} :&
    \ket{p'}\ket{x_n} \ket{0}_{c} \ket{e_n}_{p} 
    \mapsto
    \ket{p}\ket{0}_{c}\ket{e_n}_{p}  ,\\
    \nonumber
    B_{n} B_{n-1}\cdots B_{2} :&
    \ket{p'}\ket{x_n} \ket{1}_{c} \ket{e_n}_{p} 
    \mapsto
    \ket{p'}\ket{x_n} \ket{1}_{c}\ket{e_n}_{p}  ,
\end{align}
which is a desired transformation into $\ket{p}$ if $c_n=0$. 

In the second case, when $c_n=1$, the value $x_n \in \{p'_1,\ldots,p'_{n-1}\} $ is not a new value. In that case, we shall uncompute $\ket{x_n}$, i.e.\ $\ket{x_n} \mapsto \ket{0}$ based on the precomputed value $e_n$. In order to do it, we introduce the sequence of gates $B_{k}',B_{k-1}',\ldots, B_2'$, where
\begin{equation}
\label{eq:Bk2}
    B_k' : \ket{p'_k}\ket{x_n}\ket{c_n}_{u}\ket{e_n}_{p}
    \mapsto 
    \begin{cases}
    \ket{p'_k}\ket{x_n-p_k'}\ket{c_n}_{u}\ket{e_n}_{p}
     & \text{if } c_n= 1, \text{ and } k = e_n,
    \\
    \ket{p'_k}\ket{x_n}\ket{c_n}_{u}\ket{e_n}_{p}
    &\text{otherwise},
    \end{cases}
\end{equation}
is a controlled gate acting on $\ket{x_n}$ by a controlled shift $\ket{x_n}\mapsto \ket{x_n-p'_k}$ if $c_n=1$ and  $e_n=k$, see \cref{fig:subroutibe_B}. Notice that in this case, $p_k' = x_n$, and hence we achieve transformation $\ket{x_n}\mapsto \ket{0}$. Moreover, it is easy to see that at most one operator $B_k'$ will be triggered, and we have the following:
\begin{align}
\label{eq:B_2}
    B_{k-1}' B_{k-2}'\cdots B_{1}' :&
    \ket{p'}\ket{x_n} \ket{1}_{c} \ket{e_n}_{p} 
    \mapsto
    \ket{p}\ket{1}_{c}\ket{e_n}_{p}  ,\\
    \nonumber
     B_{k-1}' B_{k-2}'\cdots B_{0}' :&
    \ket{p} \ket{0}_{c} \ket{e_n}_{p} 
    \mapsto
    \ket{p}\ket{0}_{c}\ket{e_n}_{p}  ,
\end{align}
where we observe that $\ket{p} = \ket{p'}\ket{0}$ for $c_n=1$. Combining (\ref{eq:B_1}) with (\ref{eq:B_2}), we observe that 
\[
B = B_{n-1}' B_{n-2}'\cdots B_{1}' \, B_{n} B_{n-1}\cdots B_{2} 
\]
achieves transformation (\ref{eq:unitary_B}).

\textbf{Subroutine $C$}. Thirdly, we shall update register $\ket{\mu'}$ to its new value $\ket{\mu}$, where $\mu'$ is a type of a sequence $x_1,\ldots,x_{n-1}$, while $\mu$ is a type of a sequence $x_1,\ldots,x_{n}$. We will do it based on the precomputed values $e_n$, and $c_n$. Namely, we will construct the following unitary transformation: 
\begin{equation}
    \label{eq:unitary_C}
    C: 
    \ket{c_n }_{u}\ket{e_n}_{p} \ket{\mu'}\ket{0} 
    \mapsto 
    \ket{c_n }_{u}\ket{e_n}_{p} \ket{\mu}.
\end{equation}
Note that $\ket{\mu'} \in \cH_n^{n-1}$, $\ket{0}\in \cH_n^{}$, and $\ket{\mu}\in \cH_n^{n}$. Similarly to construction of the unitary (\ref{eq:unitary_B}), we distinguish two separate cases based on the binary value $c_n$. Notice that if $c_n=0$, the value $x_n \not\in \{p_1,\ldots,p_{n-1}\} $ is a new value, hence, we shall introduce a new element $\mu_{e_n}=1 $ in the sequence $\mu'=\mu_1,\ldots,\mu_{n-1}$. We will achieve it by firstly increasing the value of register $\ket{\mu_n}=\ket{0}$ by one, i.e.\ $\ket{\mu_n}=\ket{0}\mapsto\ket{1}$ by the following operator:
\begin{equation}
\label{eq:C0}
    C_0 : \ket{c_n}_{u}\ket{\mu_n}
    \mapsto 
    \begin{cases}
   \ket{c_n}_{u}\ket{\mu_n+1}
     & \text{if } c_n= 0,
    \\
    \ket{c_n}_{u}\ket{\mu_n}&\text{otherwise},
    \end{cases}
\end{equation}
which is a controlled $X$-gate acting on $\ket{\mu_n}$, and then by applying a sequence of operators $C_{n},C_{n-1},\ldots, C_2$, where 
\begin{equation}
\label{eq:Ck}
    C_k : \ket{c_n}_{u}\ket{e_n}_{p} \ket{\mu'_k}\ket{\mu'_{k-1}}
    \mapsto 
    \begin{cases}
    \ket{c_n}_{u}\ket{e_n}_{p} \ket{\mu'_{k-1}}\ket{\mu'_k}
     & \text{if } c_n= 0, \text{ and } e_n>n-k,
    \\
    \ket{c_n}_{u}\ket{e_n}_{p} \ket{\mu'_k}\ket{\mu'_{k-1}} &\text{otherwise},
    \end{cases}
\end{equation}
is a controlled SWAP-gate permuting registers $ \ket{\mu'_k}$ and $\ket{\mu'_{k-1}}$ if $c_n=0$ and  $e_n>n-k$. Here we use a convenction $\ket{\mu'_n}=\ket{0}$, see \cref{fig:subroutibe_C}. Notice that unitary operators $C_k$ do not commute, and are triggered only if $c_n=0$. Moreover, we have the following
\begin{align}
\label{eq:C_1}
    C_0C_{n} C_{n-1}\cdots C_{2} :&
    \ket{0}_{c}\ket{e_n}_{p}  \ket{\mu'}\ket{0} 
    \mapsto
    \ket{0}_{c}\ket{e_n}_{p}  \ket{\mu},\\
    \nonumber
    C_0C_{n} C_{n-1}\cdots C_{2} :&
    \ket{1}_{c}\ket{e_n}_{p}  \ket{\mu'}\ket{0} 
    \mapsto
    \ket{1}_{c}\ket{e_n}_{p}  \ket{\mu'}\ket{0} ,
\end{align}
which is a desired transformation into $\ket{\mu}$ if $c_n=0$. 

Furthermore, if $c_n=1$, we shall simply increase the value of $\ket{\mu_{e_n}}$ register, i.e.\ $\ket{\mu_{e_n}}\mapsto\ket{\mu_{e_n}+1}$. This is achieved by the sequence of gates $C_{n}',C_{n-1}',\ldots, C_2'$, where
\begin{equation}
\label{eq:Ck2}
    C_k' : \ket{c_n}_{u} \ket{e_n}_{p} \ket{\mu_k}
    \mapsto 
    \begin{cases}
    \ket{c_n}_{u} \ket{e_n}_{p} \ket{\mu_k+1}
     & \text{if } c_n= 1, \text{ and } e_n=k,
    \\
    \ket{c_n}_{u}\ket{e_n}_{p} \ket{\mu_k}
    &\text{otherwise},
    \end{cases}
\end{equation}
is a controlled $X$-gate acting on register $\ket{\mu_k}$, see \cref{fig:subroutibe_C}. Note that gates $C_k'$ commute, and their order is not important. It is easy to observe that at most one operator $C_k'$ will be triggered, and we have the following:
\begin{align}
\label{eq:C_2}
    C_{n-1}' C_{n-2}'\cdots C_{0}' :&
    \ket{1}_{c} \ket{e_n}_{p}  \ket{\mu}
    \mapsto
    \ket{0}_{c} \ket{e_n}_{p}  \ket{\mu},
    \\
    \nonumber
    C_{n-1}' C_{n-2}'\cdots C_{1}' :&
    \ket{1}_{c} \ket{e_n}_{p}  \ket{\mu'}\ket{0}
    \mapsto
    \ket{1}_{c} \ket{e_n}_{p}  \ket{\mu}.
\end{align}
Combining (\ref{eq:C_1}) with (\ref{eq:C_2}), we observe that 
\[
C=C_{n} C_{n-1} \cdots C_{2} \,C_{n-1}' C_{n-2}' \cdots C_{1}'
\]
achieves transformation (\ref{eq:unitary_C}).

\textbf{Subroutine $D$}. Fourth, we define unitary
\begin{equation}
    \label{eq:unitary_D}
    D: 
    \ket{c_n }_{u}\ket{e_n}_{p}\ket{\mu}
    \mapsto 
    \ket{0 }_{u}\ket{e_n}_{p}\ket{\mu}
\end{equation}
which controlled on the registers $\ket{\mu}$ and uncomputes position $c_n$, see \cref{fig:subroutibe_D}. We can achieve it by applying the sequence of unitary gates $D_n,\ldots,D_{1}$, where
\begin{equation}
\label{eq:Dk}
    D_k : \ket{i}_{u} \ket{e_n}_{p}\ket{\mu}
    \mapsto 
    \begin{cases}
    \ket{i+1}_{u} \ket{e_n}_{p}\ket{\mu} & \text{if } e_n= k, \text{ and } \mu_k\neq 1,
    \\
    \ket{c_n }_{u} \ket{e_n}_{p}\ket{\mu} &\text{otherwise},
    \end{cases}
\end{equation}
which is a controlled $X$-gate acting on register $\ket{\cdot }_{u}$, see \cref{fig:subroutibe_D}. Observe that at most one operator $D_k$ will be triggered, and only if $\mu_{e_n} \neq 0$ which is equivalent to the condition $c_n=1$. Therefore, 
\[
D=D_{n} D_{n-1}\cdots D_{1}
\]
achieves transformation (\ref{eq:unitary_D}).

\begin{figure}[htbp]
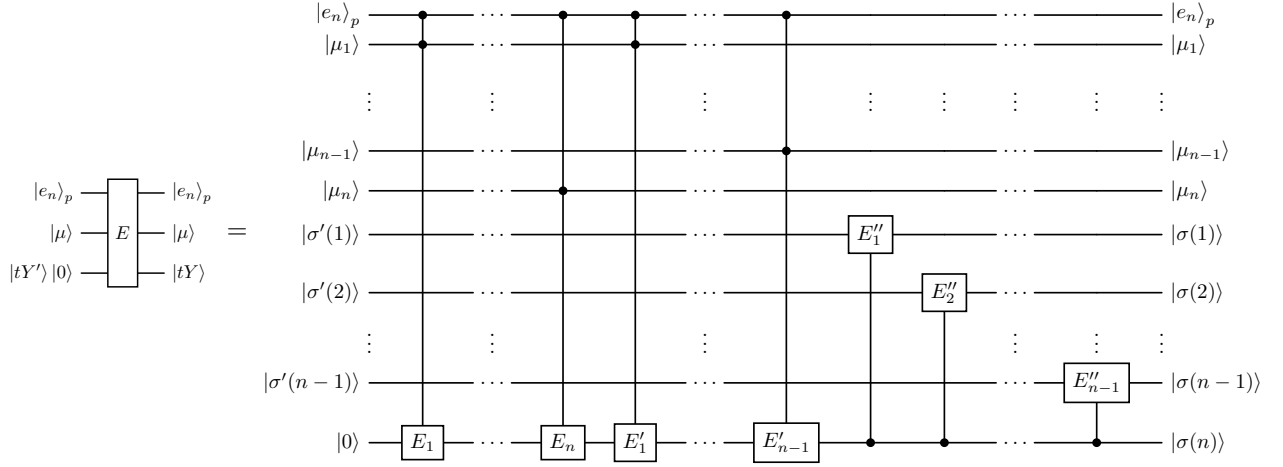
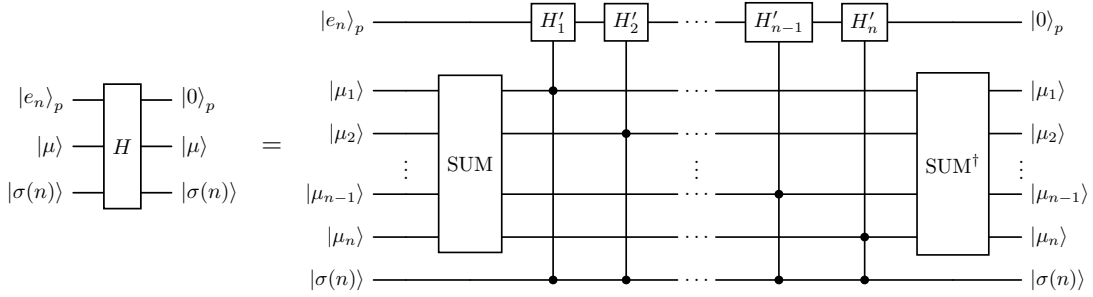

    \centering
    \begin{subfigure}{0.98\textwidth}
        \centering
        \adjincludegraphics[valign=c,width=0.17\textwidth,page=13]{figures/Prep_circuit.pdf}
        $=$
        \adjincludegraphics[valign=c,width=0.8\textwidth,page=14]{figures/Prep_circuit.pdf}
        \caption{Subroutine $E$ updates register $\ket{tY}$.  The state after gate $E_n$ is $\tfrac{1}{\sqrt{\mu_{e_n}}}\sum_{\ell=1}^{\mu_{e_n}}\ket{\ell}$, and after gate $E_{n-1}'$ it is $\tfrac{1}{\sqrt{\mu_{e_n}}}\sum_{\ell=\widetilde{\mu}_{e_n-1}+1}^{\mu_{e_n}}\ket{\ell}$.}
        \label{fig:subroutibe_E}
    \end{subfigure}
    \hfill
    \begin{subfigure}{0.95\textwidth}
        \centering
        \adjincludegraphics[valign=c,width=0.2\textwidth,page=15]{figures/Prep_circuit.pdf}
        \;$=$\;
        \adjincludegraphics[valign=c,width=0.65\textwidth,page=16]{figures/Prep_circuit.pdf}
        \caption{Subroutine $H$ uncomputes the value $e_n$ on register $\ket{\cdot}_{p}$. 
        Operator $\mathrm{SUM}$ maps $\mu$ to $\widetilde{\mu}$, see also \cref{fig:SUM_circ}.}
        \label{fig:subroutibe_F}
    \end{subfigure}
    \caption{Last two subroutines in preparation circuit $\Prep_n$ from \cref{fig:subroutibesABCDE} are sequences of controlled $X$-, and SWAP-gates.  
    Subroutine $E$ updates the register $\ket{tY'}$ to $\ket{tY}$, while subroutine $H$ uncomputes the value $e_n$ on the position register $\ket{\cdot}_p$.}
    \label{fig:subroutibesEF}
\end{figure}

\textbf{Subroutine $E$}. Fifth, we update the register $\ket{tY'}$ to $\ket{tY}$, where $\ket{tY'}$ is a transversal of the sequence $x_1,\ldots,x_{n-1}$ and $\ket{tY}$ is a transversal of the sequence $x_1,\ldots,x_{n}$. Indeed, we shall construct unitary:
\begin{equation}
    \label{eq:unitary_E}
    E: 
    \ket{e_n}_{p}\ket{\mu}\ket{tY'}\ket{0} 
    \mapsto 
    \ket{e_n}_{p}\ket{\mu}\ket{tY},
\end{equation}
notice that $\ket{tY'}\in \cH_n^{n-1}$ is supported on $n-1$ registers, the last register is initiated in $\ket{0}\in \cH_n$, and $\ket{tY}\in \cH_n^{n}$. Moreover, transversal elements are encoded in the way presented in \cref{eq:tY_ket_2}. In order to achieve the above permutation, we shall observe how $\ket{tY'}\ket{0} $ and $\ket{tY}$ differs. Notice that if $\sigma\in \S_n$ satisfies condition 
\begin{equation}
\label{eq:star}
x_{\sigma (1)}\leq\cdots\leq x_{\sigma (n)},
\end{equation}
then $\sigma (n)$ is equal to one of the following values:
\begin{equation}
    \sigma (n)=\widetilde{\mu}_{e_{n}-1}+1,\widetilde{\mu}_{e_{n}-1}+2, \ldots,   \widetilde{\mu}_{e_{n}-1} +\mu_{e_{n}}
\end{equation}
where $\widetilde{\mu}_k:=\sum_{i=1}^k \mu_i$, and notice that each of the values for $\sigma (n)$ appears for exactly the same number of permutations $\sigma\in \S_n$ which satisfy (\ref{eq:star}). Therefore, we have 
\begin{equation}
\label{eq:first_step}
\sum_{\substack{\sigma\in \S_n:\\ x_{\sigma (1)}\leq\cdots\leq x_{\sigma (n)}}} \ket{\sigma (n)} 
\propto \sum_{\ell =1}^{\mu_{e_{n}}} \ket{\widetilde{\mu}_{e_{n-1}}+ \ell} =
\sum_{\ell =\widetilde{\mu}_{e_{n-1}}+1}^{\widetilde{\mu}_{e_{n-1}}+\mu_{e_{n}}} \ket{\ell}.
\end{equation} 
Moreover, notice that we have the following factorization of elements in transversal vector $\ket{tY}$:
\begin{equation}
\label{eq:sec_step}
\ket{tY}\propto
\sum_{\substack{\sigma\in \S_n:\\ x_{\sigma (1)}\leq\cdots\leq x_{\sigma (n)}}} \ket{\sigma (1),\ldots,\sigma (n)} 
= \sum_{\substack{\sigma'\in \S_n:\\ x_{\sigma '(1)}\leq\cdots\leq x_{\sigma '(n-1)}}}
\sum_{\ell =\widetilde{\mu}_{e_{n-1}}+1}^{\widetilde{\mu}_{e_{n-1}}+\mu_{e_{n}}} 
\ket{\sigma'_\ell (1),\ldots,\sigma_\ell' (n-1)} 
\ket{\ell}.
\end{equation} 
where 
\begin{equation}
\label{eq:sec_step2}
  \sigma_\ell' (k):=
    \begin{cases}
        \sigma' (k) & \text{if } \sigma' (k) <\ell \\
        \sigma' (k)+1 & \text{otherwise } .
    \end{cases}
\end{equation}
Based on formulas (\ref{eq:first_step}) and (\ref{eq:sec_step}), we shall achieve transformation $\ket{tY'}\ket{0} \mapsto \ket{tY}$ in two steps. Firstly, based on values $\mu_k$ for $ k\in [n]$ and value $e_n$ we shall transform last register $\ket{0}$ into a correct form in accordance with (\ref{eq:first_step}). Indeed, consider a sequence of gates $E_1,\ldots, E_{n}$ where 
\begin{equation}
\label{eq:Ek}
    E_k : \ket{e_n}_{p} \ket{\mu_k}\ket{0}
    \mapsto 
    \begin{cases}
   \ket{e_n}_{p} \ket{\mu_k} \tfrac{1}{\sqrt{\mu_k}}\sum_{\ell=1}^{\mu_k} \ket{\ell}
     & \text{if } e_n= k, 
    \\
    \ket{e_n}_{p} \ket{\mu_k}\ket{0}&\text{otherwise},
    \end{cases}
\end{equation}
are transforming register $\ket{\sigma (n)}:=\ket{0}$ into an equal superposition of vectors $\tfrac{1}{\sqrt{\mu_k}}\sum_{\ell=1}^{\mu_k} \ket{\ell}$. Notice that only one gate  $E_k$ will be triggered (for $k=e_n$), and hance, we have:
\begin{align}
\label{eq:E_1}
    E_{1} E_{2}\cdots E_{n} :&
    \ket{e_n}_{p} \ket{\mu_k}\ket{0}
    \mapsto
   \ket{e_n}_{p} \ket{\mu_k} \tfrac{1}{\sqrt{\mu_{e_n}}}\sum_{\ell=1}^{\mu_{e_n}} \ket{\ell}. 
\end{align}
as presented on \cref{fig:subroutibe_E}. 
Notice that transformation (\ref{eq:Ek}) might be achieved by different means, for instance a $\text{QFT}_{\mu_k}$ transformation on $\mu_k$ dimensional subspace triggered only of $e_n= k$ and with the size of $\text{QFT}_{\mu_k}$ transformation itself is controlled on the value $\ket{\mu_k}$. Another way to achieve transformation (\ref{eq:Ek}) is by a sequence 
\begin{equation}
    \label{eq:Ek0}
    E_k=E_{k0} E_{k1}\cdots E_{kn}
\end{equation}
of controlled $2 \times 2$ gates defined as
\begin{equation}
    E_{k0}
\end{equation}
and for $j>0$ as:
\begin{equation}
\label{eq:Ek}
    E_{kj} : \ket{e_n}_{p} \ket{\mu_k}\ket{i}
    \mapsto 
    \begin{cases}
   \ket{e_n}_{p} \ket{\mu_k} \Big(\tfrac{1}{\sqrt{\mu_k+1-j}} \ket{i} + \tfrac{\sqrt{\mu_k-j}}{\sqrt{\mu_k+1-j}} \ket{i+1} \Big) 
     & \text{if } e_n= k, \, j=i, \text{ and } \mu_k>j
    \\
    \ket{e_n}_{p} \ket{\mu_k} \Big(\tfrac{\sqrt{\mu_k-j}}{\sqrt{\mu_k+1-j}} \ket{i-1} -\tfrac{1}{\sqrt{\mu_k+1-j}} \ket{i} \Big) 
     & \text{if } e_n= k, \, j=i+1, \text{ and } \mu_k>j
    \\
    \ket{e_n}_{p} \ket{\mu_k}\ket{i}&\text{otherwise},
    \end{cases}
\end{equation}
It is easy to see, that for $k=e_n$, we have:
\begin{multline}
    \ket{e_n}_{p} \ket{\mu_k}\ket{0}
\xrightarrow{E_{k0}}
    \ket{e_n}_{p} \ket{\mu_k}\ket{1}
\xrightarrow{E_{k1}}
    \ket{e_n}_{p} \ket{\mu_k}\Big( \tfrac{1}{\sqrt{\mu_k}}\ket{1} +\tfrac{\sqrt{\mu_k-1}}{\sqrt{\mu_k}}\ket{2} \Big)
\xrightarrow{E_{k2}} \\ \xrightarrow{E_{k2}}
   \ket{e_n}_{p} \ket{\mu_k}\Big( \tfrac{1}{\sqrt{\mu_k}}\ket{1} +\tfrac{1}{\sqrt{\mu_k}}\ket{2} +\tfrac{\sqrt{\mu_k-2}}{\sqrt{\mu_k}}\ket{3} \Big)
\xrightarrow{E_{k3}}  
   \cdots
\xrightarrow{E_{k( \mu_k -1)}} 
   \ket{e_n}_{p} \ket{\mu_k} \tfrac{1}{\sqrt{\mu_k}}\sum_{\ell=1}^{\mu_k} \ket{\ell}
\xrightarrow{E_{k \mu_k }} \\ \xrightarrow{E_{k \mu_k }}
   \ket{e_n}_{p} \ket{\mu_k} \tfrac{1}{\sqrt{\mu_k}}\sum_{\ell=1}^{\mu_k} \ket{\ell}
\xrightarrow{E_{k( \mu_k +1)}}   
   \cdots
\xrightarrow{E_{kn}}
\ket{e_n}_{p} \ket{\mu_k} \tfrac{1}{\sqrt{\mu_k}}\sum_{\ell=1}^{\mu_k} \ket{\ell} ,
\end{multline}
wheres for any $k\neq e_n$ all gates $E_{kj}$ acts as identity. This observation shows that (\ref{eq:Ek0}) and (\ref{eq:E_1}) holds true.

Subsequently, we shall apply a sequence of gates $E_1'\cdots E_{n-1}'$ where 
\begin{equation}
\label{eq:Ekp}
    E_k' : \ket{e_n}_{p} \ket{\mu_k}\ket{i}
    \mapsto 
    \begin{cases}
   \ket{e_n}_{p} \ket{\mu_k}\ket{i+ \mu_k}
     & \text{if } e_n> k, 
    \\
    \ket{e_n}_{p} \ket{\mu_k}\ket{i}&\text{otherwise},
    \end{cases}
\end{equation}
which is a controlled shift gate $\ket{i}\mapsto\ket{i+\mu_k}$ acting on register corresponding to $\ket{\sigma (n)}$ as presented on \cref{fig:subroutibe_E}. Notice that such gates will be triggered for all values $k<e_n$, and hence by combining it with (\ref{eq:E_1}), we obtain:
\begin{align}
\label{eq:E_1p}
    E_1'\cdots E_{n-1}'
    :&
    \ket{e_n}_{p} \ket{\mu} \tfrac{1}{\sqrt{\mu_{e_n}}}\sum_{\ell=1}^{\mu_{e_n}} \ket{\ell}
    \mapsto
   \ket{e_n}_{p} \ket{\mu} \tfrac{1}{\sqrt{\mu_{e_n}}}\sum_{\ell=\widetilde{\mu}_{e_{n-1}}+1}^{\widetilde{\mu}_{e_{n-1}}+\mu_{e_n}} 
   \ket{\ell}. 
\end{align}
Combining (\ref{eq:E_1}) with (\ref{eq:E_1p}) we obtain:
\begin{align}
\label{eq:E_1p_p}
    E_1'\cdots E_{n-1}'
 E_1\cdots E_{n}
    :&
    \ket{e_n}_{p} \ket{\mu}  \ket{0}
    \mapsto
   \ket{e_n}_{p} \ket{\mu} \tfrac{1}{\sqrt{\mu_{e_n}}}\sum_{\ell=\widetilde{\mu}_{e_{n-1}}+1}^{\widetilde{\mu}_{e_{n-1}}+\mu_{e_n}} 
   \ket{\ell}. 
\end{align}
and hence by comparing it with (\ref{eq:first_step}), we observe that we correctly computed state in register $\ket{\sigma (n)}$. 

Based on previousely computed register $\ket{\sigma (n)}$, we shall update registers $\ket{\sigma'(1),\ldots,\sigma'(n-1)}$. In order to do it, we shall use (\ref{eq:sec_step}) and (\ref{eq:sec_step2}). Indeed, we introduce another sequence of gates $E_{1}''E_{2}''\cdots E_{n-1}''$, where
\begin{equation}
\label{eq:Ek3}
    E_k'' : \ket{\sigma '(k)} \ket{\ell}
    \mapsto 
    \begin{cases}
    \ket{\sigma '(k)} \ket{\ell}
     & \text{if } j<\ell  , 
    \\
    \ket{\sigma '(k)+1} \ket{\ell}
     & \text{if } j\geq \ell, j\neq n   
    \\
    \ket{\ell} \ket{\ell}
    & \text{if }  j = n   
    \end{cases}
\end{equation}
is a controlled cyclic shift on a $(n-\ell)$ dimensional subspace acting on register $\ket{\sigma (k)'}$, see \cref{fig:subroutibe_E}. Notice that, as $\sigma(k)'<n$, the third case in (\ref{eq:Ek3}) never occure. Notice that gates $E_k''$ commute, and their order is not important. Moreover, notice that
\begin{align*}
\ket{\sigma'(1),\sigma '(2),\ldots,\sigma'(n-1)}\ket{\ell}
&
\xrightarrow{E_{1}''}
\ket{\sigma_\ell'(1),\sigma '(2),\ldots,\sigma'(n-1)}\ket{\ell}
\xrightarrow{E_{2}''}
\ket{\sigma_\ell'(1),\sigma_\ell '(2),\ldots,\sigma'(n-1)}\ket{\ell}
\xrightarrow{E_{3}''}
\\&
\cdots
\xrightarrow{E_{n-1}''}
\ket{\sigma_\ell'(1),\sigma '(2),\ldots,\sigma_\ell'(n-1)}\ket{\ell}
\end{align*}
for arbitrary $\sigma'\in \S_{n-1}$ such that $x_{\sigma'(1)}\leq x_{\sigma '(2)}\leq\cdots\leq x_{\sigma'(n-1)}$, and hence 
\begin{align}
\label{eq:E_1bis}
E_{1}''E_{2}''\cdots  E_{n-1}'' \;
    : \; &
\ket{\sigma'(1),\sigma '(2),\ldots,\sigma'(n-1)}\ket{\ell}
    \mapsto
\ket{\sigma_\ell'(1),\sigma_\ell'(2),\ldots,\sigma_\ell'(n-1)}\ket{\ell}
\end{align}
Combining (\ref{eq:E_1p_p}) with (\ref{eq:E_1bis}), we obtain
\begin{align}
\label{eq:E_1p_pp}
E_{1}''E_{2}''\cdots  E_{n-1}''
E_1'\cdots E_{n-1}'
 E_1\cdots E_{n} \;
    : \; &
    \ket{e_n}_{p} \ket{\mu} \Big( \ket{\sigma'(1),\sigma '(2),\ldots,\sigma'(n-1)}\ket{0}\Big)
    \mapsto 
    \\&
    \nonumber
   \ket{e_n}_{p} \ket{\mu} 
   \Big( \tfrac{1}{\sqrt{\mu_{e_n}}}\sum_{\ell=\widetilde{\mu}_{e_{n-1}}+1}^{\widetilde{\mu}_{e_{n-1}}+\mu_{e_n}} \ket{\sigma'(1),\sigma '(2),\ldots,\sigma'(n-1)}\ket{\ell}\Big) 
   ,
\end{align}
for arbitrary $\sigma'\in \S_{n-1}$ such that $x_{\sigma'(1)}\leq x_{\sigma '(2)}\leq\cdots\leq x_{\sigma'(n-1)}$. Combining the above expression with exact form for $\ket{tY'}$ and formula (\ref{eq:sec_step}), we obtain
\begin{align}
\label{eq:E_1p_pp}
\ket{e_n}_{p} \ket{\mu}\Big(\ket{tY'}\ket{0} \Big)=&
    \ket{e_n}_{p} \ket{\mu} 
    \Big( \tfrac{1}{\sqrt{|Y_{\mu'}|}}\sum_{\substack{\sigma'\in \S_n:\\ x_{\sigma '(1)}\leq\cdots\leq x_{\sigma '(n-1)}}}
    \ket{\sigma'(1),\sigma '(2),\ldots,\sigma'(n-1)}\ket{0}\Big)
    \xrightarrow{E_{1}''E_{2}''\cdots  E_{n-1}'' E_1'\cdots E_{n-1}' E_1\cdots E_{n}}
    \\&
    \nonumber
   \ket{e_n}_{p} \ket{\mu} 
   \Big( \tfrac{1}{\sqrt{|Y_{\mu'}|\mu_{e_n}}}\sum_{\ell=\widetilde{\mu}_{e_{n-1}}+1}^{\widetilde{\mu}_{e_{n-1}}+\mu_{e_n}} \ket{\sigma'(1),\sigma '(2),\ldots,\sigma'(n-1)}\ket{\ell}\Big) 
   =
\ket{e_n}_{p} \ket{\mu}\ket{tY},
\end{align}
and hence transformation $E=E_{1}''E_{2}''\cdots  E_{n-1}''
E_1'\cdots E_{n-1}'
 E_1\cdots E_{n}$ achieves transformation (\ref{eq:unitary_E}).

\textbf{Subroutine $H$}. Lastly, we define unitary
\begin{equation}
    \label{eq:unitary_F}
    H: 
    \ket{e_n}_{p}\ket{\mu}\ket{\sigma (n)}
    \mapsto 
    \ket{0}_{p}\ket{\mu}\ket{\sigma (n)}
\end{equation}
which uncomputes the position value $e_n$ on the position register $\ket{\cdot}_p$ controlled on $\ket{\mu} $ and $\sigma (n)\in \cH_n$, which is the last sub-register in $\ket{Yt} \in \cH_n^n$. In that way, unitary $H$ uncomputes position $e_n$, see \cref{fig:subroutibe_F}. We achieve it by first applying the sequence of controlled shift-gates denoted as $H_1,\ldots,H_{n-1}$, where
\begin{equation}
\label{eq:Fk}
    H_k :\ket{\mu_k}\ket{\mu_{k+1}}
    \mapsto 
    \ket{\mu_k}\ket{\mu_{k+1}+\mu_k}
\end{equation}
is controlled on register $\ket{\mu_k}$, see \cref{fig:subroutibe_F}. Notice that gates $H_k$ do not commute and sequence $H_{n-1}\cdots H_1$ effectively applies the following transformation to register $\ket{\mu}$:
\begin{equation}
\label{eq:Fkk}
    \mathrm{SUM}:\ket{\mu}
    \mapsto 
    \ket{\widetilde{\mu}}
\end{equation}
where 
\begin{equation}
    \ket{\widetilde{\mu}} =\ket{\widetilde{\mu}_1,\ldots,\widetilde{\mu}_n}, 
    \quad \quad \text{where} \quad 
    \widetilde{\mu}_k:=\sum_{i=1}^k \mu_i.
\end{equation}
In accordance to (\ref{eq:first_step}), register $\ket{\sigma (n)}$ is in the following superposition:
\begin{equation}\ket{\sigma (n)} 
 =
\sum_{\ell =\widetilde{\mu}_{e_{n-1}}+1}^{\widetilde{\mu}_{e_{n}}} \ket{\ell} .
\end{equation}
In other words, the value $\sigma (n)$ stored in register $\ket{\sigma (n)}$ satisfies the following inequalities:
\begin{equation}
\label{eq:Fin}
 \widetilde{\mu}_{e_n-1}  <\sigma (n)\leq  \widetilde{\mu}_{e_n} ,
\end{equation}
and as $\widetilde{\mu_k} $ is a non-decreasing series, $\sigma (n) \leq  \widetilde{\mu}_{k}$ for all $k\leq e_n$, while $\sigma (n) >  \widetilde{\mu}_{k}$ for all $k> e_n$. 
This observation allows us to uncompute the value $e_n$. Indeed, we can achieve this by the following sequence of gates $H_1',\ldots,H_n'$, where 
\begin{equation}
\label{eq:Fkprim}
    H_k': \ket{e_n}_{p} \ket{\widetilde{\mu}_k}\ket{\sigma (n)}
    \mapsto 
    \begin{cases}
   \ket{e_n-1}_{p} \ket{\widetilde{\mu}_k}\ket{\sigma (n)}
     & \text{if } \sigma (n)\leq  \widetilde{\mu}_k  , 
    \\
     \ket{e_n}_{p} \ket{\widetilde{\mu}_k}\ket{\sigma (n)}
     &\text{otherwise},
    \end{cases}
\end{equation}
is a shift $\ket{e_n}_{p}\mapsto\ket{e_n-1}_{p}$ on the position register $\ket{\cdot}_{p}$ triggered if $\sigma (n)\geq  \widetilde{\mu}_k$. Based on observation (\ref{eq:Fin}), such a controlled shift will be triggered for all $k\leq e_n$, as a result
\begin{equation}
    H_n',\cdots H_1':
    \ket{e_n}_{p} \ket{\widetilde{\mu}}\ket{\sigma (n)}
    \mapsto
    \ket{0}_{p} \ket{\widetilde{\mu}}\ket{\sigma (n)}
\end{equation}
correctly uncomputes the register $\ket{\cdot}_p$. 

Lastly, we define
\begin{equation}
\label{eq:Fkbis}
    H_k'' :\ket{\mu_k}\ket{\mu_{k+1}}
    \mapsto 
    \ket{\mu_k}\ket{\mu_{k+1}-\mu_k}
\end{equation}
is controlled on register $\ket{\mu_k}$, see \cref{fig:subroutibe_F}. Notice that operation $H_k''$ is an inverse of operation $H_k$ defined in (\ref{eq:Fk}). Therefore, sequence $H_{n-1}''\cdots H_1''=\mathrm{SUM}^\dagger$, which effectively applies the following transformation to register $\ket{\mu}$:
\begin{equation}
   \mathrm{SUM}^\dagger:\ket{\widetilde{\mu}}
    \mapsto 
   \ket{\mu} 
\end{equation}
see (\ref{eq:Fk}). 

Summarizing, transformation $H \defeq \mathrm{SUM}^\dagger \cdot 
\,H_{n}'\cdots H_1'\cdot 
\mathrm{SUM} $ achieves transformation (\ref{eq:unitary_F}), as presented on \cref{fig:subroutibe_F}. 

\subsection{Computational complexity}
\label{sec:complexity_Prep}

In this section, we argue that the computational complexity of the preparation circuit $ \Prep$ is $\widetilde{O}(n^3)$. Firstly, notice that such a circuit is defined recursively as $\mathrm{P} = \Prep_1 \Prep_2 \cdots \Prep_n$, see \cref{eq:Perp_induction} and \cref{fig:Prep}. Hence it is enough to show that $\Prep_n$ has computational complexity $\widetilde{O}(n^2)$. 

In fact, the circuit $\Prep_n$ consists of six subroutines $A,B,C,D,E$ and $H$, as presented in \cref{fig:subroutibesABCDE}. 

Subroutines $A,B,C,D$ consist of a linear number of gates $A_k,A_k',B_k,B_k',C_k,C_k' $ and $D_k$. All of those gates perform a clasical computation, as addition, subtraction, or shift, all controlled on the constant (one or two) number of registers. As such, they can be prepared with $\widetilde{O}(1)$ complexity each. This shows that $A,B,C,D$ has linear total gate and depth complexity $\widetilde{O}(n)$. 

On the other hand, gates $E_k$ are defined as a sequence $E_k=E_{k0}\cdots E_{kn}$, see \cref{eq:Ek0}. Each $E_{ki}$ defines a controlled Givens rotation, hance can be implemented with $\widetilde{O}(n)$ complexity. 
Gates $E_k', E_k''$ can be implemented with $\widetilde{O}(1)$ by the same argument in previous paragraph. 
This shows that $E$ has quadratic total gate and depth complexity $\widetilde{O}(n^2)$.

Lastly, subroutine $H$ consists of the SUM subroutine and the linear number of controlled shift gates. From \cref{fig:SUM_circ} we see that complexity of the SUM gate is $\widetilde{O}(n)$, which leads to $\widetilde{O}(n)$ complexity of $H$. 

For completeness, we also note that P uses only constant amount of ancillas in total (since each subroutine is sequential and we can reuse the ancillas), so its space complexity is $O(1)$.

\subsection{Alternative encoding for permutations}

In this section, we present an alternative encoding for arbitrary element $\sigma\in \S_n$, and a quantum circuit subroutine which switches between a standard encoding and the alternative one. 

Observe that arbitrary permutation $\sigma\in \S_n$ is uniquely characterized by the sequence $(i_n,\ldots, i_2)$ where $i_k\in \mathbb{Z}_k$ via following correspondence:
\begin{equation}
\label{eq:perm_correspo}
\sigma 
=
c_{(1\cdots n)}^{i_n} \cdot
c_{(1\cdots n-1)}^{i_{n-1}} \cdots
c_{(12)}^{i_2}
\end{equation}
where $c_{(1\cdots k)}$ denotes a cyclic shift $1\rightarrow 2$, $2\rightarrow 3$, $\ldots$, $k\rightarrow 1$. Notice that the number of distinct permutations $\sigma\in \S_n$ coincides with the number of tuples $(i_n,\ldots, i_2)$ where $i_k\in \mathbb{Z}_k$. 

Consider now arbitrary permutation $\sigma\in \S_n$ determined by numbers $\sigma (1),\ldots,\sigma (n)$. In the following, we present how to determine touple $(i_n,\ldots, i_2)$ corresponding to $\sigma$ via (\ref{eq:perm_correspo}). 
Firstly, notice that
\begin{equation}
\label{eq:perm_E}
i_n=\sigma (n) .
\end{equation}
Secondly, based on the value $i_n$ and permutation $\sigma$, we define the following permutation 
\begin{align}
\label{eq:perm_C}
\sigma' :=&
c_{(1\cdots n)}^{n-i_n} \sigma =
c_{(1\cdots n-1)}^{i_{n-1}} \cdots
c_{(12)}^{i_2} .
\end{align}
Consequtively, based on the form of permutation $\sigma'$ in (\ref{eq:perm_C}), one reads
\begin{equation}
\label{eq:perm_D}
i_{n-1}=\sigma' (n-1) .
\end{equation}
Furthermore, based on the value $i_{n-1}$ and permutation $\sigma'$, we define the following permutation 
\begin{align}
\label{eq:perm_F}
\sigma'' :=&
c_{(1\cdots n-1)}^{(n-1)-i_{n-1}} \sigma '=
c_{(1\cdots n-2)}^{i_{n-2}} \cdots
c_{(12)}^{i_2} .
\end{align}
Following this procedure, we recursively encode tuple $(i_n,\ldots, i_2)$ corresponding to $\sigma$ via (\ref{eq:perm_correspo}). 

Based on the observations in previous paragraph, we define the following  quantum gates:
\begin{align}
\label{eq:perm_A}
G_k: 
\ket{i}\ket{j}
&
\mapsto
\ket{i}\ket{j+i \, (\text{mod }k)  }
\\
\label{eq:perm_B}
G_k': 
\ket{i}\ket{j}
&
\mapsto
\ket{i+k-j}\ket{j}
\end{align}
for arbitrary $i\in [n]$ and $j\in [k]$. Notice that gates $G_k,G_k'$ are controlled shift gates acting on the second and the first register respectively. As we shall see, arranging those gates as presented on \cref{Fig:perm} achieves the following transformation:
\begin{equation}
\label{eq:perm_H}
\ket{\sigma (1)}\cdots \ket{\sigma (n)}
\ket{0}\cdots\ket{0}
\mapsto
\ket{1}\cdots \ket{ n}
\ket{i_2}\cdots\ket{i_n}
\end{equation} 
where tuple $(i_n,\ldots, i_2)$ corresponds to $\sigma$ via (\ref{eq:perm_correspo}). 

Indeed,  by applying controlled gate $G_n$, see \cref{eq:perm_A}, to the last register, we achieve transformation 
\begin{align*}
&
\ket{\sigma (1)}\cdots \ket{\sigma (n)}
\ket{0}\cdots\ket{0}\ket{0}
\mapsto
\\&
\ket{\sigma (1)}\cdots \ket{\sigma (n)}
\ket{0}\cdots\ket{0}\ket{i_n}
\end{align*}
which follows form (\ref{eq:perm_E}). 
Furthermore, applying gate $G_n'$ to first $n$ registers controlled on the last register, see \cref{eq:perm_B},  we achieve transformation 
\begin{align*}
&
\ket{\sigma (1)}\cdots \ket{\sigma (n-1)}\ket{\sigma (n)}
\ket{0}\cdots\ket{0}\ket{i_n}
\mapsto
\\&
\ket{\sigma' (1)}\cdots \ket{\sigma '(n-1)}\ket{n}
\ket{0}\cdots\ket{0}\ket{i_n}
\end{align*}
where permutation $\sigma' \in \S_{n-1}$ is determined from $\sigma$ via \cref{eq:perm_C}, in particular
\begin{align*}
\sigma' =
c_{(1\cdots n-1)}^{i_{n-1}} \cdots
c_{(12)}^{i_2} .
\end{align*}
Consecutively, by applying controlled $G_{n-1}$ gate, we achieve transformation
\begin{align*}
&
\ket{\sigma' (1)}\cdots \ket{\sigma '(n-1)}\ket{n}
\ket{0}\cdots\ket{0}\ket{0}\ket{i_n}
\mapsto
\\&
\ket{\sigma' (1)}\cdots \ket{\sigma '(n-1)}\ket{n}
\ket{0}\cdots\ket{0}\ket{i_{n-1}}\ket{i_n}
\end{align*}
which follows form (\ref{eq:perm_D}). 
Furthermore, applying gate $G_{n-1}'$ to first $n-1$ registers, see \cref{eq:perm_B}, we achieve transformation 
\begin{align*}
&
\ket{\sigma' (1)}\cdots \ket{\sigma '(n-1)}\ket{n-1}\ket{n}
\ket{0}\cdots\ket{0}\ket{i_{n-1}}\ket{i_n}
\mapsto
\\&
\ket{\sigma'' (1)}\cdots \ket{\sigma ''(n-2)}\ket{n-1}\ket{n}
\ket{0}\cdots\ket{0}\ket{i_{n-1}}\ket{i_n}
\end{align*}
where permutation $\sigma'' \in \S_{n-2}$ is determined from $\sigma'$ via \cref{eq:perm_F}. 
One can see that this recursive procedure leads to transformation (\ref{eq:perm_H}).

\section{Quantum circuits for step 3: implementation of the isometry $V$}\label{sec:step_3_circuits}

In this section, we build a quantum circuit for the isometry $V$, defined as in \cref{eq:V_blocks} and \cref{def:V_matrix_elements}, such that $V_{\lambda,\mu}^{\dagger}$ compresses the $T$ register (elements of which correspond to standard Young tableaux) formed after the $\mathrm{QFT_{S_n}}$, into the $\widetilde{M}$ register (elements of which correspond to Gelfand--Tsetlin patterns). Recall that this compression is possible because we have a guarantee that our quantum state after $\mathrm{QFT_{S_n}}$ lies within $\mathrm{im}(V)$, so the circuit is reversible.

\begin{figure}[H]
    \centering
    \adjincludegraphics[valign=c,width=1.0\textwidth,page=1]{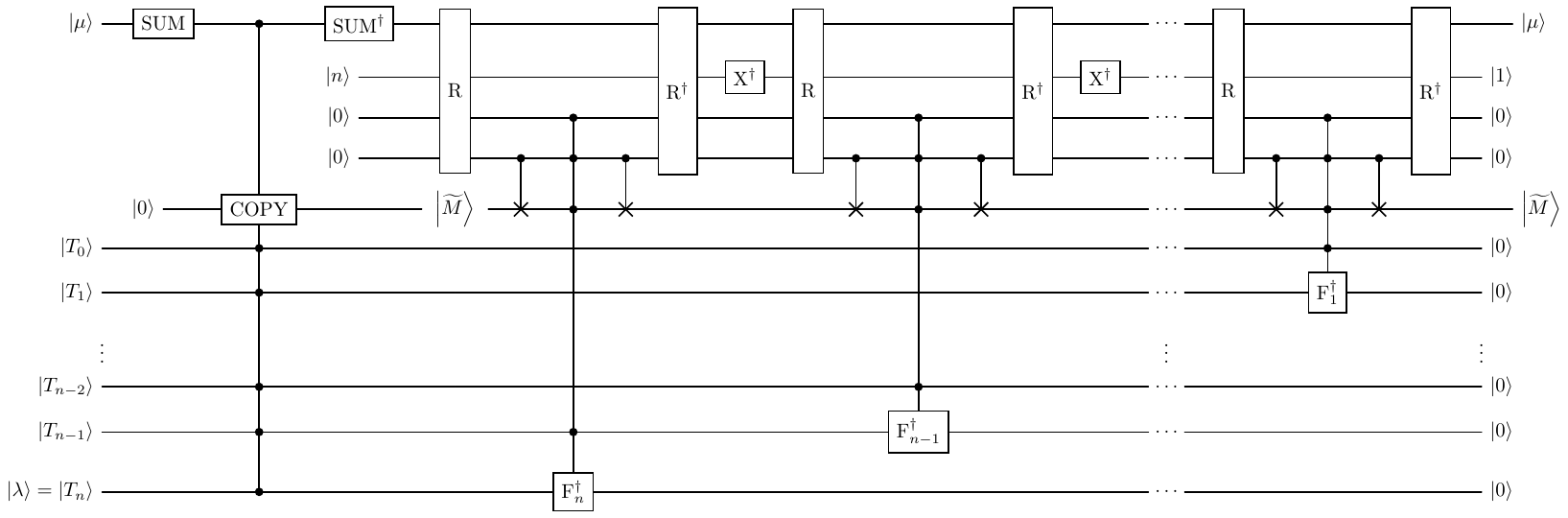}
    \caption{Circuit for isometry $V$. The R, COPY, SUM, and controlled-SWAP gates are defined in \cref{fig:R_circ}, \cref{fig:PREPM_circ}, \cref{fig:SUM_circ}, \cref{fig:CSWAP_circ}. The $\mathrm{X}$ gate is the generalized Pauli-$X$, and as such $\mathrm{X}^{\dagger}$ is decrementing the numerical entry of the input state by 1 cyclically. The controlled $\mathrm{F}_l$ gates, $l \in \{1,\dots,n\}$, should be read from right to left, as converting from the split basis to the Young--Yamanouchi basis, so in our notation from left to right we put their complex conjugate. Here the entries $\sof{F^{a,b,c}_{d}}^f_e$ of the $F$-unitaries in this circuit are recovered from explicitly controlling on the wires containing the information about the $a,c,d,f$ indices from \cref{def:V_matrix_elements} ($b$ is always a single box), with $e$ as the output.}
    \label{fig:V_circ}
\end{figure}

The isometry $V$ could be implemented recursively using controlled unitaries $F^{a,b,c}_{d}$, as shown in \cref{fig:V_circ}. 
Recall that to convert from GT patterns to SYT, one can describe a transform between the split basis and the Young--Yamanouchi basis, see \cref{fig:V_tn_def}. 
Switching between these two different coupling schemes is a change in the fusion order, and as such can be done using $F$-moves (see \cref{fig:F_move}), which are in turn described by $F$-symbols. 
In our case, the $F$-symbols take on a special simple form as in \cref{eq:lemma_F}. Notice that if in $\sof*{F^{(\mu_k-j),(1),T^{k-1}_{j-1}}_{\widetilde{M}_k}}^{T^{k-1}_j}_{(\mu_k-j+1)}$ from \cref{def:V_matrix_elements} we treat $T_j^{k-1}$ as the output wire of the unitary $F^{a,b,c}_{d}$, then it is easy to see that the isometry $V$ has a recursive structure in $T_j^{k}$, since any $F$-move unitary producing $T_j^{k-1}$ depends only on the previous entry of the SYT, namely $T_{j-1}^{k-1}$.

\begin{figure}[H]
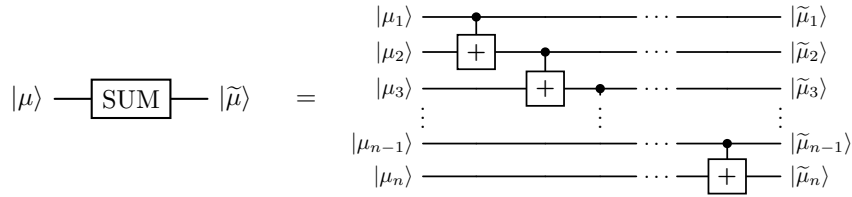

    \centering
    \adjincludegraphics[valign=c,width=0.20\textwidth,page=2]{figures/Isometry_V.pdf}
    \quad$=$\quad
    \adjincludegraphics[valign=c,width=0.4\textwidth,page=3]{figures/Isometry_V.pdf}
    \caption{Circuit for the SUM gate.}
    \label{fig:SUM_circ}
\end{figure}

\textbf{$\widetilde{M}$ register preparation.} In the circuit for isometry $V$, the $\mathrm{F}_l$ gates are used to prepare ($\mathrm{F}_l^{\dagger}$ to unprepare) the SYT $T$ in the Young--Yamanouchi basis, using the composition $\mu$ and the GT pattern $\widetilde{M}$ as control, for $l \in \{1,\dots,n\}$. Reading the circuit \cref{fig:V_circ} from left to right, we first need to prepare the $\widetilde{M}$ register to be able to control. We can prepare $\widetilde{M}$ using controlled gates on the $T$ register and the register $\widetilde{\mu}$. Recall that $\widetilde{\mu}$ is how we denote the register containing the sums of the first $k$ entries of $\mu$, $k \in \{1,\dots,n\}$, that is:
\begin{equation}
    \widetilde{\mu}_k:=\sum_{t=1}^k \mu_t.
\end{equation}
It is easy to prepare $\widetilde{\mu}$ from $\mu$ using the SUM gate, which takes all $\mu$ as input and outputs the sum $\sum_{t=1}^i \mu_t$ on each wire $i$. It is defined in \cref{fig:SUM_circ}.

\begin{figure}[H]
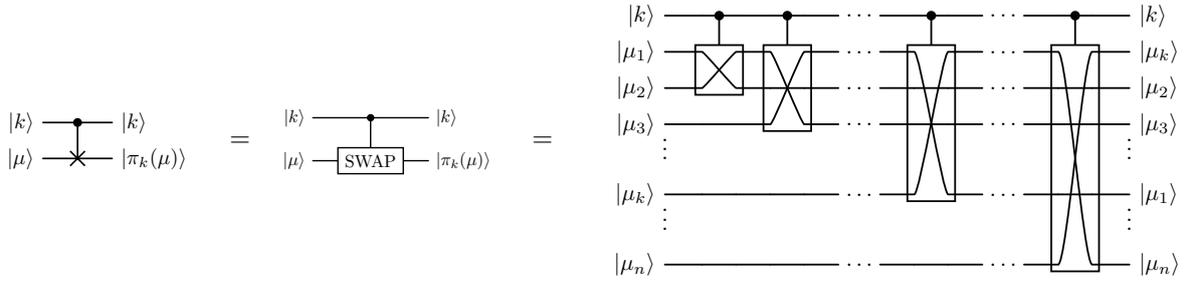

    \centering
    \adjincludegraphics[valign=c,width=0.15\textwidth,page=6]{figures/Isometry_V.pdf}
    \quad$=$\quad
    \adjincludegraphics[valign=c,width=0.17\textwidth,page=7]{figures/Isometry_V.pdf}
    \quad$=$\qquad
    \adjincludegraphics[valign=c,width=0.45\textwidth,page=8]{figures/Isometry_V.pdf}
    \caption{Circuit for the controlled SWAP gate. Each control on step $i$ of this circuit activates if and only if the condition $c_q = i$ is satisfied, where $c_q$ is the control qubit state (in our case, $c_q = k$).}
    \label{fig:CSWAP_circ}
\end{figure}

In order to then prepare $\widetilde{M}$, we use the property from \cref{def:V_matrix_elements} that $T^k := T_{\widetilde{\mu}_k} = \widetilde{M}_k$. It suffices to select these $T^k$ from the $T$ register and copy them into the respective places in the $\widetilde{M}$ register. To do that, we select using swap gates on the $T$ register controlled from the $\widetilde{\mu}$ register. This gives the COPY gate preparing $\widetilde{M}$, given in \cref{fig:PREPM_circ}. The controlled swap is defined in \cref{fig:CSWAP_circ}.

\begin{figure}[H]
    \centering
    \adjincludegraphics[valign=c,width=0.7 \textwidth,page=11]{figures/Isometry_V.pdf}
    \caption{Circuit for the controlled $\mathrm{COPY}$ gate. 
    Here we use that $T^k \defeq T_{\widetilde{\mu}_k} = \widetilde{M}_k$ according to \cref{def:V_matrix_elements}.}
    \label{fig:PREPM_circ}
\end{figure}

\textbf{Main part of the circuit.} Now that we have access to $\widetilde{M}$, the main idea to make the control on the $\mathrm{F}_l$ gates explicit is to take the pair of composition and step $(\mu, i)$, where $i \in [n]$ is indexing the rows of SYT $T$, and convert this pair into the $(k,j)$ indices pair, which is labelling $T^k_j$ from \cref{def:V_matrix_elements} and \cref{eq:T_convention} as well as $\mu_k-j$ and $\widetilde{M}_k$. Here, $k$ is denoting nonzero rows of $\mu$ and $j$ is defined as: 
\begin{equation}
    j = i-\sum_{t=1}^{k-1} \mu_t = i - \widetilde{\mu}_{k-1}.
\end{equation}
\begin{example}
    Consider $\mu = (4,2,1,1,0,0,0,0)$ for $n=8$. Then for this $\mu$, to any $i \in \{1,2,\dots,8\}$ we can find a corresponding pair $(k,j)$, and in our case the $(k,j)$ look like:
    \begin{center}
        \begin{tabular}{ c|c|c|c|c|c|c|c|c } 
         i     & 1   & 2   & 3   & 4   & 5   & 6   & 7   & 8 \\ 
         \hline
         (k,j) &(1,1)&(1,2)&(1,3)&(1,4)&(2,1)&(2,2)&(3,1)&(4,1) \\ 
        \end{tabular}
    \end{center}
\end{example}

\begin{figure}[H]
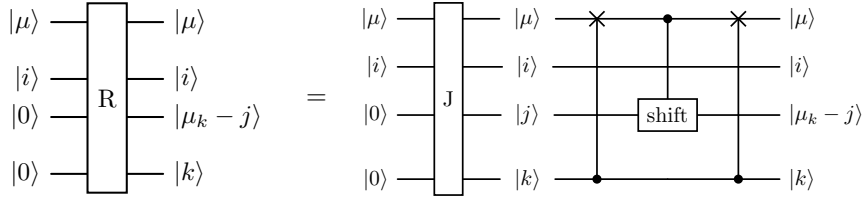

    \centering
    \adjincludegraphics[valign=c,width=0.21\textwidth,page=4]{figures/Isometry_V.pdf}
    \quad$=$\quad
    \adjincludegraphics[valign=c,width=0.4\textwidth,page=5]{figures/Isometry_V.pdf}
    \caption{Circuit for the R gate, where $i \in [n]$ is the row of the SYT $T$.}
    \label{fig:R_circ}
\end{figure}

We can prepare $(k,j)$ from $(\mu,i)$ using the R gate, defined in \cref{fig:R_circ}. Specifically, within the R gate we use the J gate to prepare the $(k,j)$ pair from $(\mu,i)$, and we then use the controlled swap and simple algebraic gates to prepare $(k,\mu_k-j)$ from $(k,j)$ controlled on $\mu$. The controlled shift gate is defined as taking the control value and modifying the target value by copying the control and decrementing it by the target entry, as such:
\begin{equation}
    \mathrm{shift}: \ket{c}\ket{t} \mapsto \ket{c}\ket{c-t}
\end{equation}

\begin{figure}[H]
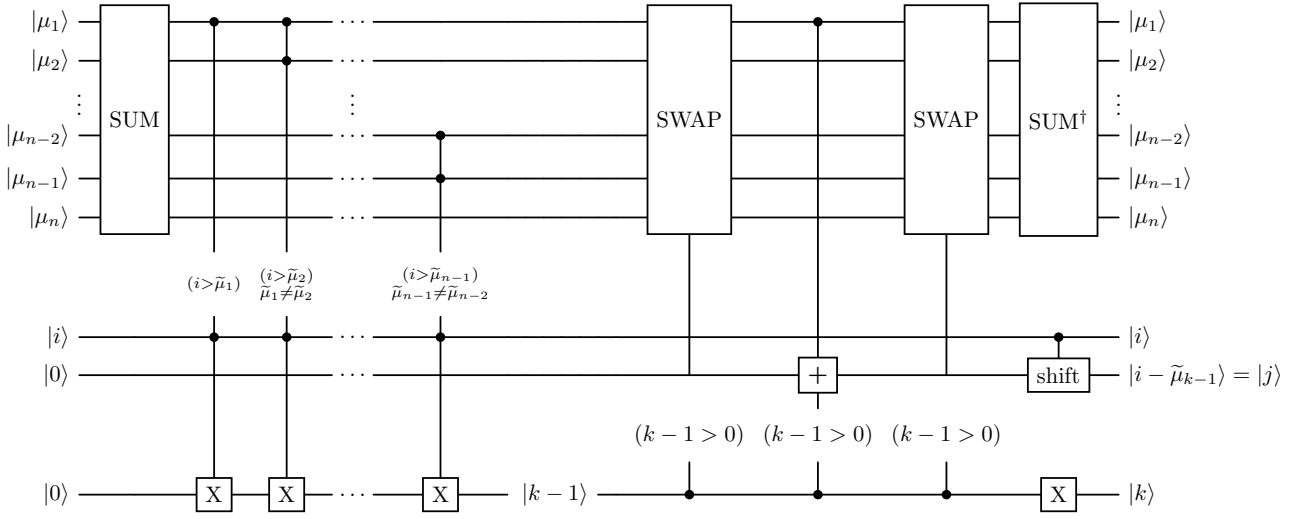

    \centering
    \adjincludegraphics[valign=c,width=1.0\textwidth,page=10]{figures/Isometry_V.pdf}
    \caption{Circuit for the J gate. Here the double controls after the SUM gate on the neighbouring wires $\widetilde{\mu}_{t}$ and $\widetilde{\mu}_{t-1}$ for every $t \in \{2,\dots,n-1\}$ are needed to check whether the entries in these wires are equal or not.  If the states are different, the control gate is performed, otherwise the control gate is not performed. The controlled SWAP is defined as identity if the control qubit is zero (which is showcased in the circuit using brackets on a wire). The brackets on a control wire denote the control condition for this wire (for example: $(i>\mu_1)$ denotes that controlled increment is executed only if this condition between the control nodes $i$ and $\mu_1$ is satisfied). Notice that the controlled $\boxed{+}$ gate between the swaps adds the value from the $\widetilde{\mu}$ register, not from the $k$ register.}
    \label{fig:J_circ}
\end{figure}

The J gate is defined as shown in \cref{fig:J_circ}. We prepare $j$ by definition, using the $i$ index register and the $\widetilde{\mu}$ register. Notice that this circuit also uses a constant amount of hidden ancillas for each control, which doesn't increase the space complexity much.

\subsection{Computational complexity}
\subsubsection*{Total gate and time complexity.} First, note that in the $V$ isometry most of the subroutines are sequential, and we can perform at most a constant amount of gates in parallel. Therefore our time complexity is the same as the total gate complexity. 

To implement a given unitary $\mathrm{F}_l$, it is enough to use at most $O(n^2)$ Givens rotations, if we precompute the $F$-symbols. Computation of $F$-symbols using \cref{eq:lemma_F} is clearly very efficient, with complexity polylogarithmic in $n$. Therefore, since it is enough to use at most $O(n^2)$ Givens rotations for each $F$-unitary and the computation of each $F$-symbol is polylog in $n$, it should give us the complexity of $\widetilde{O}(n^2)$ to implement one unitary $F$.

From \cref{fig:CSWAP_circ} we can determine that complexity of a single controlled SWAP is $O(n)$. From \cref{fig:SUM_circ} we see that complexity of the SUM gate is $\widetilde{O}(n)$. In the implementation of the J gate, given in \cref{fig:J_circ}, the simple algebra gates for control checks each have constant or polylogarithmic complexity in $n$, so the total complexity of control checks is $\widetilde{O}(n)$; the $\mathrm{X}$ gates controlled on $\mu,i$ each have constant complexity, so the total complexity of controlled $\mathrm{X}$ gates with the checks is $\widetilde{O}(n)$; two controlled swaps give $O(n)$, and two SUM gates are $\widetilde{O}(n)$. Therefore the total complexity of the J gate is $\widetilde{O}(n)$. Using this, we deduce that complexity of the R gate is also $\widetilde{O}(n)$.

Therefore, the complexity of one recursive block in \cref{fig:V_circ}, containing two R gates, two controlled swaps, a generalized Pauli $\mathrm{X}^{\dagger}$, and a controlled $F$ unitary, is $\widetilde{O}(n^2) + 2\widetilde{O}(n) + 2\widetilde{O}(n) + const = \widetilde{O}(n^2)$. We have $n$ of these blocks.

Finally, the complexity of the controlled $\mathrm{COPY}$ gate, shown in \cref{fig:PREPM_circ}, is $2n\widetilde{O}(n) + \widetilde{O}(n) = \widetilde{O}(n^2)$. With this, the total complexity of $V$ is given by: $2\widetilde{O}(n) + \widetilde{O}(n^2) + n\widetilde{O}(n^2) = \widetilde{O}(n^3)$.

\subsubsection*{Space complexity.} The input and output of the circuit for the isometry $V$ include the $\mu$, $\widetilde{M}$, and $T$ registers, as well as some constant number of extra inputs of size $d=n$. We describe the space complexity in terms of the number of wires (states) of size $d=n$. The registers $\mu$ and $T$ appear before $V^{\dagger}$ is executed, but we still describe their space complexity here, for completeness. The register $\mu$ is comprised of $n$ inputs of size $n$, so it has complexity $O(n)$. The register $T$ is comprised of $n$ components (plus the 1-dimensional empty set component $T_0$) $T=(T_0,T_1,\dots,T_n)$, where each component $T_i$ is a partition comprised of (at worst) $i$ inputs of size $n$, giving the complexity of $\frac{n(n+1)}{2}=O(n^2)$ for the register $T$. Besides $\mu$ and $T$, the input to $V^{\dagger}$ requires an ancillary register that will turn into $\widetilde{M}$ on the output from $V$. This register must have the same size as $\widetilde{M}$, and since it is a GT pattern of length $n$ and entry dimension $d=n$, we again have $n$ components with each component $\widetilde{M_i}$ a string of $i$ inputs of size $n$, thus the complexity of $\widetilde{M}$ is $\frac{n(n+1)}{2}=O(n^2)$. Therefore, before considering the ancillas needed for the gates, the space complexity from input/output is $O(n^2)$.

The SUM, controlled SWAP, and (consequently) the COPY gate need no ancillas, so they do not increase the space complexity. After precomputing the $F$ symbols, controlled $F$-unitaries require a constant amount of ancillas, since the Givens rotations require a constant amount of ancillas. Precomputing the $F$ symbols requires at most $\widetilde{O}(n)$ ancillas to have access to copies of $\mu, \widetilde{M}_k, T^k, \lambda$, and we can reuse these ancillas after every precomputation, since the $F$-unitaries are performed sequentially. Within the R gate, only the J gate requires ancillas. For the J gate, each control needs a constant number of ancillas to do the inequality checks from \cref{fig:J_circ}, but since all the inequality checks are mutually independent we do not need to store any information on the ancilla register between the sequential steps. Therefore, we can refurbish the ancillas after every use (i.e.\ do control checks, use the ancilla, then reset by inverting the control checks). Thus the J gate only needs a constant number of ancillas. So, in total, we need $\widetilde{O}(n)$ gate ancillas.

Finally, combining the two calculations above implies that the total space complexity of the isometry $V$ is: $O(n^2) + \widetilde{O}(n) = O(n^2)$.


\section{High-dimensional BCH Schur transform}

The first approach to the Schur transform, provided by Bacon, Chuang, and Harrow in Ref.~\cite{Bacon2006a}, is based on sequential applications of the Clebsch--Gordan transforms, see \cref{fig:bch_circuit}. 
The complexity of this implementation was upper bounded by $\widetilde{O}(n  \poly(d))$ in the number of subsystems $n$ and their local dimension $d$, which roughly comes from $n$ uses of the Clebsch--Gordan transform, each requiring $\widetilde{O}(\poly(d))$ operations. 
Recently, this upper bound for Clebsch--Gordan transform was refined to $\widetilde{O}(d^4)$ and for the Schur transform to $\widetilde{O}(d^4n)$ \cite{Nguyen2024}.

In his PhD thesis \cite{Harrow2005}, Aram Harrow outlined how to achieve $\poly(n,\log d)$ complexity\footnote{See the footnote at the beginning of Section 8.1.2 in Ref.~\cite{Harrow2005}.}. 
Unfortunately, the details of this construction were never fully developed. 
The key idea is to use a pre-processing compression map that reduces the local dimension of the main part of the circuit from arbitrary $d$ to $d = n$, and then apply a sequence of $n$ Clebsch--Gordan transforms with local dimension $n$ instead of $d$.

In this section, we explore this idea and provide an exact circuit for the pre-processing compression map, which indeed allows one to lift the Schur transform in Ref.~\cite{Bacon2006a} to achieve $\widetilde{O}(\min(n^5, d^4 n))$ complexity.

The aforementioned key idea by Aram Harrow can be described as follows. A vector $x = (x_1,\dotsc,x_n) \in [d]^n$ of $n$ symbols, each chosen from $d$ possible characters, can always be encoded as a vector $e = (e_1,\dotsc,e_n) \in [n]^n$ of $n$ symbols, each chosen from only $n$ characters, together with an additional decoding vector $p = (p_1,\dotsc,p_n) \in [d]^n$. Formally, the vector $x = (x_1,\dotsc,x_n) \in [d]^n$ is retrieved by
\begin{equation}
\label{eq:retrive}
x_i = p_{e_i}
\end{equation}
for all $i \in [n]$.

As an example, consider the vector $x = (9,2,5,9,11,2) \in [11]^6$. It can be encoded as $e = (3,1,2,3,4,1) \in [6]^6$, together with the decoding vector $p = (2,5,9,11,\cdot,\cdot) \in [11]^6$, where $\cdot \in [11]$ denotes arbitrary values. Note that the original vector $x$ can be recovered from $e$ and $p$ through \eqref{eq:retrive}.

\begin{figure}[ht]
    \centering
    
    \begin{subfigure}[b]{1\textwidth}
          \centering
      \adjincludegraphics[valign=c,width=0.17\textwidth,page=9]{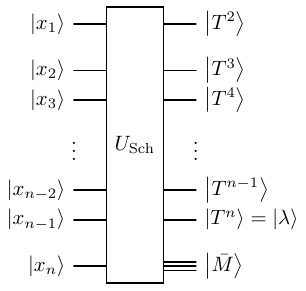}
    \quad$=$\quad
    \adjincludegraphics[valign=c,width=0.4\textwidth,page=10]{figures/bch_circuits.pdf}
    \caption{Recursive decomposition of preparation isometry $\widehat{\mathrm{P}}$. In $i$th step, a new register $\ket{x_i}$ is added to previousely encoded registers $\ket{x_1,\ldots,x_{i-1}}$.}
    \label{fig:p_hat_recursive}
    \end{subfigure}
    
    \vspace{0.5cm}
    
    \begin{subfigure}[b]{1\textwidth}
         \centering
    \adjincludegraphics[valign=c,width=0.2\textwidth,page=11]{figures/bch_circuits.pdf}
    \;$=$
    \adjincludegraphics[valign=c,width=0.75\textwidth,page=12]{figures/bch_circuits.pdf}
    \caption{A quantum circuit for an isometry $\widehat{\Prep}_n$ utilizes two auxiliary registers ($\ket{\cdot }_{p}\in \C^n$ and $\ket{\cdot }_{u}\in \C^2$) that store information about the position $e_n$ and uniqueness $c_n$ of $x_n$. 
    It can be further decomposed into four subroutines: $A, B, C$ and $D$. \Cref{fig:subroutibesABCD} presents the quantum circuits for the aforementioned subroutines.}
    \label{fig:prep_hat}
    \end{subfigure}
    
    \caption{Preparation circuit $\widehat{\Prep}$ encodes vector $x \in [d]^n$ via a reduced vector $e \in [n]^n$ and a decoding vector $p \in [d]^n$. This idea was outlined in Aram Harrow’s PhD thesis \cite{Harrow2005}, while the present the construction in full detail.}
    \label{fig:combined}
\end{figure}

\begin{figure}[H]
    \centering
    
    \begin{subfigure}[b]{1\textwidth}
    \centering
    \begin{minipage}[t]{0.50\textwidth}
        \centering
        \includegraphics[width=1\textwidth,page=3]{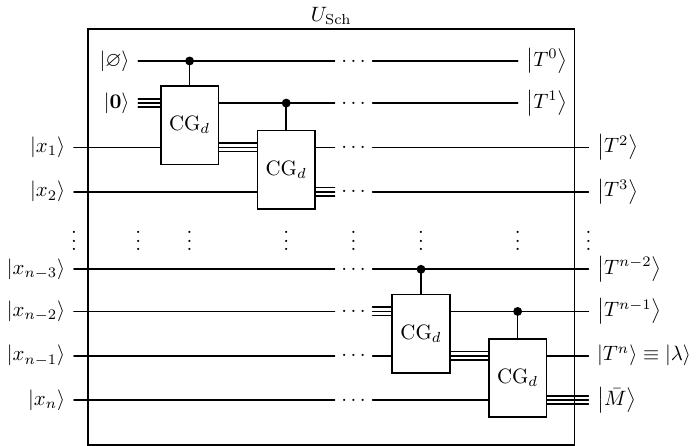}
    \end{minipage}%
    \hfill
    \begin{minipage}[t]{0.45\textwidth}
        \centering
        \includegraphics[width=1.05\textwidth,page=4]{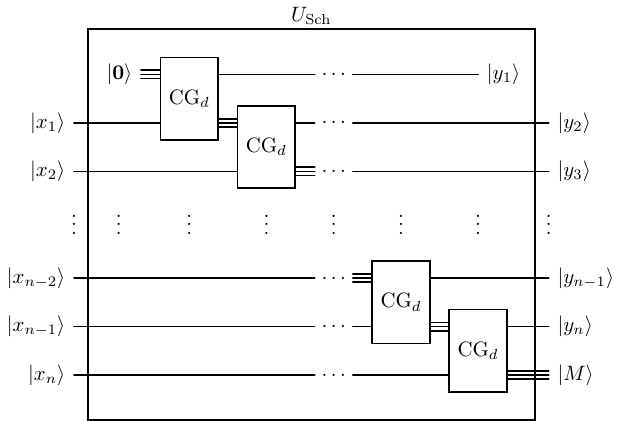}
    \end{minipage}
    \caption{BCH quantum Schur transform in standard encoding (left) and in Yamanouchi encoding (right). 
    $T = (T^0,\dotsc,T^n \equiv \lambda) \in \SYT(\lambda)$ encodes an a standard Young Tableaux of shape $\lambda$ and $\bar{M}$ represents $d-1$ bottom rows of a Gelfand--Tseltin pattern $M \in \GT(\lambda)$, which has the top row $\lambda$.
    $y_i$ represents a row number of a box added to the Young diagram $T^{i-1}$ to obtain $T^i$.}
    \label{fig:bch_circuit}
    \end{subfigure}
    
    \vspace{0.5cm}
    
    \begin{subfigure}[b]{1\textwidth}
 \centering
    \includegraphics[width=0.6\textwidth,page=8]{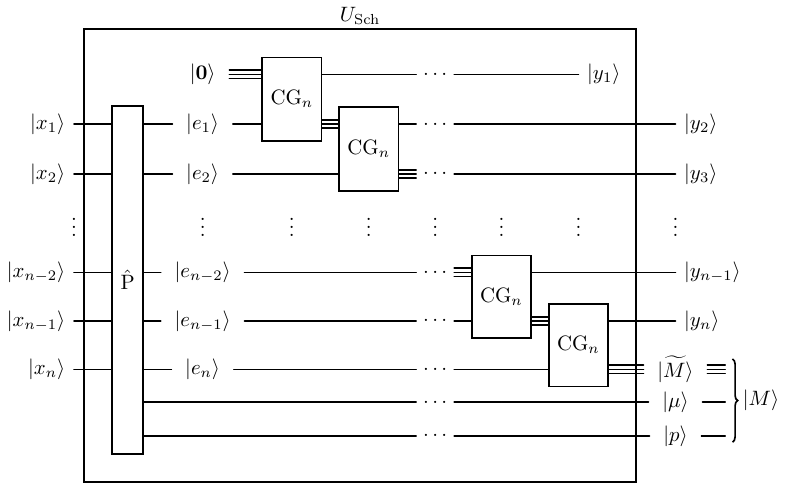}
    \caption{High-dimensional version of BCH Schur transform. 
    The main trick behind this implementation is the use of preprocessing operation $\hat{\mathrm{P}}$, which compresses the input data similarly to the gate $\mathrm{P}$ from \cref{sec:step_1_circuits}.
    This compression allows to use smaller CG transforms $\CG_n$ instead of $\CG_d$, which is more efficient in the regime $d \geq n$.}
    \label{fig:bch_high_dim}
    \end{subfigure}
    
    \caption{Comparison of the first approach to the Schur transform, provided by Bacon, Chuang, and Harrow in Ref.~\cite{Bacon2006a}, with its enhancement via a compression circuit. With this enhancement, the $d$-dimensional Clebsch--Gordan transformations are replaced by $n$-dimensional ones, improving the computational complexity from $\widetilde{O}(d^4 n)$ to $\widetilde{O}(\min(n^5, d^4 n))$.}
    \label{fig:BCH_compared}
\end{figure}

We use this idea and present a quantum circuit for the following isometry 
\begin{equation}
    \label{eq:prep2}
    \hat{\mathrm{P}}: \ket{x}
    \mapsto\ket{p}\ket{\mu}\ket{e}
    ,
\end{equation}
where $\ket{x}, \ket{p} \in \cH_d^{n}$, $\ket{\mu}, \ket{e} \in \cH_n^{n}$, and $e$ and $p$ are vectors related to $x$ by \cref{eq:retrive}. 
The vector $\ket{\mu}$ is the type vector associated with both $x$ and $e$. 
Such an isometry (\ref{eq:prep2}) can in fact be defined recursively. Suppose that we have a transformation:
\begin{equation}
    \label{eq:prep_prim2}
    \widehat{\Prep}': \ket{x'}
    \mapsto\ket{e'}\ket{\mu'}\ket{p'}
\end{equation}
which transforms computational basis vector $\ket{x'}=\ket{x_1,\ldots,x_{n-1}}\in \cH_d^{n-1}$ into related compressed vector $\ket{e'}=\ket{e_1,\ldots,e_{n-1}}\in \cH_{n-1}^{n-1}$, alphabet vector $\ket{p'}=\ket{p_1',\ldots,p_{n-1}'}\in \cH_d^{n-1}$, type vector $\ket{\mu'}=\ket{\mu_1',\ldots,\mu_{n-1}'}\in \cH_n^{n-1}$. 
We shall define the following map: 
\begin{equation}
    \label{eq:prep_prim2}
    \widehat{\Prep}_n: 
   \ket{e'}\ket{\mu'}\ket{p'}\otimes \ket{x_n}
   \mapsto
   \ket{e}\ket{\mu}\ket{p}
   .
\end{equation}
In that way, transformation (\ref{eq:prep2}) can be achieved by
\begin{equation}
    \label{eq:Perp_induction2}
    \hat{\mathrm{P}} = \widehat{\Prep}_1 \widehat{\Prep}_2 \cdots \widehat{\Prep}_n
\end{equation}
as presented on \cref{fig:p_hat_recursive}.

In fact, a circuit for $\widehat{\Prep}_n$ can be directly composed from the subroutines $A, B, C,$ and $D$ presented in \cref{sec:step_1_circuits} (see \cref{fig:prep_hat}). 
According to the analysis of computational complexity in \cref{sec:complexity_Prep}, the subroutines $A, B, C, D$ each have linear total gate and time complexity $\widetilde{O}(n)$. 
As a result, the computational complexity of the preparation circuit $\hat{\mathrm{P}}$ is $\widetilde{O}(n^2)$, since it is recursively defined via \cref{eq:Perp_induction2}.
It also requires linear number $\widetilde{O}(n)$ of ancillas.

In the following part of this section, we explain how the preparation circuit $\hat{\mathrm{P}}$ lifts the computational complexity of the Schur transform from $\widetilde{O}(d^4 n)$ to $\widetilde{O}(\min(n^5, d^4 n))$, building on the original approach of Bacon, Chuang, and Harrow in Ref.~\cite{Bacon2006a}.

The quantum circuit for the Schur transform in Ref.~\cite{Bacon2006a} is based on sequential applications of the Clebsch--Gordan transform. \cref{fig:bch_circuit} present this circuit, which consists of $n$ successive Clebsch--Gordan transforms. We refer to \cite{Bacon2006a} and the more recent works \cite{23Nguyen,23Grinko} for details. Since the $d$-dimensional Clebsch--Gordan transform can be implemented with complexity $\widetilde{O}(d^4)$, the overall complexity of this version of the Schur transform is $\widetilde{O}(nd^4)$. 
It is worth noting that, depending on the encoding of Gelfand--Tsetlin patterns, the space requirements are $\widetilde{O}(d(n+d))$ for the standard encoding and $\widetilde{O}(d^2)$ for the more efficient Yamanouchi encoding; see Refs.~\cite{23Nguyen,23Grinko} for details. A sketch of the circuit is shown in \cref{fig:bch_circuit}.

The key idea, originally outlined by Aram Harrow \cite{Harrow2005}, is to compress an arbitrary vector $\ket{x} \in [d]^n$ into $e = (e_1,\dotsc,e_n) \in [n]^n$ and then apply a sequence of $n$-dimensional Clebsch--Gordan transforms to $e$, as illustrated in \cref{fig:bch_high_dim}. 
Since the preparation circuit can be implemented with space complexity $\widetilde{O}(n^2)$ and each $n$-dimensional Clebsch--Gordan transform requires $\widetilde{O}(n^4)$ gates and depth, the overall gate and time complexity of this enhanced version of the Schur transform is $\widetilde{O}(\min(n^5, nd^4))$. 
For simplicity, \cref{fig:BCH_compared} sketches the circuit only for the more efficient Yamanouchi encoding.
As final step in out argument, in the next section we describe why replacing $\CG_d$ transform by $\CG_n$ does not break the correctness of the circuit.

\subsection{Reduced Wigner unitary}

In this section, we explain why our high-dimensional circuit implements the same Schur transform unitary as the original BCH Schur transform.

\begin{figure}[H]
    \centering
    \includegraphics[width=\textwidth,page=7]{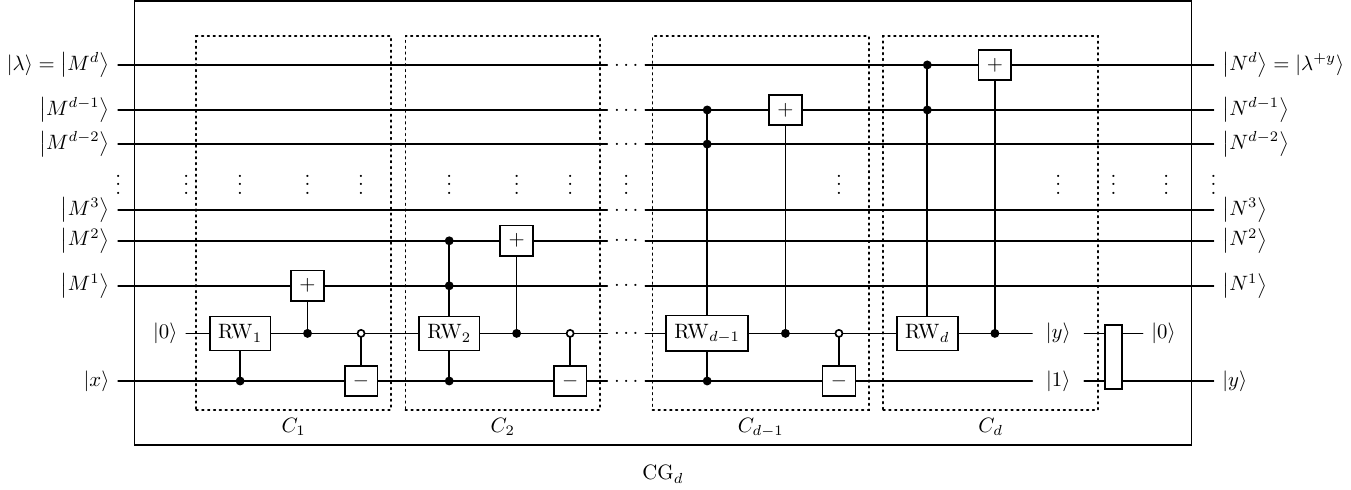}
    \caption{Clebsch--Gordan transform $\CG_d$ in Yamanouchi encoding for decomposing tensor product of an arbitrary irrep $T^{k-1}$ with the defining irrep $\square$ of $\U{d}$. 
    To obtain standard encoding output a simple operation is needed to compute $T^{k-1}$ from $T^k$ and Yamanouchi symbol $y_k$ (encoding the row of added box).}
    \label{fig:cg_transform}
\end{figure}

\begin{figure}[H]
    \centering
    \includegraphics[width=0.2\textwidth,page=13]{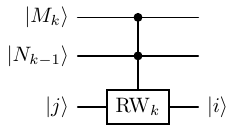}
    \caption{Reduced Wigner unitary used in the construction of CG transform, see \cref{fig:cg_transform}}
    \label{fig:rw_unitary}
\end{figure}

The reason for that is the building blocks of Clebsch--Gordan transform are so-called Reduced Wigner unitaries $\mathrm{RW}^{\lambda,\nu}_d$, which have matrix entries that do not depend to the local dimension, once $d \geq n$. 
To see that, we recall from \cite{Vilenkin1992} that matrix entries of $\mathrm{RW}_d$ for $j \in \set{0, 1,\dotsc,d-1}$, $i \in \set{1,\dotsc,d}$ are (see also \cref{fig:rw_unitary})
    \begin{align*}
    \bra{i}\mathrm{RW}_d^{M_d,N_{d-1}}\ket{j}
    &=\begin{cases}
    S(i, j)\left|\frac{\prod_{k \neq j}\left(\ell_{k, d-1} - \ell_{i, d} - 1 \right) \prod_{k \neq i}\left(\ell_{k, d}-\ell_{j, d-1}\right)}{\prod_{k \neq i}\left(\ell_{k,d}-\ell_{i, d}\right) \prod_{k \neq j}\left(\ell_{k, d-1}-\ell_{j, d-1}-1\right)}\right|^{1/2}, &\text{ if } j \in \set{1,\dotsc,d-1} \\
    \left|\frac{\prod_{k=1}^{d-1}\left(\ell_{k, d-1}-\ell_{i,d}-1\right)}{\prod_{k \neq i}\left(\ell_{k,d}-\ell_{i,d}\right)}\right|^{1/2}, &\text{ if } j = 0
    \end{cases}
  \end{align*}
  where 
\begin{align*}
    \ell_{k,d} &\defeq M^d_{k} - k, \quad \quad
    \ell_{k,d-1} \defeq N^{d-1}_{k} - k - \delta_{k,j} \quad \quad
    S(i,j) \defeq 
    \begin{cases}
      1  &\text{ if }  i \leq j \\
      -1 &\text{ if } i > j.
    \end{cases}
\end{align*}
From these formulas we directly see, that the ratios under square roots do not depend on the actual local dimension, but rather on the shape of Yound diagrams $M_d$ and $N_{d-1}$.
Therefore, when multiplying $\mathrm{RW}$ unitaries in \cref{fig:cg_transform}, the resulting transformation will only depend on the Young diagram shapes which correspond to the rows of the Gelfand--Tsetlin pattern and not on how many zeros are padded to the right in any given row.

\section*{Acknowledgements}

D.G.\ and M.O.\ acknowledge support by NWO grant NGF.1623.23.025 (“Qudits in theory and experiment”) and NWO Vidi grant (Project No.\ VI.Vidi.192.109). M.L.\ acknowledges support from Beyond Moore’s Law project.

\printbibliography

@Inbook{Vilenkin1992,
  author="Vilenkin, Naum Ya. and Klimyk, Anatoli U.",
  title="Representations in the Gel'fand-Tsetlin Basis and Special Functions",
  bookTitle="Representation of Lie Groups and Special Functions: Volume 3: Classical and Quantum Groups and Special Functions",
  year="1992",
  publisher="Springer Netherlands",
  address="Dordrecht",
  pages="361--446",
  isbn="978-94-017-2881-2",
  doi="10.1007/978-94-017-2881-2_5"
}

@article{Krovi2019,
   title={An efficient high dimensional quantum Schur transform},
   volume={3},
   ISSN={2521-327X},
   DOI={10.22331/q-2019-02-14-122},
   journal={Quantum},
   publisher={Verein zur Forderung des Open Access Publizierens in den Quantenwissenschaften},
   author={Krovi, Hari},
   year={2019},
   month=feb, 
   pages={122} 
}

@article{le1987new,
  title={New perspective on the $U(n)$ Wigner-Racah calculus. II. Elementary reduced Wigner coefficients for $U(n)$},
  author={Le Blanc, R and Hecht, Karl T},
  journal={Journal of Physics A: Mathematical and General},
  volume={20},
  number={14},
  pages={4613},
  year={1987},
  publisher={IOP Publishing},
  doi={10.1088/0305-4470/20/14/009}
}

@article{kawano2016quantum,
  title={Quantum Fourier transform over symmetric groups—improved result},
  author={Kawano, Yasuhito and Sekigawa, Hiroshi},
  journal={Journal of Symbolic Computation},
  volume={75},
  pages={219--243},
  year={2016},
  publisher={Elsevier},
  doi={10.1016/j.jsc.2015.11.016}
}

@article{Harrow2005,
  author        = {Harrow, Aram W.},
  journal       = {Ph.D thesis, Massachusetts Institute of Technology, Cambridge, MA, 2005},
  title         = {Applications of coherent classical communication and the Schur transform to quantum information theory},
  year          = {2005},
  archiveprefix = {arXiv},
  eprint        = {quant-ph/0512255},
  primaryclass  = {quant-ph}
}

@Article{Bacon2006a,
  author    = {Bacon, Dave and Chuang, Isaac L. and Harrow, Aram W.},
  journal   = {Physical Review Letters},
  title     = {Efficient Quantum Circuits for Schur and Clebsch-Gordan Transforms},
  year      = {2006},
  issn      = {1079-7114},
  month     = oct,
  number    = {17},
  pages     = {170502},
  volume    = {97},
  doi       = {10.1103/physrevlett.97.170502},
  publisher = {American Physical Society (APS)},
  archivePrefix={arXiv},
  eprint={quant-ph/0407082},
}

@Article{Kirby2018,
  author    = {Kirby, William M. and Strauch, Frederick W.},
  journal   = {Quantum Information and Computation},
  title     = {A practical quantum algorithm for the Schur transform},
  year      = {2018},
  issn      = {1533-7146},
  month     = aug,
  number    = {9 \& 10},
  pages     = {721--742},
  volume    = {18},
  doi       = {10.26421/qic18.9-10-1},
  publisher = {Rinton Press},
  archivePrefix={arXiv},
  eprint={1709.07119},
}

@PhdThesis{Wright2016,
  author = {John Wright},
  school = {Carnegie Mellon University},
  title  = {How to Learn a Quantum State},
  year   = {2016},
  url    = {http://reports-archive.adm.cs.cmu.edu/anon/2016/abstracts/16-108.html},
}

@Article{Keyl2001,
  author    = {Keyl, M. and Werner, R. F.},
  journal   = {Physical Review A},
  title     = {Estimating the spectrum of a density operator},
  year      = {2001},
  issn      = {1094-1622},
  month     = oct,
  number    = {5},
  pages     = {052311},
  volume    = {64},
  doi       = {10.1103/physreva.64.052311},
  publisher = {American Physical Society (APS)},
  archivePrefix={arXiv},
  eprint={quant-ph/0102027},
}

@Article{Acharya2020,
  author    = {Acharya, Jayadev and Issa, Ibrahim and Shende, Nirmal V. and Wagner, Aaron B.},
  journal   = {IEEE Journal on Selected Areas in Information Theory},
  title     = {Estimating Quantum Entropy},
  year      = {2020},
  issn      = {2641-8770},
  month     = aug,
  number    = {2},
  pages     = {454--468},
  volume    = {1},
  doi       = {10.1109/jsait.2020.3015235},
  publisher = {Institute of Electrical and Electronics Engineers (IEEE)},
  archivePrefix={arXiv},
  eprint={1711.00814},
}

@Article{KEYL2006,
  author    = {Keyl, M.},
  journal   = {Reviews in Mathematical Physics},
  title     = {Quantum state estimation and large deviations},
  year      = {2006},
  issn      = {1793-6659},
  month     = feb,
  number    = {01},
  pages     = {19--60},
  volume    = {18},
  doi       = {10.1142/s0129055x06002565},
  publisher = {World Scientific Pub Co Pte Lt},
  archivePrefix={arXiv},
  eprint={quant-ph/0412053},
}

@Article{Haah2017,
  author    = {Haah, Jeongwan and Harrow, Aram W. and Ji, Zhengfeng and Wu, Xiaodi and Yu, Nengkun},
  journal   = {IEEE Transactions on Information Theory},
  title     = {Sample-optimal tomography of quantum states},
  year      = {2017},
  issn      = {1557-9654},
  pages     = {1--1},
  doi       = {10.1109/tit.2017.2719044},
  publisher = {Institute of Electrical and Electronics Engineers (IEEE)},
  archivePrefix={arXiv},
  eprint={1508.01797},
}

@InProceedings{O’Donnell2016,
  author     = {O’Donnell, Ryan and Wright, John},
  booktitle  = {Proceedings of the forty-eighth annual ACM symposium on Theory of Computing},
  title      = {Efficient quantum tomography},
  year       = {2016},
  month      = jun,
  pages      = {899--912},
  publisher  = {ACM},
  series     = {STOC ’16},
  collection = {STOC ’16},
  doi        = {10.1145/2897518.2897544},
  archivePrefix={arXiv},
  eprint={1508.01907},
}

@InProceedings{O’Donnell2017,
  author     = {O’Donnell, Ryan and Wright, John},
  booktitle  = {Proceedings of the 49th Annual ACM SIGACT Symposium on Theory of Computing},
  title      = {Efficient quantum tomography II},
  year       = {2017},
  month      = jun,
  pages      = {962--974},
  publisher  = {ACM},
  series     = {STOC ’17},
  collection = {STOC ’17},
  doi        = {10.1145/3055399.3055454},
  archivePrefix={arXiv},
  eprint={1612.00034},
}

@Misc{Buhrman2022,
  author    = {Buhrman, Harry and Linden, Noah and Mančinska, Laura and Montanaro, Ashley and Ozols, Maris},
  title     = {Quantum majority vote},
  year      = {2022},
  copyright = {Creative Commons Attribution 4.0 International},
  doi       = {10.48550/ARXIV.2211.11729},
  publisher = {arXiv},
}

@Article{Zheng2023,
  author    = {Zheng, Han and Li, Zimu and Liu, Junyu and Strelchuk, Sergii and Kondor, Risi},
  journal   = {PRX Quantum},
  title     = {Speeding Up Learning Quantum States Through Group Equivariant Convolutional Quantum Ansätze},
  year      = {2023},
  issn      = {2691-3399},
  month     = may,
  number    = {2},
  pages     = {020327},
  volume    = {4},
  doi       = {10.1103/prxquantum.4.020327},
  publisher = {American Physical Society (APS)},
  archivePrefix={arXiv},
  eprint={2112.07611},
}

@article{Cotler2019,
  title = {Quantum Virtual Cooling},
  author = {Cotler, Jordan and Choi, Soonwon and Lukin, Alexander and Gharibyan, Hrant and Grover, Tarun and Tai, M. Eric and Rispoli, Matthew and Schittko, Robert and Preiss, Philipp M. and Kaufman, Adam M. and Greiner, Markus and Pichler, Hannes and Hayden, Patrick},
  journal = {Phys. Rev. X},
  volume = {9},
  issue = {3},
  pages = {031013},
  numpages = {11},
  year = {2019},
  month = {Jul},
  publisher = {American Physical Society},
  doi = {10.1103/PhysRevX.9.031013},
  url = {https://link.aps.org/doi/10.1103/PhysRevX.9.031013}
}

@article{Huggins2021,
  title = {Virtual Distillation for Quantum Error Mitigation},
  author = {Huggins, William J. and McArdle, Sam and O'Brien, Thomas E. and Lee, Joonho and Rubin, Nicholas C. and Boixo, Sergio and Whaley, K. Birgitta and Babbush, Ryan and McClean, Jarrod R.},
  journal = {Phys. Rev. X},
  volume = {11},
  issue = {4},
  pages = {041036},
  numpages = {25},
  year = {2021},
  month = {Nov},
  publisher = {American Physical Society},
  doi = {10.1103/PhysRevX.11.041036},
  url = {https://link.aps.org/doi/10.1103/PhysRevX.11.041036}
}

@Article{Christandl2004,
  author        = {Christandl, Matthias and Mitchison, Graeme},
  journal       = {Commun. Math. Phys., Vol. 261, No. 3, pp. 789-797 (2006)},
  title         = {The Spectra of Density Operators and the Kronecker Coefficients of the Symmetric Group},
  year          = {2004},
  issn          = {1432-0916},
  month         = oct,
  number        = {3},
  pages         = {789--797},
  volume        = {261},
  archiveprefix = {arXiv},
  copyright     = {Assumed arXiv.org perpetual, non-exclusive license to distribute this article for submissions made before January 2004},
  date          = {2004-09-02},
  doi           = {10.1007/s00220-005-1435-1},
  eprint        = {quant-ph/0409016},
  primaryclass  = {quant-ph},
  publisher     = {Springer Science and Business Media LLC},
}

@Article{ODonnell2021,
  author    = {O’Donnell, Ryan and Wright, John},
  journal   = {Communications in Mathematical Physics},
  title     = {Quantum Spectrum Testing},
  year      = {2021},
  issn      = {1432-0916},
  month     = aug,
  number    = {1},
  pages     = {1--75},
  volume    = {387},
  doi       = {10.1007/s00220-021-04180-1},
  publisher = {Springer Science and Business Media LLC},
}

@Article{Hayashi2002a,
  author    = {Hayashi, Masahito},
  journal   = {Journal of Physics A: Mathematical and General},
  title     = {Optimal sequence of quantum measurements in the sense of Stein s lemma in quantum hypothesis testing},
  year      = {2002},
  issn      = {0305-4470},
  month     = dec,
  number    = {50},
  pages     = {10759--10773},
  volume    = {35},
  doi       = {10.1088/0305-4470/35/50/307},
  publisher = {IOP Publishing},
}

@Article{Hayashi2024,
  author        = {Hayashi, Masahito},
  title         = {Another quantum version of Sanov theorem},
  year          = {2024},
  month         = jul,
  archiveprefix = {arXiv},
  copyright     = {Creative Commons Attribution 4.0 International},
  doi           = {10.48550/ARXIV.2407.18566},
  eprint        = {2407.18566},
  primaryclass  = {quant-ph},
  publisher     = {arXiv},
}

@Article{Cheng2024,
  author        = {Cheng, Hao-Chung and Datta, Nilanjana and Liu, Nana and Nuradha, Theshani and Salzmann, Robert and Wilde, Mark M.},
  journal       = {npj Quantum Information, volume 11, Article number 94, June 2025},
  title         = {An invitation to the sample complexity of quantum hypothesis testing},
  year          = {2024},
  issn          = {2056-6387},
  month         = jun,
  number        = {1},
  volume        = {11},
  archiveprefix = {arXiv},
  copyright     = {arXiv.org perpetual, non-exclusive license},
  date          = {2024-03-26},
  doi           = {10.1038/s41534-025-00980-8},
  eprint        = {2403.17868},
  primaryclass  = {quant-ph},
  publisher     = {Springer Science and Business Media LLC},
}

@Article{Audenaert2007,
  author    = {Audenaert, K. M. R. and Calsamiglia, J. and Muñoz-Tapia, R. and Bagan, E. and Masanes, Ll. and Acin, A. and Verstraete, F.},
  journal   = {Physical Review Letters},
  title     = {Discriminating States: The Quantum Chernoff Bound},
  year      = {2007},
  issn      = {1079-7114},
  month     = apr,
  number    = {16},
  pages     = {160501},
  volume    = {98},
  doi       = {10.1103/physrevlett.98.160501},
  publisher = {American Physical Society (APS)},
}

@Article{Noetzel2014,
  author    = {Nötzel, J},
  journal   = {Journal of Physics A: Mathematical and Theoretical},
  title     = {Hypothesis testing on invariant subspaces of the symmetric group: part I. Quantum Sanov’s theorem and arbitrarily varying sources},
  year      = {2014},
  issn      = {1751-8121},
  month     = may,
  number    = {23},
  pages     = {235303},
  volume    = {47},
  doi       = {10.1088/1751-8113/47/23/235303},
  publisher = {IOP Publishing},
}

@Article{Hayashi1997,
  author        = {Hayashi, Masahito},
  journal       = {J.Phys.A34:3413-3419,2001},
  title         = {Asymptotics of Quantum Relative Entropy From Representation Theoretical Viewpoint},
  year          = {1997},
  issn          = {1361-6447},
  month         = apr,
  number        = {16},
  pages         = {3413--3419},
  volume        = {34},
  archiveprefix = {arXiv},
  copyright     = {Assumed arXiv.org perpetual, non-exclusive license to distribute this article for submissions made before January 2004},
  date          = {1997-04-24},
  doi           = {10.1088/0305-4470/34/16/309},
  eprint        = {quant-ph/9704040},
  primaryclass  = {quant-ph},
  publisher     = {IOP Publishing},
}

@misc{Montanaro2018,
      title={A Survey of Quantum Property Testing}, 
      author={Ashley Montanaro and Ronald de Wolf},
      year={2018},
      eprint={1310.2035},
      archivePrefix={arXiv},
      primaryClass={quant-ph},
      url={https://arxiv.org/abs/1310.2035}, 
}

@misc{Hu2024,
      title={Sample Optimal and Memory Efficient Quantum State Tomography}, 
      author={Yanglin Hu and Enrique Cervero-Martín and Elias Theil and Laura Mančinska and Marco Tomamichel},
      year={2024},
      eprint={2410.16220},
      archivePrefix={arXiv},
      primaryClass={quant-ph},
      url={https://arxiv.org/abs/2410.16220}, 
}

@Article{Bravyi2017,
  author        = {Bravyi, Sergey and Gambetta, Jay M. and Mezzacapo, Antonio and Temme, Kristan},
  title         = {Tapering off qubits to simulate fermionic Hamiltonians},
  year          = {2017},
  month         = jan,
  archiveprefix = {arXiv},
  copyright     = {arXiv.org perpetual, non-exclusive license},
  doi           = {10.48550/ARXIV.1701.08213},
  eprint        = {1701.08213},
  primaryclass  = {quant-ph},
  publisher     = {arXiv},
}

@Article{Gu2021,
  author    = {Gu, Shouzhen and Somma, Rolando D. and Şahinoğlu, Burak},
  journal   = {Quantum},
  title     = {Fast-forwarding quantum evolution},
  year      = {2021},
  issn      = {2521-327X},
  month     = nov,
  pages     = {577},
  volume    = {5},
  doi       = {10.22331/q-2021-11-15-577},
  publisher = {Verein zur Forderung des Open Access Publizierens in den Quantenwissenschaften},
}

@Article{Lacroix2023,
  author    = {Lacroix, Denis and Ruiz Guzman, Edgar Andres and Siwach, Pooja},
  journal   = {The European Physical Journal A},
  title     = {Symmetry breaking/symmetry preserving circuits and symmetry restoration on quantum computers: A quantum many-body perspective},
  year      = {2023},
  issn      = {1434-601X},
  month     = jan,
  number    = {1},
  volume    = {59},
  doi       = {10.1140/epja/s10050-022-00911-7},
  publisher = {Springer Science and Business Media LLC},
}

@Article{Burkat2025,
  author        = {Burkat, Jędrzej and Fitzpatrick, Nathan},
  title         = {The Quantum Paldus Transform: Efficient Circuits with Applications},
  year          = {2025},
  month         = jun,
  archiveprefix = {arXiv},
  copyright     = {arXiv.org perpetual, non-exclusive license},
  doi           = {10.48550/ARXIV.2506.09151},
  eprint        = {2506.09151},
  primaryclass  = {quant-ph},
  publisher     = {arXiv},
}

@Article{Alagic2019,
  author        = {Alagic, Gorjan and Majenz, Christian and Russell, Alexander},
  journal       = {Advances in Cryptology - EUROCRYPT 2020. EUROCRYPT 2020. Lecture Notes in Computer Science, vol 12107. Springer, Cham},
  title         = {Efficient simulation of random states and random unitaries},
  year          = {2019},
  issn          = {1611-3349},
  month         = oct,
  pages         = {759--787},
  archiveprefix = {arXiv},
  booktitle     = {Advances in Cryptology – EUROCRYPT 2020},
  copyright     = {arXiv.org perpetual, non-exclusive license},
  doi           = {10.1007/978-3-030-45727-3_26},
  eprint        = {1910.05729},
  isbn          = {9783030457273},
  primaryclass  = {quant-ph},
  publisher     = {Springer International Publishing},
}

@article{Larocca2022,
  title = {Group-Invariant Quantum Machine Learning},
  author = {Larocca, Mart\'{\i}n and Sauvage, Fr\'ed\'eric and Sbahi, Faris M. and Verdon, Guillaume and Coles, Patrick J. and Cerezo, M.},
  journal = {PRX Quantum},
  volume = {3},
  issue = {3},
  pages = {030341},
  numpages = {25},
  year = {2022},
  month = {Sep},
  publisher = {American Physical Society},
  doi = {10.1103/PRXQuantum.3.030341},
  url = {https://link.aps.org/doi/10.1103/PRXQuantum.3.030341}
}

@misc{Ragone2023,
      title={Representation Theory for Geometric Quantum Machine Learning}, 
      author={Michael Ragone and Paolo Braccia and Quynh T. Nguyen and Louis Schatzki and Patrick J. Coles and Frederic Sauvage and Martin Larocca and M. Cerezo},
      year={2023},
      eprint={2210.07980},
      archivePrefix={arXiv},
      primaryClass={quant-ph},
      url={https://arxiv.org/abs/2210.07980}, 
}

@article{Nguyen2024,
  title = {Theory for Equivariant Quantum Neural Networks},
  author = {Nguyen, Quynh T. and Schatzki, Louis and Braccia, Paolo and Ragone, Michael and Coles, Patrick J. and Sauvage, Fr\'ed\'eric and Larocca, Mart\'{\i}n and Cerezo, M.},
  journal = {PRX Quantum},
  volume = {5},
  issue = {2},
  pages = {020328},
  numpages = {40},
  year = {2024},
  month = {May},
  publisher = {American Physical Society},
  doi = {10.1103/PRXQuantum.5.020328},
  url = {https://link.aps.org/doi/10.1103/PRXQuantum.5.020328}
}

@misc{Zheng2025,
      title={Towards Super-polynomial Quantum Speedup of Equivariant Quantum Algorithms with SU($d$) Symmetry}, 
      author={Han Zheng and Zimu Li and Sergii Strelchuk and Risi Kondor and Junyu Liu},
      year={2025},
      eprint={2207.07250},
      archivePrefix={arXiv},
      primaryClass={quant-ph},
      url={https://arxiv.org/abs/2207.07250}, 
}

@Article{Botero2017,
  author    = {Botero, Alonso},
  journal   = {Revista Colombiana de Matemáticas},
  title     = {Quantum Information and the Representation Theory of the Symmetric Group},
  year      = {2017},
  issn      = {0034-7426},
  month     = jan,
  number    = {2},
  pages     = {191},
  volume    = {50},
  doi       = {10.15446/recolma.v50n2.62210},
  publisher = {Universidad Nacional de Colombia},
}

@InProceedings{Childs2007,
author="Childs, Andrew M.
and Harrow, Aram W.
and Wocjan, Pawe{\l}",
editor="Thomas, Wolfgang
and Weil, Pascal",
title="Weak Fourier-Schur Sampling, the Hidden Subgroup Problem, and the Quantum Collision Problem",
booktitle="STACS 2007",
year="2007",
publisher="Springer Berlin Heidelberg",
address="Berlin, Heidelberg",
pages="598--609",
isbn="978-3-540-70918-3"
}

@misc{Soleimanifar2022,
      title={Testing matrix product states}, 
      author={Mehdi Soleimanifar and John Wright},
      year={2022},
      eprint={2201.01824},
      archivePrefix={arXiv},
      primaryClass={quant-ph},
      url={https://arxiv.org/abs/2201.01824}, 
}

@Article{Cervero2023,
  author        = {Cervero, Enrique and Mančinska, Laura},
  title         = {Weak Schur sampling with logarithmic quantum memory},
  year          = {2023},
  month         = sep,
  archiveprefix = {arXiv},
  copyright     = {arXiv.org perpetual, non-exclusive license},
  doi           = {10.48550/ARXIV.2309.11947},
  eprint        = {2309.11947},
  primaryclass  = {quant-ph},
  publisher     = {arXiv},
}

@misc{Cervero2024,
      title={A memory and gate efficient algorithm for unitary mixed Schur sampling}, 
      author={Enrique Cervero-Martín and Laura Mančinska and Elias Theil},
      year={2024},
      eprint={2410.15793},
      archivePrefix={arXiv},
      primaryClass={quant-ph},
      url={https://arxiv.org/abs/2410.15793}, 
}

@Article{Hayashi2002,
  author    = {Hayashi, Masahito and Matsumoto, Keiji},
  journal   = {Physical Review A},
  title     = {Quantum universal variable-length source coding},
  year      = {2002},
  issn      = {1094-1622},
  month     = aug,
  number    = {2},
  pages     = {022311},
  volume    = {66},
  doi       = {10.1103/physreva.66.022311},
  publisher = {American Physical Society (APS)},
}

@InProceedings{Hayashi2003,
  author    = {Hayashi, M. and Matsumoto, K.},
  booktitle = {IEEE International Symposium on Information Theory, 2003. Proceedings.},
  title     = {Simple construction of quantum universal variable-length source coding},
  year      = {2003},
  pages     = {459},
  publisher = {IEEE},
  doi       = {10.1109/isit.2003.1228476},
}

@Article{Yang2016,
  author    = {Yang, Yuxiang and Chiribella, Giulio and Hayashi, Masahito},
  journal   = {Physical Review Letters},
  title     = {Optimal Compression for Identically Prepared Qubit States},
  year      = {2016},
  issn      = {1079-7114},
  month     = aug,
  number    = {9},
  pages     = {090502},
  volume    = {117},
  doi       = {10.1103/physrevlett.117.090502},
  publisher = {American Physical Society (APS)},
}

@Article{Jozsa1998,
  author    = {Jozsa, Richard and Horodecki, Michał and Horodecki, Paweł and Horodecki, Ryszard},
  journal   = {Physical Review Letters},
  title     = {Universal Quantum Information Compression},
  year      = {1998},
  issn      = {1079-7114},
  month     = aug,
  number    = {8},
  pages     = {1714--1717},
  volume    = {81},
  doi       = {10.1103/physrevlett.81.1714},
  publisher = {American Physical Society (APS)},
}

@Article{Schumacher1995,
  author    = {Schumacher, Benjamin},
  journal   = {Physical Review A},
  title     = {Quantum coding},
  year      = {1995},
  issn      = {1094-1622},
  month     = apr,
  number    = {4},
  pages     = {2738--2747},
  volume    = {51},
  doi       = {10.1103/physreva.51.2738},
  publisher = {American Physical Society (APS)},
}

@Article{Matsumoto2007,
  author    = {Matsumoto, Keiji and Hayashi, Masahito},
  journal   = {Physical Review A},
  title     = {Universal distortion-free entanglement concentration},
  year      = {2007},
  issn      = {1094-1622},
  month     = jun,
  number    = {6},
  pages     = {062338},
  volume    = {75},
  doi       = {10.1103/physreva.75.062338},
  publisher = {American Physical Society (APS)},
}

@Article{BlumeKohout2014,
  author    = {Blume-Kohout, Robin and Croke, Sarah and Gottesman, Daniel},
  journal   = {IEEE Transactions on Information Theory},
  title     = {Streaming Universal Distortion-Free Entanglement Concentration},
  year      = {2014},
  issn      = {1557-9654},
  month     = jan,
  number    = {1},
  pages     = {334--350},
  volume    = {60},
  doi       = {10.1109/tit.2013.2292135},
  publisher = {Institute of Electrical and Electronics Engineers (IEEE)},
}

@Article{Harrow2013,
  author        = {Harrow, Aram W.},
  title         = {The Church of the Symmetric Subspace},
  year          = {2013},
  month         = aug,
  archiveprefix = {arXiv},
  copyright     = {arXiv.org perpetual, non-exclusive license},
  doi           = {10.48550/ARXIV.1308.6595},
  eprint        = {1308.6595},
  primaryclass  = {quant-ph},
  publisher     = {arXiv},
}

@Article{Keyl1999,
  author    = {Keyl, M. and Werner, R. F.},
  journal   = {Journal of Mathematical Physics},
  title     = {Optimal cloning of pure states, testing single clones},
  year      = {1999},
  issn      = {1089-7658},
  month     = jul,
  number    = {7},
  pages     = {3283--3299},
  volume    = {40},
  doi       = {10.1063/1.532887},
  publisher = {AIP Publishing},
}

@Article{Cirac1999,
  author    = {Cirac, J. I. and Ekert, A. K. and Macchiavello, C.},
  journal   = {Physical Review Letters},
  title     = {Optimal Purification of Single Qubits},
  year      = {1999},
  issn      = {1079-7114},
  month     = may,
  number    = {21},
  pages     = {4344--4347},
  volume    = {82},
  doi       = {10.1103/physrevlett.82.4344},
  publisher = {American Physical Society (APS)},
}

@Article{Keyl2001a,
  author    = {Keyl, M. and Werner, R.F.},
  journal   = {Annales Henri Poincaré},
  title     = {The Rate of Optimal Purification Procedures},
  year      = {2001},
  issn      = {1424-0661},
  month     = feb,
  number    = {1},
  pages     = {1--26},
  volume    = {2},
  doi       = {10.1007/pl00001027},
  publisher = {Springer Science and Business Media LLC},
}

@Article{Li2024,
  author        = {Li, Zhaoyi and Fu, Honghao and Isogawa, Takuya and Chuang, Isaac},
  title         = {Optimal Quantum Purity Amplification},
  year          = {2024},
  month         = sep,
  archiveprefix = {arXiv},
  copyright     = {arXiv.org perpetual, non-exclusive license},
  doi           = {10.48550/ARXIV.2409.18167},
  eprint        = {2409.18167},
  primaryclass  = {quant-ph},
  publisher     = {arXiv},
}

@Article{Childs2025,
  author    = {Childs, Andrew M. and Fu, Honghao and Leung, Debbie and Li, Zhi and Ozols, Maris and Vyas, Vedang},
  journal   = {Quantum},
  title     = {Streaming quantum state purification},
  year      = {2025},
  issn      = {2521-327X},
  month     = jan,
  pages     = {1603},
  volume    = {9},
  doi       = {10.22331/q-2025-01-21-1603},
  publisher = {Verein zur Forderung des Open Access Publizierens in den Quantenwissenschaften},
}

@Article{Schatzki2024,
  author    = {Schatzki, Louis and Larocca, Martín and Nguyen, Quynh T. and Sauvage, Frédéric and Cerezo, M.},
  journal   = {npj Quantum Information},
  title     = {Theoretical guarantees for permutation-equivariant quantum neural networks},
  year      = {2024},
  issn      = {2056-6387},
  month     = jan,
  number    = {1},
  volume    = {10},
  doi       = {10.1038/s41534-024-00804-1},
  publisher = {Springer Science and Business Media LLC},
}

@Article{Horodecki2002,
  author    = {Horodecki, Paweł and Ekert, Artur},
  journal   = {Physical Review Letters},
  title     = {Method for Direct Detection of Quantum Entanglement},
  year      = {2002},
  issn      = {1079-7114},
  month     = aug,
  number    = {12},
  pages     = {127902},
  volume    = {89},
  doi       = {10.1103/physrevlett.89.127902},
  publisher = {American Physical Society (APS)},
}

@Article{Brun2004,
  author    = {Brun, T.A.},
  journal   = {Quantum Information and Computation},
  title     = {Measuring polynomial functions of states},
  year      = {2004},
  issn      = {1533-7146},
  month     = sep,
  number    = {5},
  pages     = {401--408},
  volume    = {4},
  doi       = {10.26421/qic4.5-6},
  publisher = {Rinton Press},
}

@Article{Subasi2019,
  author    = {Subaşı, Yiğit and Cincio, Lukasz and Coles, Patrick J},
  journal   = {Journal of Physics A: Mathematical and Theoretical},
  title     = {Entanglement spectroscopy with a depth-two quantum circuit},
  year      = {2019},
  issn      = {1751-8121},
  month     = jan,
  number    = {4},
  pages     = {044001},
  volume    = {52},
  doi       = {10.1088/1751-8121/aaf54d},
  publisher = {IOP Publishing},
}

@article{Gisin97,
  title = {Optimal Quantum Cloning Machines},
  author = {Gisin, N. and Massar, S.},
  journal = {Phys. Rev. Lett.},
  volume = {79},
  issue = {11},
  pages = {2153--2156},
  numpages = {0},
  year = {1997},
  month = {Sep},
  publisher = {American Physical Society},
  doi = {10.1103/PhysRevLett.79.2153},
  url = {https://link.aps.org/doi/10.1103/PhysRevLett.79.2153}
}

@inproceedings{97Beals,
author = {Beals, Robert},
title = {Quantum computation of Fourier transforms over symmetric groups},
year = {1997},
isbn = {0897918886},
publisher = {Association for Computing Machinery},
address = {New York, NY, USA},
url = {https://doi.org/10.1145/258533.258548},
doi = {10.1145/258533.258548},
booktitle = {Proceedings of the Twenty-Ninth Annual ACM Symposium on Theory of Computing},
pages = {48–53},
numpages = {6},
location = {El Paso, Texas, USA},
series = {STOC '97}
}

@misc{24Wills,
      title={Generalised Coupling and An Elementary Algorithm for the Quantum Schur Transform}, 
      author={Adam Wills and Sergii Strelchuk},
      year={2024},
      eprint={2305.04069},
      archivePrefix={arXiv},
      primaryClass={quant-ph} 
}

@misc{23Grinko,
      title={Gelfand-Tsetlin basis for partially transposed permutations, with applications to quantum information}, 
      author={Dmitry Grinko and Adam Burchardt and Maris Ozols},
      year={2023},
      eprint={2310.02252},
      archivePrefix={arXiv},
      primaryClass={quant-ph}
}

@misc{23Nguyen,
      title={The mixed Schur transform: efficient quantum circuit and applications}, 
      author={Quynh T. Nguyen},
      year={2023},
      eprint={2310.01613},
      archivePrefix={arXiv},
      primaryClass={quant-ph}
}

@article{25Bastidas,
  title = {Unification of finite symmetries in the simulation of many-body systems on quantum computers},
  author = {Bastidas, Victor M. and Fitzpatrick, Nathan and Joven, K. J. and Rossi, Zane M. and Islam, Shariful and Van Voorhis, Troy and Chuang, Isaac L. and Liu, Yuan},
  journal = {Phys. Rev. A},
  volume = {111},
  issue = {5},
  pages = {052433},
  numpages = {27},
  year = {2025},
  month = {May},
  publisher = {American Physical Society},
  doi = {10.1103/PhysRevA.111.052433},
  url = {https://link.aps.org/doi/10.1103/PhysRevA.111.052433}
}

@misc{grinko2023port,
      title={Efficient quantum circuits for port-based teleportation}, 
      author={Dmitry Grinko and Adam Burchardt and Maris Ozols},
      year={2023},
      eprint={2312.03188},
      archivePrefix={arXiv},
      primaryClass={quant-ph}
}

@PHDThesis{25Dmitry,
 title                = {Mixed Schur-Weyl duality in quantum information},
 author               = {Grinko, Dmitry},
 year                 = 2025,
 month                = feb,
 url                  = {http://hdl.handle.net/11245.1/d9e16c26-ef20-40c5-b847-53c13d1a8a1a}
}

@misc{23Fei,
      title={Efficient Quantum Algorithm for Port-based Teleportation}, 
      author={Jiani Fei and Sydney Timmerman and Patrick Hayden},
      year={2023},
      eprint={2310.01637},
      archivePrefix={arXiv},
      primaryClass={quant-ph},
      url={https://arxiv.org/abs/2310.01637}, 
}

@inproceedings{24Fei_QIP,
  author    = {Jiani Fei and Sydney Timmerman and Patrick Hayden},
  title     = {Quantum Algorithm for Reducing Induced Representations with Applications to Port-based Teleportation},
  booktitle = {Quantum Information Processing (QIP 2024)},
  year      = {2024},
  month     = jan,
  address   = {Taipei, Taiwan},
  note      = {Contributed talk, Jan 15, 16:30--17:00},
  url       = {https://www.youtube.com/watch?v=PhoEYpTXHqI},
  howpublished = {Conference program: Afternoon/Parallel Session}
}

@book{simon2023topological,
  title={Topological quantum},
  author={Simon, Steven H},
  year={2023},
  publisher={Oxford University Press}
}

@misc{etingof2011reptheory,
      title={Introduction to representation theory}, 
      author={Pavel Etingof and Oleg Golberg and Sebastian Hensel and Tiankai Liu and Alex Schwendner and Dmitry Vaintrob and Elena Yudovina},
      year={2011},
      eprint={0901.0827},
      archivePrefix={arXiv},
      primaryClass={math.RT},
      url={https://arxiv.org/abs/0901.0827}, 
}

@book{mac1998categories,
  title={Categories for the working mathematician},
  author={Mac Lane, Saunders},
  year={1998},
  publisher={Springer Science \& Business Media},
  url={https://math.mit.edu/~hrm/palestine/maclane-categories.pdf}
}

\end{document}